\documentclass[12pt,authoryear]{elsarticle}
\usepackage{subfigure}
\usepackage{hyperref}
\usepackage[english]{babel}
\usepackage{amssymb}
\usepackage{amsfonts}
\usepackage{dsfont}
\usepackage{amsmath}
\usepackage{multirow}
\usepackage[dvipsnames]{xcolor}
\usepackage{array}

\newcommand{\be}{\begin{equation}}
\newcommand{\ee}{\end{equation}}
\newcommand{\bq}{\begin{eqnarray}}
\newcommand{\eq}{\end{eqnarray}}

\newcommand{\bc}{\begin{center}}
\newcommand{\ec}{\end{center}}
\newcommand{\beq}{\begin{equation}}
\newcommand{\eeq}{\end{equation}}
\newcommand{\bea}{\begin{eqnarray}}
\newcommand{\eea}{\end{eqnarray}}

\newcommand{\ket}[1]{\ensuremath{| #1 \rangle}} 
\newcommand{\bra}[1]{\ensuremath{\langle #1 |}}

\journal{Advances in Atomic, Molecular and Optical Physics}

\begin{document}

\begin{frontmatter}

\title{Laser cooled molecules}

\author{N. J. Fitch}
\ead{n.fitch@imperial.ac.uk}
\author{M. R. Tarbutt}
\ead{m.tarbutt@imperial.ac.uk}
\address{Centre for Cold Matter, Blackett Laboratory, Imperial College London, Prince Consort Road, London SW7 2AZ, United Kingdom}
 
\begin{abstract}
The last few years have seen rapid progress in the application of laser cooling to molecules.  In this review, we examine what kinds of molecules can be laser cooled, how to design a suitable cooling scheme, and how the cooling can be understood and modelled. We review recent work on laser slowing, magneto-optical trapping, sub-Doppler cooling, and the confinement of molecules in conservative traps, with a focus on the fundamental principles of each technique. Finally, we explore some of the exciting applications of laser-cooled molecules that should be accessible in the near term.  
\end{abstract}

\begin{keyword}
laser cooling 
\sep
ultracold molecules
\end{keyword}

\end{frontmatter}

DRAFT: \today

\maketitle

\tableofcontents

\section{Introduction}
\label{sec:intro}

During the last decade, there has been a major effort to extend laser cooling techniques from atoms to molecules. This effort is motivated by a wide variety of interesting applications, which stem from the rotational and vibrational structure of molecules, their large polarizabilities, and their strong coupling to microwave fields. Molecules are already used to measure the electric dipole moments of electrons and protons, study nuclear parity violation, and search for varying fundamental constants. Cooling the molecules to low temperature can increase the number of molecules in these experiments, increase the relevant coherence times, and improve the degree of control over the molecules. All these factors will improve the precision of these tests of fundamental physics. Ordered arrays of molecules, all interacting through long-range dipole-dipole interactions, are well suited to the study of strongly-correlated many-body quantum systems and the emergent phenomena they exhibit, with applications across condensed matter physics, nuclear and particle physics, cosmology, and chemistry. In the field of quantum information processing, the long-lived rotational and vibrational states of molecules make interesting qubits, the dipole-dipole interaction can be used to implement quantum gates, and the strong coupling to microwave photons may be used to interface molecular qubits with photonic or solid-state systems. Ultracold molecules also bring a new dimension to the study of collisions and chemical reactions. In this regime, it becomes possible to engineer the outcome of an encounter through the choice of initial quantum state, the number of partial waves involved, the orientations of the molecules, or by applying external electric or magnetic fields.

In many ways, laser cooling of molecules has followed the route pioneered with atoms decades earlier. First, a beam of molecules was laser-cooled in the transverse dimension~\citep{Shuman2010,Hummon2013}, and then these molecules were decelerated to low speed using radiation pressure~\citep{Barry2012}. Subsequently, magneto-optical trapping was demonstrated for a few molecular species~\citep{Barry2014,Truppe2017b, Anderegg2017, Collopy2018}. Sub-Doppler cooling methods have been used to bring these trapped molecules into the ultracold regime~\citep{Truppe2017b, Anderegg2018, McCarron2018}, reaching temperatures of just a few $\mu$K~\citep{Cheuk2018, Caldwell2019, Ding2020}. Laser-cooled molecules have been confined in magnetic traps~\citep{Williams2018, McCarron2018}, optical dipole traps~\citep{Anderegg2018}, and tweezer traps~\citep{Anderegg2019}, and the coherent control of their internal states has been studied~\citep{Williams2018, Blackmore2018, Caldwell2020}. Recently, collisions between pairs of laser-cooled molecules have been investigated~\citep{Cheuk2020,Anderegg2021}, and the first studies of collisions between laser-cooled atoms and molecules have been reported~\citep{Jurgilas2021}. The methods first developed for diatomic molecules are also now being extended to polyatomic molecules~\citep{Kozyryev2017, Augenbraun2020, Mitra2020, Baum2020}. From this brief overview, we can see that a vibrant field of research has emerged that is now reaching out in many interesting directions.

In this article, we explain how laser cooling and trapping techniques are applied to molecules, review recent progress in the field, and highlight several areas of science where laser cooled molecules are likely to have a significant impact in the near future. We note that there are other methods to produce ultracold molecules which have also been tremendously successful and productive, notably atom  association~\citep{Ni2008}, optoelectric Sisyphus cooling~\citep{Prehn2016}, and electrodynamic slowing and focussing~\citep{Chen2016}. We do not cover those topics in this review. We assume the reader is familiar with the basic elements of molecular structure described in many text books, e.g. \citet{Bransden2003}, and with the principal ideas of laser cooling and trapping of atoms~\citep{Metcalf1999}. Readers may also like to consult other recent reviews on laser cooling of molecules~\citep{McCarron2018b, Tarbutt2018, Isaev2020} and on ultracold molecules more generally~\citep{Krems2009, Carr2009, Quemener2012, Bohn2017,Rios2020}.

\section{Choosing molecules and designing laser cooling schemes}
\label{sec:transition}

In this section, we consider in detail those aspects of molecular structure that are important to the design of a laser cooling scheme.

\subsection{Desirable properties}

In an early work on the topic, \citet{DiRosa2004} considered the criteria for laser cooling to work for molecules. Laser cooling relies on rapid scattering of a large number of photons. This calls for a strong transition with nearly diagonal vibrational branching ratios so that the scattering rate will be high and only a few vibrational branches need to be addressed. It is also desirable to avoid decay to intermediate states, since this complicates laser cooling. Molecules with relatively simple ground-state hyperfine structure tend to be preferable, though this is not a decisive factor since there are many ways to generate the sidebands needed to address multiple hyperfine components.  We note that, while a strong transition is needed to cool molecules from a high initial temperature, narrow transitions can also be useful for cooling to much lower temperatures~\citep{Collopy2015,Truppe2019}, as is often done for atoms such as Yb and Sr. Until recently, molecules with transitions deep in the ultraviolet were thought to be unsuitable for laser cooling, since it is difficult to generate the laser power required. However, laser technology continues to advance rapidly, making an ever increasing range of species accessible, including those with transitions in the ultraviolet.  In this context, interesting examples that are currently being pursued are AlF and AlCl, which have strong transitions, extremely favourable vibrational branching ratios, and large values of the photon recoil momentum.  It is also possible to generate ultraviolet cooling light by using broadband pulsed lasers and addressing the cooling transition with complementary pairs of teeth in an optical frequency comb~\citep{Jayich2016}. 

Table \ref{tab:laser-cooled-molecules} presents some relevant parameters for a selection of molecules amenable to laser cooling. In addition to some basic molecular constants, we have included a parameter $N_{\rm L}^{5.0}$, our estimate of the minimum number of lasers needed to keep a molecule in the cooling cycle in 99.999\% of all photon scattering events.  Similarly, we have included $N_{\rm L}^{100}$, our estimate of the minimum number of lasers needed to scatter enough photons to alter the molecule's velocity by 100~m/s.  We stress that this list is not exhaustive and that new candidate molecules are being identified and characterized at a significant pace, both in terms of their suitability for laser cooling and their potential applications.

\begin{table}[!tb]
    \footnotesize
    \centering
    \def\arraystretch{1.3} 
    \setlength\tabcolsep{3pt} 
    \begin{tabular}{cccccccccc}
        \hline \hline
         \multirow{2}{*}{species} & cycling & $\lambda$  & $\Gamma/2\pi$ & recoil & \multirow{2}{*}{$b_{0,0}$} & \multirow{2}{*}{$N_{\rm L}^{5.0}$} & \multirow{2}{*}{$N_{\rm L}^{100}$} & $|\mu_{\rm{E}}|$ & $B$
         \\
         & transition(s) & (nm) & (MHz) & (cm/s) & & & & (Debye) & (GHz)
         \\ \hline
         \multirow{2}{*}{CaF} & 
         $A^{2}\Pi_{1/2}-X^{2}\Sigma^{+}$ & 
         606 & 8.3 & 1.12 & 0.97 & 4 & 3 & \multirow{2}{*}{3.1} & \multirow{2}{*}{10.27}
         \\
         & $B^{2}\Sigma^{+}-X^{2}\Sigma^{+}$ & 
         531 & 1.27 & 6.3 & 0.998 & 3 & 2 & & 
         \\ \hline
         SrF & 
         $A^{2}\Pi_{1/2}-X^{2}\Sigma^{+}$ & 
         663 & 7 & 0.57 & 0.98 & 3 & 3 & 3.5 & 7.49 
         \\ \hline
         YbF & 
         $A^{2}\Pi_{1/2}-X^{2}\Sigma^{+}$ & 
         552 & 5.7 & 0.38 & 0.93 & 5 & 4 & 3.9 & 7.22 
         \\ \hline
         BaF & 
         $A^{2}\Pi_{1/2}-X^{2}\Sigma^{+}$ & 
         860 & $\sim$3 & 0.30 & 0.95 & 5 & 4 & 3.2 & 6.47 
         \\ \hline
         RaF & 
         $A^{2}\Pi_{1/2}-X^{2}\Sigma^{+}$ & 
         753 & 4  & 0.22 & 0.97 & 5 & 5 & 4.1 & 5.40
         \\ \hline
         \multirow{2}{*}{BaH} & 
         $A^{2}\Pi_{1/2}-X^{2}\Sigma^{+}$ & 
         1061 & 1.17 & 0.27 & 0.988 & 5 & 4 & \multirow{2}{*}{2.7} & \multirow{2}{*}{101.4}
         \\
         & $B^{2}\Sigma^{+}-X^{2}\Sigma^{+}$ & 
         905 & 1.27 & 0.32 & 0.953 & 6 & 5 & &  
         \\ \hline
         AlF & $A^{1}\Pi-X^{1}\Sigma^{+}$ & 
         228 & 84 & 3.81 & 0.996 & 3 & 2 & 1.5 & 16.64
         \\ \hline
         YO & $A^{2}\Pi_{1/2}-X^{2}\Sigma^{+}$ &
         614 & 5.3 & 0.62 & 0.992 & 4 & 4 & 4.5 & 11.61
         \\ \hline
         MgF & $A^{2}\Pi_{1/2}-X^{2}\Sigma^{+}$ 
         & 360 & 22 & 2.56 & 0.998 & 4 & 3 & 3.1 & 15.50 
         \\ \hline
         AlCl & $A^{1}\Pi-X^{1}\Sigma^{+}$ 
         & 262 & 31.8 & 2.44 & 0.999 & 3 & 3 & 1.6 & 7.32 
         \\ \hline
         BH & $A^{1}\Pi-X^{1}\Sigma^{+}$ & 
         433 & 1.3 & 7.81 & 0.986 & 4 & 3 & 1.7 & 354.2 
         \\ \hline
         TlF & $B^{3}\Pi-X^{1}\Sigma^{+}$ & 
         272 & 1.6 & 0.66 & 0.989 & 6 & 5 & 4.2 & 6.69 
         \\ \hline
         \multirow{2}{*}{CH} & 
         $A^{2}\Delta-X^{2}\Pi$ & 
         431 & 0.3 & 7.12 & 0.991 & \multirow{2}{*}{12} & 8 & \multirow{2}{*}{1.5} & \multirow{2}{*}{433.4} 
         \\
         & $B^{2}\Sigma^{-}-X^{2}\Pi$ & 
         389 & 0.4 & 7.89 & 0.902 & & 7 & &
         \\ \hline
         \multirow{2}{*}{SrOH} & $\Tilde{A}^{2}\Pi_{1/2}-\Tilde{X}^{2}\Sigma^{+}$ & 
         688 & \multirow{2}{*}{$\sim$7} & 0.56 & 0.960 & \multirow{2}{*}{6} & \multirow{2}{*}{5} & \multirow{2}{*}{1.9} & \multirow{2}{*}{7.47}
         \\
         & $\Tilde{B}^{2}\Sigma^{+}-\Tilde{X}^{2}\Sigma^{+}$ & 
         611 & & 0.63 & 0.979 & & & 
         \\ \hline
         CaOH & $\Tilde{A}^{2}\Pi_{1/2}-\Tilde{X}^{2}\Sigma^{+}$ & 
         626 & ~1 & 1.12 & 0.957 & 7 & 5 & 1.0 & 10.02
         \\ \hline
         YbOH & $\Tilde{A}^{2}\Pi_{1/2}-\Tilde{X}^{2}\Sigma^{+}$ & 
         577 & 8 & 0.36 & 0.897 & 7 & 6 & 1.9 & 7.35 
         \\ \hline
         CaOCH$_{3}$ & $\Tilde{A}^{2}E_{1/2}-\Tilde{X}^{2}A_{1}$ & 
         630 & $\sim$6 & 0.93 & 0.931 & 9 & 7 & 1.6 & 3.49 
         \\ \hline \hline
    \end{tabular}
    \caption{Relevant parameters for some of the molecule species that have been laser cooled or have been discussed in the literature in this context.  The columns correspond to the main laser-cooling transition, its wavelength, natural linewidth, the recoil velocity when a cooling photon is scattered, the $v'=0 \rightarrow v=0$ branching ratio, our estimate of the minimum number of cooling lasers needed to reduce optical leaks to below $10^{-5}$ (exceed 99.999\% cycling closure), our estimate of the minimum number of cooling lasers needed to alter the molecule's velocity by 100~m/s, the electric dipole moment ($\mu_{\rm{E}}$) in the molecule-fixed frame, and the rotational constant ($B$) in the ground rovibronic state.  The $N_{\rm L}$ value(s) for CH include the multiple lasers needed to close the rotational structure, as discussed in section~\ref{sec:rotationalTransitions}.}
    \label{tab:laser-cooled-molecules}
\end{table}

\subsection{Notation for molecular structure}

Molecular structure is discussed in detail in several text books, e.g. \citet{BrownCarrington2003}. Here, we briefly outline some salient points and introduce our notation. We use the following angular momentum operators: $\vec{L}$ is the total orbital angular momentum of the electrons, $\vec{S}$ is the total electron spin, $\vec{I}=\sum_i \vec{I}_i$ is the total nuclear spin and $\vec{I}_i$ is the spin of nucleus $i$, $\vec{R}$ is the rotational angular momentum of the molecule, $\vec{N}=\vec{L}+\vec{R}$ is the total angular momentum neglecting spin, $\vec{J} = \vec{N} + \vec{S}$ is the total angular momentum apart from nuclear spin, and $\vec{F} = \vec{J} + \vec{I}$ is the total angular momentum. The quantum numbers for the projections of $\vec{L}$, $\vec{S}$ and $\vec{J}$ onto the internuclear axis are $\Lambda$, $\Sigma$ and $\Omega$, respectively. The quantum number for the projection of angular momentum $X$ onto the space-fixed $z$-axis is $m_X$. We use $p$ to denote the parity. Primes are attached to quantum numbers of excited states. Molecular term symbols,\footnote{In this work, we consider only heteronuclear molecules.  Homonuclear molecules have an extra reflection symmetry, which is denoted as an additional $g/u$ subscript.} used to denote a particular electronic state, are expressed as $^{2S+1}\Lambda_{\Omega}^{+/-}$ where $\Lambda = \{0,1,2,\ldots\}$ are denoted as $\{\Sigma,\Pi,\Delta,\ldots\}$ respectively and the $+/-$ superscript, used only for $\Sigma$ states, gives the reflection symmetry through a plane containing the internuclear axis.\footnote{Electronic states with $\Lambda=0$ ($\Sigma$ states) should not be confused with the projection of $\vec{S}$ onto the internuclear axis, also denoted $\Sigma$.  The distinction should be clear by context.}  Electronic states are conventionally given an additional single-letter label where $X$ denotes the ground state, e.g. $X^{2}\Sigma^{+}$.  Excited states of the same (different) spin multiplicity as the ground state are labelled $A,B,\ldots$ ($a,b,\ldots$), nominally in order of ascending energy.  For polyatomic molecules, the letter label is usually decorated as $\tilde{A},\tilde{B},\ldots$ to avoid confusion with symmetry groups of the same notation.  We note that these conventions are only loosely followed and that the alphabetical labelling sequence often goes awry, especially for heavy molecules.

Recall that in the Born-Oppenheimer approximation the energy eigenfunctions of a diatomic molecule are written as a product of an electronic part $\phi_{n}(\vec{R}_N;\{\vec{r}_i\})$, a vibrational part $\frac{1}{R_N}f_v(R_N)$, and a rotational part described by the spherical harmonic functions $Y_{R,m_R}(\Theta,\Phi)$:
\begin{equation}
    \psi_{n,v,R,m_R}(\vec{R}_N;\{\vec{r}_i\}) = \frac{1}{R_N} 
    \, 
    \phi_n(\vec{R}_N;\{\vec{r}_i\}) 
    \, 
    f_v(R_N) 
    \,
    Y_{R,m_R}(\Theta, \Phi).
   \label{eqn:BO_wavefunction}
\end{equation}
Here, $n$ is the set of quantum numbers labelling the electronic state, $v$ labels the vibrational state, the vector $\vec{R}_N$ describes the relative displacement of the nuclei and has spherical polar components $(R_N,\Theta,\Phi)$, and we use $\{\vec{r}_i\}$ to refer to the set of all electronic coordinates. The $\phi_n$ satisfy the electronic Schr\"odinger equation obtained by fixing the nuclei in place, so is a function of the $\{\vec{r}_i\}$ and depends on $\vec{R_N}$ as a parameter. 

In a molecule, the various angular momenta are often coupled together, and the most appropriate choice of basis states depends on the relative coupling strengths. In the limits where certain couplings dominate the interaction, we obtain the well-known Hund's coupling cases~\citep{HerzbergI}.  For a state with $\Lambda \ne 0$ and $S \ne 0$, the Hamiltonian describing the spin-orbit interaction and the rotational energy is $H_{\rm rso} = A \vec{L}\cdot\vec{S} + B \vec{R}^2$. When $A \gg B$, the state splits into spin-orbit manifolds of well-defined $\Omega$, each of which has a set of rotational levels labelled by $J$. This is Hund's case (a), and we write the eigenstates using the notation $|n,v,J,\Omega,m_J\rangle$. When $A \ll B$, the state splits into rotational levels labelled by $N$, each of which is then split into spin-orbit levels labelled by $J$. This is Hund's case (b) and we write the eigenstates using the notation $|n,v,N,J,m_J\rangle$. For $\Sigma$ states with $S \ne 0 $ there is no spin-orbit interaction but there is an equivalent spin-rotation interaction, $\gamma \vec{N}\cdot\vec{S}$, which again splits rotational states into levels labelled by $J$. When $\Lambda \ne 0$, the states of definite parity are linear combinations of $+\Lambda$ and $-\Lambda$. Due to their interaction with $\Sigma$ states, these parity eigenstates are not quite degenerate. This is $\Lambda$-doubling.

\subsection{Transition strengths and selection rules}

The intensity of an electric dipole transition between initial and final states is proportional to $\omega_{if}^3|\vec{d}_{if}|^2$, where $\omega_{if}$ is the transition frequency and
\begin{equation}
    \vec{d}_{if} = \langle n',v',J',\Omega',m_J' | \vec{d} | n,v,J,\Omega,m_J \rangle.
    \label{eqn:transitionDipole}
\end{equation}
Here, $\vec{d}$ is the dipole moment operator, which we can write as $\vec{d} = \sum_{p} d_p\hat{e}_p^*$ where $\hat{e}$ are the spherical basis vectors. 

To evaluate this matrix element, it is helpful to rotate the coordinate system so that the axes align with those of the molecule (for a diatomic molecule, the new $z$ axis will lie along the internuclear axis). We introduce the dipole moment operator in this rotated frame, $\vec{\mu}$, which is related to $\vec{d}$ by
\begin{equation}
    d_p = \sum_{q=-1}^1 D^{(1)*}_{p,q}(\Theta,\Phi)\mu_q.
\end{equation}
Here, $p$ and $q$ refer to spherical tensor components and $D^{(1)}$ is the rotation operator. This gives us
\begin{equation}
    \vec{d}_{if} = \sum_p \hat{e}_p^* \sum_q M_{p,q}\langle n',v'| \mu_q |n,v\rangle
    \label{eqn:transitionDipoleFactorized}
\end{equation}
where 
\begin{equation}
    M_{p,q}=\langle J',\Omega',m_J' | D^{(1)*}_{p,q} | J,\Omega,m_J \rangle
    \label{eqn:rotationalME}
\end{equation}
depends only on angular coordinates and dictates the angular momentum selection rules, and
\begin{equation}
    \langle n',v'| \mu_q |n,v\rangle = \int \left(\frac{1}{R_N} \phi_{n'}^*f_{v'}^*\right)
    \, \mu_q \, 
    \left(\frac{1}{R_N}\phi_{n}f_v\right)
    \,
    R_N^2 
    \, 
    dR_N d\vec{r}_1\,d\vec{r}_2...d\vec{r}_n.
    \label{eqn:electronicME}
\end{equation}
The integral form in equation (\ref{eqn:electronicME}) makes the coordinate dependence clear and highlights that $\langle n',v'| \mu_q |n,v\rangle$ does not depend on $\Theta$ and $\Phi$. Next, we introduce the quantity
\begin{equation}
    \mu_{q}^{n',n}(R_N) = \langle n' | \mu_{q} | n \rangle = \int \, \phi_{n'}^*(\vec{R}_N;\{\vec{r}_i\}) \, 
    \mu_q
    \, 
    \phi_{n}(\vec{R}_N;\{\vec{r}_i\})
    \,
    d\vec{r}_1\,d\vec{r}_2...d\vec{r}_n.
    \label{eqn:dipoleFunction}
\end{equation}
When $n'=n$, this is the dipole moment for state $n$, which we write as $\mu_{q}^{n}(R_N)$ without the second superscript. Its value at the equilibrium separation, $R_0$, is usually called the permanent dipole moment of the molecule in state $n$. When $n'\ne n$, equation (\ref{eqn:dipoleFunction}) is the transition dipole moment function for the transition between $n$ and $n'$.  The transition dipole moment can be calculated using a variety of quantum chemistry techniques, e.g. \citet{Hickey2014}, or can be determined experimentally by measuring the transition rate. 

Equation (\ref{eqn:transitionDipoleFactorized}) can now be written as
\begin{align}
    \vec{d}_{if} &= \sum_p \hat{e}_p^*\sum_q M_{p,q} \int f_{v'}^*(R_N) \mu_q^{n',n}(R_N) f_{v}(R_N) dR_N.
    \label{eqn:d_if_full}
\end{align}
We can expand $\mu_q^{n',n}$ around the equilibrium separation $R_0$ giving
\begin{equation}
    \mu_q^{n',n}(R_N) \simeq \mu_q^{n',n}(R_0) + \left.\frac{d \mu_q^{n',n}}{d R_N}\right|_{R_0}(R_N-R_0) + \ldots.
    \label{eqn:transitionDipoleExpansion}
\end{equation}
The first term usually dominates in this expansion, in which case we have
\begin{equation}
    \vec{d}_{if} \simeq \sum_p \hat{e}_p^*\sum_q M_{p,q} \times \mu_q^{n',n}(R_0) \times \int f_{v'}^* f_{v} dR_N.
    \label{eqn:d_if}
\end{equation}
The last factor in equation (\ref{eqn:d_if}) is the overlap integral between vibrational state $v'$ in electronic state $n'$ and vibrational state $v$ in electronic state $n$. The square of this overlap integral,
\begin{equation}
    q_{v',v} = \left|\int f_{v'}^* f_{v} dR_N\right|^2,
    \label{eqn:FCfactor}
\end{equation}
is the Franck-Condon factor.

\subsection{Vibrational branching ratios}

\begin{figure}[!tb]
\centering
\includegraphics[width=0.7\textwidth]{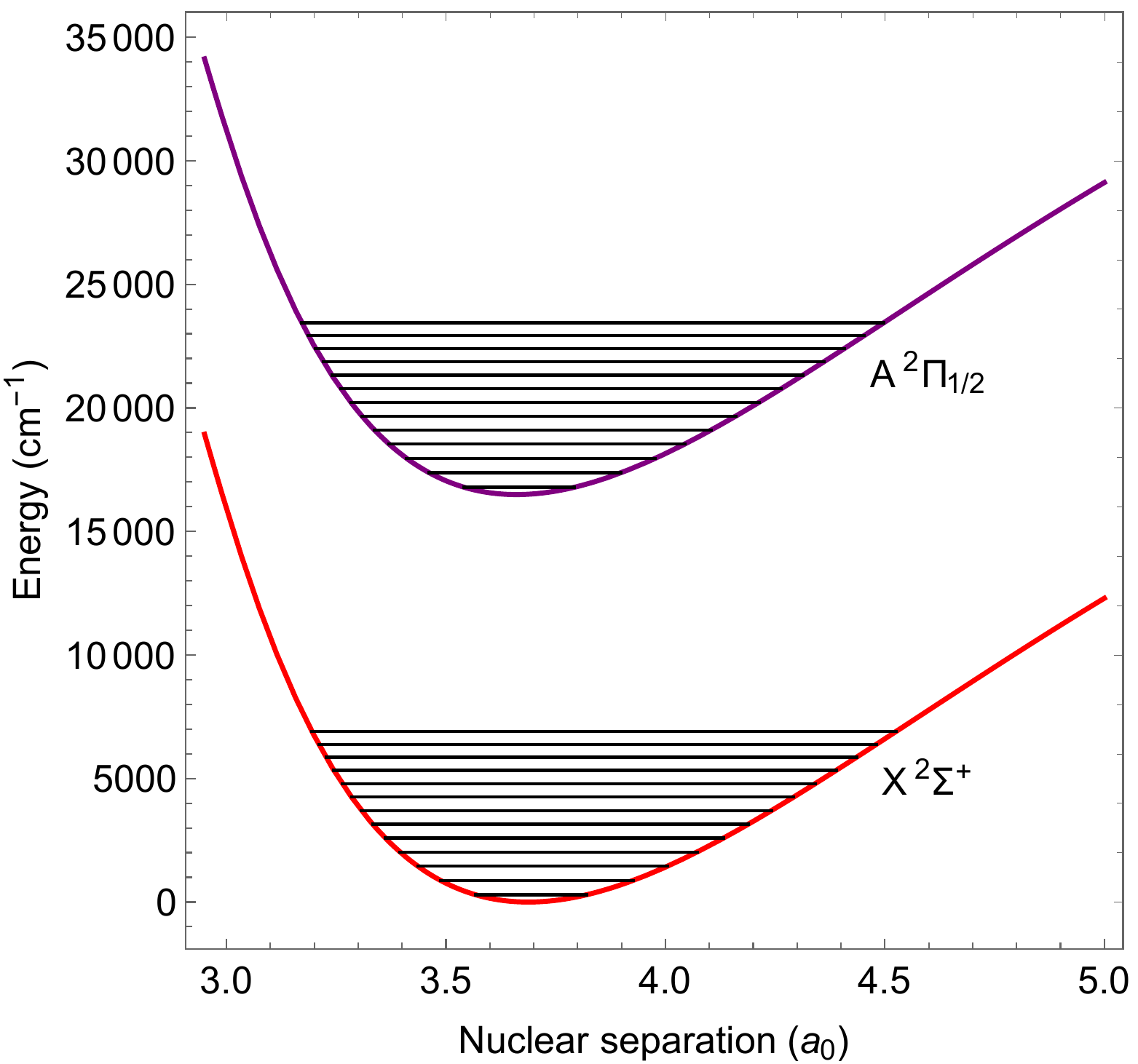}
\caption{RKR potentials for the $X ^{2}\Sigma^{+}$ and $A ^{2}\Pi_{1/2}$ states of CaF, calculated using the molecular constants given in \citet{Kaledin1999}. The low-lying vibrational states are also shown.}
\label{fig:CaF_RKRl}
\end{figure}

To a good approximation, the vibrational branching ratio for the transition from $v'$ to $v$ when the molecule decays from $n'$ to $n$ is
\begin{equation}
    b_{v',v} = \frac{q_{v',v} \omega_{v',v}^3}{\sum_{v''}\,q_{v',v''} \omega_{v',v''}^3}
    \label{eqn:vibrational_branching}
\end{equation}
where $\omega_{v',v''}$ is the transition frequency. However, it is worth noting that the terms in equation (\ref{eqn:transitionDipoleExpansion}) beyond the first can sometimes alter the branching ratios significantly. Laser cooling is easier when the number of vibrational branches that need to be addressed is small. For most experiments, the leak out of the cooling cycle must be reduced below $10^{-5}$ or $10^{-6}$. The best systems have $b_{v',v=v'}\approx 1$, with branching ratios that diminish very rapidly as $|v-v'|$ increases. This occurs when the potentials for the two electronic states are nearly the same. Examples include the alkaline earth monofluorides and monohydroxides, where the valence electron is localized on the metal atom and has little influence on the bonding.

The branching ratios between low-lying vibrational states, where anharmonic effects are small, can be estimated using a harmonic oscillator approximation. Here, harmonic oscillator eigenfunctions are used to evaluate the overlap integral in equation (\ref{eqn:FCfactor}). For each electronic state, the eigenfunctions are characterised by just two parameters, the harmonic frequency $\omega_{\rm e}$, and the equilibrium separation $R_0$. The first is identical to the vibrational constant, and the second can be found from the rotational constant $B_{\rm e} = \hbar^2/(2m_{\rm r}R_0^2)$ where $m_{\rm r}$ is the reduced mass. When $B_{\rm e}$ is similar for the two electronic states, and so too is $\omega_{\rm e}$, the branching ratios will be favourable. Table \ref{tab:branchingRatios} shows the branching ratios calculated for the $A^{2}\Pi_{1/2}(v') \rightarrow X^{2}\Sigma^{+}(v)$ transitions of CaF. The first (top) entry in each case is the value calculated using the harmonic oscillator approximation. The values on the diagonals are close to 1, and the branching ratios diminish rapidly away from the diagonals, showing that this is a favourable system for laser cooling.

\begin{table}[!tb]
\small
\begin{center}
\begin{tabular}{c | c  c  c  c  c }
\hline\hline
 & $v=0$ & $v=1$ & $v=2$ & $v=3$ & $v=4$ \\
 \hline
 & $9.80\times 10^{-1}$ & $1.96\times 10^{-2}$ & $1.18\times 10^{-4}$ & $1.42\times 10^{-7}$ & $2.27\times 10^{-10}$ \\ 
 $v'=0$  & $9.80\times 10^{-1}$ & $1.95\times 10^{-2}$ & $6.00\times 10^{-4}$ & $2.22\times 10^{-5}$ & $9.60\times 10^{-7}$ \\
 & $9.77\times 10^{-1}$ & $2.21\times 10^{-2}$ & $7.15\times 10^{-4}$ & $2.75\times 10^{-5}$ & $1.22\times 10^{-6}$ \\
 \hline
 & $2.39\times 10^{-2}$ & $9.37\times 10^{-1}$ & $3.86\times 10^{-2}$ & $3.53\times 10^{-4}$ & $5.75\times 10^{-7}$ 
 \\
 $v'=1$ & $2.49\times 10^{-2}$ & $9.37\times 10^{-1}$ & $3.68\times 10^{-2}$ & $1.70\times 10^{-3}$ & $8.42\times 10^{-5}$ \\
 & $2.17\times 10^{-2}$ & $9.34\times 10^{-1}$ & $4.19\times 10^{-2}$ & $2.04\times 10^{-3}$ & $1.05\times 10^{-4}$ \\
 \hline
 & $4.37\times 10^{-4}$ & $4.65\times 10^{-2}$ & $8.95\times 10^{-1}$ & $5.69\times 10^{-2}$ & $7.03\times 10^{-4}$ 
 \\
 $v'=2$ & $1.59\times 10^{-5}$ & $4.86\times 10^{-2}$ & $8.96\times 10^{-1}$ & $5.21\times 10^{-2}$ & $3.22\times 10^{-3}$ \\
 & $4.95\times 10^{-6}$ & $4.24\times 10^{-2}$ & $8.94\times 10^{-1}$ & $5.95\times 10^{-2}$ & $3.87\times 10^{-3}$ \\
  \hline
 & $6.60\times 10^{-6}$ & $1.28\times 10^{-3}$ & $6.78\times 10^{-2}$ & $8.55\times 10^{-1}$ & $7.47\times 10^{-2}$ 
 \\
 $v'=3$ & $4.29\times 10^{-7}$ & $4.19\times 10^{-5}$ & $7.11\times 10^{-2}$ & $8.58\times 10^{-1}$ & $6.56\times 10^{-2}$ \\
 & $3.89\times 10^{-7}$ & $1.19\times 10^{-5}$ & $6.20\times 10^{-2}$ & $8.56\times 10^{-1}$ & $7.52\times 10^{-2}$ \\
  \hline
 & $8.88\times 10^{-8}$ & $2.59\times 10^{-5}$ & $2.51\times 10^{-3}$ & $8.79\times 10^{-2}$ & $8.16\times 10^{-1}$ 
 \\
 $v'=4$ & $1.16\times 10^{-9}$ & $1.72\times 10^{-6}$ & $7.28\times 10^{-5}$ & $9.26\times 10^{-2}$ & $8.23\times 10^{-1}$ \\
 & $1.87\times 10^{-9}$ & $1.55\times 10^{-6}$ & $1.84\times 10^{-5}$ & $8.08\times 10^{-2}$ & $8.21\times 10^{-1}$ \\
 \hline\hline
\end{tabular}
\end{center}
\caption{Vibrational branching ratios, $b_{v',v}$ for the $A ^{2}\Pi_{1/2}(v') \rightarrow X ^{2}\Sigma^{+}(v)$ transitions of CaF. There are three entries for each value. The first is calculated using the harmonic oscillator approximation, the second using the RKR potential and only the first term in equation (\ref{eqn:transitionDipoleExpansion}), and the third using the RKR potential and the transition dipole moment function, $\mu^{n',n}(R_N)$ given in \cite{Pelegrini2005}. }
\label{tab:branchingRatios}
\end{table}

A more accurate set of branching ratios can be determined using Rydberg-Klein-Rees (RKR) potentials. The RKR potential is constructed from measured band spectra, which are typically parameterized using a small number of molecular constants. The procedure, as described for example by \citet{Rees1947} and \cite{LeRoy2016}, is straightforward to implement. As an example, figure~\ref{fig:CaF_RKRl} shows the RKR potentials for the $X^{2}\Sigma^{+}$ and $A^{2}\Pi_{1/2}$ states of CaF, calculated using the molecular constants given in \citet{Kaledin1999}. The eigenfunctions of the Hamiltonian can then be calculated numerically for these potentials, followed by the overlap integral of equation (\ref{eqn:FCfactor}). The second entries in the set of branching ratios given in table \ref{tab:branchingRatios} are the values calculated using this procedure. Comparing to the harmonic oscillator results, we see that the fractional change is small for the larger values of $b_{v',v}$, but sometimes very large for the small values. For example, $b_{0,3}$ increases by two orders of magnitude and becomes non-negligible for laser cooling. This highlights the importance of using accurate potentials to estimate branching ratios, or measuring the branching ratios directly, in designing any laser cooling scheme. If the $R_N$-dependence of the transition dipole moment is known, even higher accuracy can be obtained for the branching ratios by evaluating the integral appearing in equation (\ref{eqn:d_if_full}) instead of its approximate form. The third entries in table \ref{tab:branchingRatios} are obtained using this approach. In this example, the inclusion of the $R_N$-dependence of $\mu^{n',n}$ makes relatively minor changes to most of the branching ratios. 

\subsection{Closed rotational transitions}
\label{sec:rotationalTransitions}

The angular factor in equation (\ref{eqn:d_if_full}) is
\begin{align}
     M_{p,q}&=\langle J',\Omega',m_J' | D^{(1)*}_{p,q} | J,\Omega,m_J \rangle \nonumber\\
     &= (-1)^{2 J'-m_J'-\Omega'}\sqrt{(2J+1) (2J'+1)}  \nonumber\\
     &\quad \times 
     \left(
    \begin{array}{ccc}
    J' & 1 & J \\
    -m_J' & p & m_J \\
    \end{array}
    \right) 
    \left(
    \begin{array}{ccc}
    J' & 1 & J \\
    -\Omega ' & q & \Omega  \\
    \end{array}
    \right).
\label{eqn:Mpq}
\end{align}
This equation gives the relative intensities of rotational branches between case (a) basis states. A similar expression can be written down for case (b) basis states~\citep{BrownCarrington2003}. We see from equation (\ref{eqn:Mpq}) that the selection rule on $J$ is $\Delta J=J-J'=0,\pm 1$, except in the case where $\Omega'=\Omega=0$ where it is  restricted to $\Delta J = \pm 1$. In addition, we have the usual selection rule that parity must change in an electric dipole transition. 

\begin{figure}[!tb]
\centering
\includegraphics[width=\textwidth]{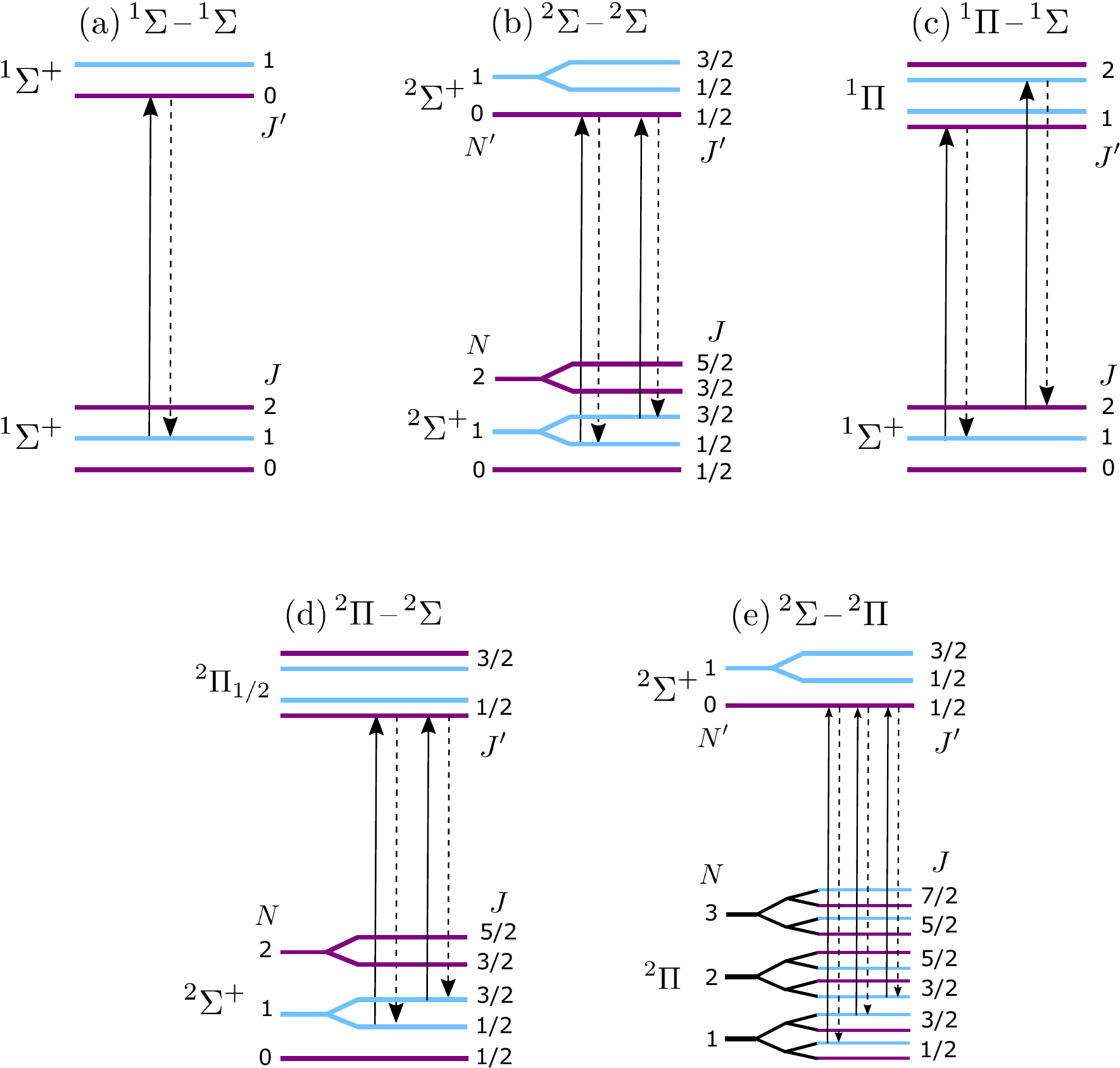}
\caption{Rotationally-closed components of various transitions, including excitations (solid vertical arrows) and decays (dashed vertical arrows).  Energy levels are colour coded according to their parity $p$, being light blue for negative parity ($p=-1$) and dark purple for positive parity ($p=1$). (a) $^1\Sigma-{}^1\Sigma$. (b) $^2\Sigma-{}^2\Sigma$. Both components shown have to be addressed. (c) $^1\Pi-{}^1\Sigma$. Either of the transitions indicated, or any of the others in the same sequence, can be used. (d) $^2\Pi-{}^2\Sigma$. Both components shown have to be addressed. (e) $^2\Sigma-{}^2\Pi$. All three components shown have to be addressed, unless $A \ll B$ in which case the rightmost transition is forbidden.}
\label{fig:rotationalBranching}
\end{figure}

The selection rule on $J$ implies that an excited state can decay on up to three rotational branches. However, a careful choice of transition can limit this to just a single rotational branch~\citep{Stuhl2008}. Figure~\ref{fig:rotationalBranching} shows how to avoid rotational branching for several electronic transitions. The simplest is a $^1\Sigma-{}^1\Sigma$ transition, as shown in (a). Each rotational level of the upper state can decay to two levels in the lower state, following the $\Delta J = \pm 1$ selection rule. The exception is $J'=0$, which can only decay to $J=1$, so the transition $^1\Sigma(J'=0)-{}^1\Sigma(J=1)$ is rotationally closed. Figure~\ref{fig:rotationalBranching}(b) shows a $^2\Sigma-{}^2\Sigma$ transition. Here, the spin-rotation interaction splits each rotational level (apart from $N=0$) into a pair of levels with $J=N \pm 1/2$. The laser cooling transition is the same as in (a), but the excited state ($J'=1/2$) can decay to both the $J=1/2$ and $J=3/2$ components of $N=1$. Both components have to be addressed by the laser light, but the spin-rotation splitting is usually small enough that this can easily be done by adding an rf sideband to the laser. An example, already used for laser cooling~\citep{Truppe2017}, is the $B^{2}\Sigma^{+}-X^{2}\Sigma^{+}$ transition of CaF.

The situation is different for a $^1\Pi-{}^1\Sigma$ transition, illustrated in figure~\ref{fig:rotationalBranching}(c). Here, $\Lambda$-doubling in the upper state leads to a pair of states of opposite parity for each value of $J'$. For every $J'$, one parity component can decay to two lower levels, while the other can only decay to a single lower level. It follows that the $^1\Pi(J'=J)-{}^1\Sigma(J)$ transitions (i.e. Q-branch transitions) are rotationally closed for all $J$. Thus, all lower levels other than $J=0$ could be used for laser cooling. AlCl and AlF are good examples of laser-coolable molecules with this structure, and experiments to cool these species are currently being built~\citep{Truppe2019,Hemmerling2020}. Note that heavy molecules such as TlF can be cooled using a $^3\Pi_1-{}^1\Sigma$ transition~\citep{Hunter2012}, which has the same structure as the $^1\Pi-{}^1\Sigma$ case. For lighter molecules such as AlF, the $^3\Pi_1$-- $^1\Sigma$ transition could be used for narrow-line cooling to temperatures near the recoil limit~\citep{Truppe2019}.

Figure~\ref{fig:rotationalBranching}(d) shows a $^{2}\Pi$ --$^{2}\Sigma$ transition. Here, we have assumed that the spin-orbit splitting of the excited state is much larger than the rotational splitting ($A\gg B$), and we focus on the $^{2}\Pi_{1/2}$ manifold that is of most interest for laser cooling. $\Lambda$-doubling splits the rotational levels of $^{2}\Pi_{1/2}$ into states of opposite parity, and the spin-rotation interaction splits the rotational levels of $^{2}\Sigma$ into pairs of the same parity. The transition $^{2}\Pi_{1/2}(J'=1/2,p=-1)-{}^{2}\Sigma^{+}(N=1)$ is rotationally closed. As before, both spin-rotation components of the transition have to be addressed. All the molecules laser cooled so far have this structure.

As a final example, figure~\ref{fig:rotationalBranching}(e) shows a $^{2}\Sigma-{}^{2}\Pi$ transition. For convenience, we have used a case (b) notation where rotational levels, distinguished by $N$, are split by the spin-orbit interaction into pairs labelled by $J$, and these are split further into opposite parity $\Lambda$-doublets. In this case, the transition $^{2}\Sigma(N'=0)-{}^{2}\Pi(N=1)$ is rotationally closed, provided both spin-orbit components are addressed. However, $^{2}\Pi$ states are often closer to case (a) states, or have similar values of $A$ and $B$ so do not conform to either coupling case. In this case, the transition to the state labelled as $(N=2,J=3/2)$ is allowed and must also be addressed for laser cooling to work. This case is awkward because the splittings are typically too large to be addressable with modulators, calling for three separate lasers for each vibrational branch. The $B^{2}\Sigma^{-}-X^{2}\Pi$ transition of CH is an interesting example. The ground state has $A/B = 1.984$ and the branching ratios for the three transitions shown in figure~\ref{fig:rotationalBranching}(e) are in the ratio (from left to right) 0.333:0.623:0.044. The weak branch falls below $10^{-3}$ for $A/B < 0.35$ and below $10^{-4}$ for $A/B < 0.11$.
 
\subsection{Hyperfine structure}
\label{sec:hyperfine}

An understanding of the hyperfine structure is important for designing a laser cooling scheme. The hyperfine states can also be a useful resource for controlling collisions and for quantum simulation and information processing. 

\subsubsection{Hyperfine interactions}

The relevant terms in the hyperfine Hamiltonian depend on the electronic state and on the nuclear spins. A thorough treatment of these terms together with expressions for their matrix elements can be found in \citet{BrownCarrington2003}. The main terms, and the ones most relevant to our discussion, are

\begin{subequations}
\label{eq:HypHam}
\begin{align}
     H_{SN} &= \gamma \vec{S}\cdot\vec{N}, \label{eq:HSN}\\
     H_{IL} &= \sum_i a^i \vec{I_i} \cdot \vec{L}, \label{eq:HIL}\\
     H_{IS} &= \sum_i b_{\rm F}^i \vec{I_i}\cdot\vec{S} + t^i \sqrt{6}T^2(C)\cdot T^2(\vec{I_i},\vec{S}), \label{eq:HIS} \\
     H_{IN} &= \sum_i c_{I}^i \vec{I_i}\cdot\vec{N}, \label{eq:HIN}\\
     H_{Q} &= \sum_i -e T^{2}(\nabla \vec{E}) \cdot T^{2}({\bf Q}_i), \label{eq:HQ}\\
     H_{II} &= c_4 \vec{I_1}\cdot\vec{I_2} - c_3 \sqrt{6}T^2(C)\cdot T^2(\vec{I_1},\vec{I_2})\label{eq:HII}.
\end{align}
\end{subequations}
The summation is over the nuclei.  $H_{SN}$ is the electron spin-rotation interaction. It is not strictly a hyperfine interaction since it does not involve the nuclear spins, but it is often of a similar magnitude to other terms in the hyperfine Hamiltonian so must be treated together with them. $H_{IL}$ is the orbital hyperfine interaction describing the interaction of the nuclear magnetic moments with the magnetic field at the nuclei produced by the orbital motion of the electrons. It is relevant for electronic states with $\Lambda>0$. The interaction between the electron and nuclear magnetic moments, $H_{IS}$  has two parts, the Fermi contact part and the electron-nuclear spin dipolar interaction. These terms are sometimes expressed using the parameters $b$ and $c$ introduced by \citet{Frosch1952}, the relation being $b_{\rm F} = b +c/3$ and $t=c/3$. The dipolar interaction in equation (\ref{eq:HIS}) is written as the scalar product of two rank-2 spherical tensors; $T^2(\vec{I},\vec{S})$ is the one formed from $\vec{I}$ and $\vec{S}$, while $T^2(C)$ is a spherical tensor whose components are the renormalised spherical harmonics $C_q^2 = \sqrt{4 \pi/5}\,Y_{2,q}(\Theta, \Phi)$.  $H_{IN}$ is the nuclear spin-rotation interaction. $H_{Q}$ is the interaction between the electric quadrupole moments of the nuclei and the electric field gradient at the nuclei. For each nucleus, two parameters appear in the matrix elements of $H_{Q}$, $e q_0 Q$ and $e q_2 Q$. Here, $e Q$ is the nuclear quadrupole moment, $q_0$ is the electric field gradient in the direction of the internuclear axis, and $q_2$ is the gradient in the perpendicular direction and is only relevant for states with $\Lambda > 0$. Finally, $H_{II}$ represents the tensor and scalar interactions between the nuclear dipole moments associated with the two nuclear spins $I_1$ and $I_2$.

\subsubsection{Examples of hyperfine structure}

\begin{figure}[!tb]
\centering
\includegraphics[width=\textwidth]{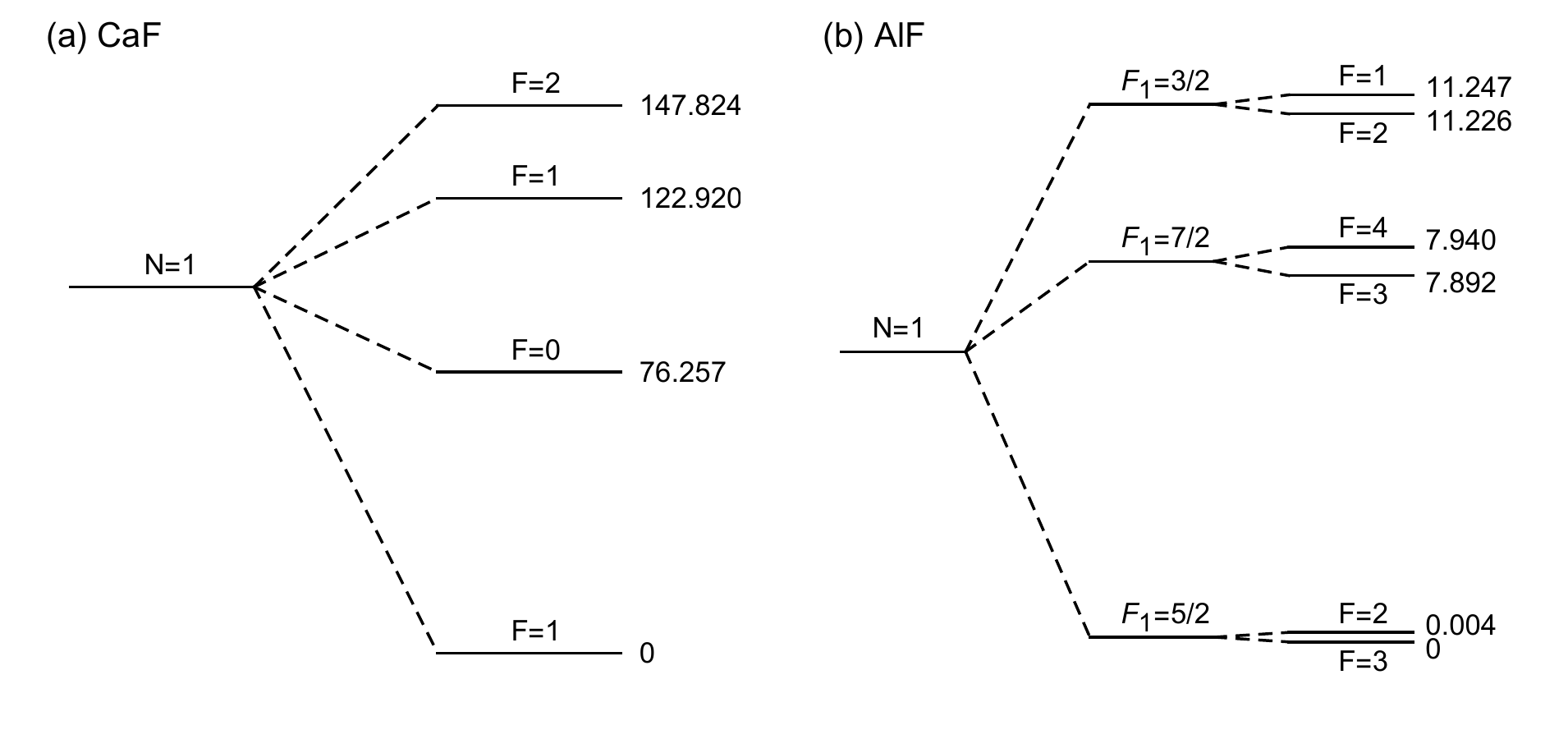}
\caption{Hyperfine structure in the $N=1$ levels of (a) CaF $X^2\Sigma^{+}(v=0)$ and (b) AlF $X^1\Sigma^{+}(v=0)$. The energies are given in MHz relative to the lowest hyperfine level.}
\label{fig:hyperfine}
\end{figure}

Now let us consider some relevant examples. We first look at a $^2\Sigma$ diatomic molecule having one nucleus of zero spin and the other with spin $I=1/2$. This is the structure of ground-state alkaline-earth monohydrides and monofluorides. The hyperfine Hamiltonian is
\begin{equation}
    H_{\rm hyp} = H_{SN} + H_{IS} + H_{IN}.
    \label{eq:hypDoubletSigma}
\end{equation}
Figure~\ref{fig:hyperfine}(a) shows the example of CaF in the state $X ^2\Sigma^+(v=0,N=1)$, which is the lower level of the main laser cooling transition. To calculate the energies, we have used equation (\ref{eq:hypDoubletSigma}) and the parameters from \citet{Childs1981}. The spin-rotation interaction splits $N=1$ into two levels with $J=1/2$ and $J=3/2$, and these are further split by the interaction with the nuclear spin, resulting in the four hyperfine levels shown. Since $H_{SN}$ and $H_{IS}$ have similar magnitudes, the two $F=1$ states are mixtures of $J=1/2$ and 3/2.

Our second example is a $^{1}\Sigma$ diatomic molecule. Here, because there is no electron spin, the hyperfine structure is much smaller and is typically dominated by the electric quadrupole interaction. The hyperfine Hamiltonian is
\begin{equation}
    H_{\rm hyp} = H_{Q} + H_{IN} + H_{II}.
    \label{eq:hypSingletSigma}
\end{equation}
Figure~\ref{fig:hyperfine}(b) shows the example of AlF in the state $X ^1\Sigma^+(v=0,N=1)$, the lower level of the laser cooling transition. To calculate the energies, we have used equation (\ref{eq:hypSingletSigma}) and the parameters from \citet{Truppe2019}. The main splitting is due to the electric quadrupole moment of the Al nucleus, which has $I_1=5/2$. This splits $N=1$ into three levels with $F_1=3/2,5/2,7/2$, where we have introduced the intermediate angular momentum $\vec{F_1} = \vec{N} + \vec{I}_1$. The contribution to this splitting from the term $\vec{I_1}.\vec{N}$ is very much smaller. The F nucleus has no quadrupole moment, since it has $I_{2}=1/2$. Each $F_1$ level is split in two by the term $\vec{I_2}.\vec{N}$, resulting in the structure shown. Because these latter splittings are much smaller, all the states can be labelled by $F_1$ and $F$. The contribution of $H_{II}$ is even smaller and is most relevant for the ground rotational state, $N=0$, where both $H_{Q}$ and $H_{IN}$ are absent.

For laser cooling, all hyperfine components of the rotationally closed transition need to be addressed for each vibrational branch. Because the hyperfine splittings are fairly small, this can usually be done by using acousto-optic and/or electro-optic modulators to add the relevant sidebands to the lasers. In cases where the splitting exceeds 1~GHz this becomes more difficult and it may be necessary to use separate lasers.

\subsubsection{Hyperfine-induced transitions}

The electron-nuclear spin dipolar interaction in $H_{IS}$ has matrix elements coupling states of the same $F$ in rotational states $N$ and $N \pm 2$. This can result in a small leak out of the cooling cycle. To estimate the size of this leak, consider again the case of a $^{2}\Sigma$ molecule with a single nuclear spin $I=1/2$. The state $\ket{N,J,F}=\ket{3,5/2,2}$ acquires a small admixture of $\ket{1,3/2,2}$ due to the hyperfine interaction,
\begin{equation}
    \ket{3,5/2,2} = \ket{3,5/2,2}_0 + \epsilon \ket{1,3/2,2}_0,
\end{equation}
where the subscript 0 indicates the unperturbed state and
\begin{equation}
    \epsilon = \frac{_0\bra{1,3/2,2}H_{IS}\ket{3,5/2,2}_0}{E_{3}-E_{1}}=\frac{3}{50}\sqrt{\frac{3}{2}} \frac{t}{B}.
\end{equation}
Here, $E_N$ is the unperturbed energy of rotational manifold $N$. The probability of decaying to $\ket{3,5/2,2}$ is
$\epsilon^2 = \frac{27}{5000}\left(\frac{t}{B}\right)^2$ times the probability of decaying to $\ket{1,3/2,2}$. For CaF, this is $9 \times 10^{-8}$, which is negligible in most situations. The electric quadrupole interaction (when it exists) also results in a leak to $N=3$ but this is typically even smaller. Hyperfine interactions in the excited state have a similar influence and are likely to dominate in the case of a $^{1}\Pi - {}^{1}\Sigma$ transition since the hyperfine interaction is usually much larger in the $^{1}\Pi$ state.


\subsection{Dark states}
\label{sec:dark}

As illustrated in figure~\ref{fig:rotationalBranching}, laser cooling of molecules always involves transitions where $F' \le F$.\footnote{Transitions where $F'>F$ are called type-I and those where $F' \le F$ are type-II.} In this situation there are dark states, which are states of the molecule that do not couple to the light field. Stated another way, they are stationary states of the combined molecule-light system that have no excited state component. A molecule will be optically pumped into a dark state after scattering only a small number of photons. This is an impediment to laser cooling that has to be overcome by destabilizing the dark states. Conversely, dark states can be a useful resource for cooling molecules to very low temperature because they play a key role in sub-Doppler cooling mechanisms and because a molecule in a dark state is not heated by photon scattering. In this case, rather than destabilizing the dark states, it is useful to engineer ones that are robust.

To identify the dark states, let us consider a molecule with a set of ground states $\ket{\gamma,F,M}$ with energies $\hbar \omega_{F}$ and a set of excited states $\ket{\gamma',F',m'}$ with energies $\hbar \omega_{F'}$. Here, $\gamma$ and $\gamma'$ represent other quantum numbers needed to label the states. The Hamiltonian is $H = H_0 + \sum_{i} H_i$ where $H_0 = \sum_{F,m} \hbar \omega_{F}\ket{\gamma,F,m}\bra{\gamma,F,m}+\sum_{F',m'} \hbar \omega_{F'}\ket{\gamma',F',m'}\bra{\gamma',F',m'}$ and $H_i= - \vec{d}\cdot\vec{\cal{E}}^i$. Here, $\vec{\cal{E}}^i$ is the field due to a laser of frequency $\omega_i$ and wavevector $\vec{k}_{i}$. We expand the wavefunction in terms of the field-free eigenstates,
\begin{equation}
    \ket{\psi(t)} =\sum_{F,m} a_{F,m}(t) e^{-i \omega_F t} \ket{\gamma,F,m} + \sum_{F',m'} a_{F',m'}(t) e^{-i \omega_{F'} t} \ket{\gamma',F',m'}.
\end{equation}
The Schr\"odinger equation is 
\begin{align}
    &i \hbar \sum_{F,m} \frac{d a_{F,m}}{d t} e^{-i \omega_F t}\ket{\gamma,F,m} + i \hbar \sum_{F',m'} \frac{d a_{F',m'}}{d t} e^{-i \omega_{F'} t}\ket{\gamma',F',m'} \nonumber\\&= \sum_{i,F,m} H_i a_{F,m} e^{-i \omega_F t} \ket{\gamma,F,m} + \sum_{i,F',m'} H_i a_{F',m'} e^{-i \omega_{F'} t} \ket{\gamma',F',m'}. 
\end{align}
A dark state must have $a_{F',m'}=0$ for all $F'$ and $m'$. Applying this condition, and multiplying from the left by $\bra{\gamma',F',m'} e^{i \omega_{F'}t}$, we obtain
\begin{equation}
    \sum_{i,F,m} a_{F,m} e^{i\left(\omega_{F'} - \omega_{F}\right)t} \bra{\gamma',F',m'}H_i\ket{\gamma,F,m} = 0.
\end{equation}
Writing $\vec{\cal{E}}^i = \vec{E}^i \cos(\omega_i t)$, then making the rotating wave approximation and introducing $\Delta_{i,F,F'} = \omega_i - (\omega_{F'} - \omega_{F})$, this becomes
\begin{equation}
    \sum_{i,F,m} a_{F,m} e^{-i \Delta_{i,F,F'} t} \bra{\gamma',F',m'}\vec{d}\cdot\vec{E}^i\ket{\gamma,F,m} = 0.
\end{equation}
Expanding the scalar product and using the Wigner-Eckart theorem, we reach the condition
\begin{equation}
    \sum_{i,F,m,q} a_{F,m} e^{-i \Delta_{i,F,F'} t}(-1)^{q+F'-m'} \left(
\begin{array}{ccc}
 F' & 1 & F \\
 -m' & q & m \\
\end{array}
\right)
\bra{\gamma',F'}|d|\ket{\gamma,F} E^i_{-q} = 0,
\end{equation}
for all $F',m'$. This condition can be expressed in the form $\vec{A} \cdot \vec{a} = 0$, where $\vec{a}$ is the vector of the coefficients $a_{F,m}$ and $\vec{A}$ is the matrix whose elements are
\begin{equation}
    A_{(F',m'),(F,m)} = \sum_{i,q} e^{-i \Delta_{i,F,F'} t}(-1)^{q+F'-m'} \left(
\begin{array}{ccc}
 F' & 1 & F \\
 -m' & q & m \\
\end{array}
\right)
\bra{\gamma',F'}|d|\ket{\gamma,F} E^i_{-q}.
\end{equation}
This is equivalent to finding the set of vectors that span the null space of $\vec{A}$. This procedure can be used to find the dark states for any set of ground and excited states and any light field. Note, however, that it does not guarantee that the dark states are time-independent; this has to be checked separately.

Let us consider the simple case of a molecule with just a single ground level, $F$, and a single excited level, $F'$, driven by a single frequency of light. Table \ref{tab:dark} presents the results for certain special choices of polarization. Columns 3 and 4 correspond to light that drives $\sigma^-$ and $\pi$ transitions respectively, and here the dark states are the obvious ones. In column 5, the light is linearly polarized along $x$ when $\theta=\pi/2$, linearly polarized along $y$ when $\theta=0$, and corresponds to the standard one-dimensional $\sigma^{+}\sigma^{-}$ configuration when $\theta = k z$. This is an important case since it is frequently used to model sub-Doppler cooling processes. The final column, with $\theta = k z$, corresponds to a pair of beams counter-propagating along $z$ and linearly polarized at an arbitrary angle $\phi$ to one another. This is another important case in the context of sub-Doppler cooling, and is discussed in section \ref{sec:Sisyphus1D}. We note that in the last two cases the dark states are position dependent. A dark molecule moving sufficiently slowly through the light field will adiabatically follow the changing polarization and remain dark, but if the molecule is moving too quickly, it will not stay in the dark state.  We can also extend this discussion to a pair of laser beams with different frequencies. If the two frequency components have the same polarization there will still be stationary dark states (the same number as for a single frequency). There are typically no stationary dark states when the polarizations of the two frequency components are different. There are exceptions to this rule however. For example, if both components have $E_{-1}=0$, or both have $E_{+1}=0$, there will be a dark state when $F'=F-1$.

\begin{table}[!tb]
    \footnotesize
    \renewcommand{\arraystretch}{1.4}
    \centering
    \begin{tabular}{>{\centering\arraybackslash}p{0.13cm} >{\centering\arraybackslash}p{0.2cm} | >{\centering\arraybackslash}p{1.5cm} >{\centering\arraybackslash}p{1.5cm} >{\centering\arraybackslash}p{3.7cm} >{\centering\arraybackslash}p{4.2cm}}
        \hline \hline
        & & \multicolumn{4}{c}{$(E_{-1},E_{0},E_{+1})$} \\
        \cline{3-6}
         $F$ & $F'$ & $(0,0,1)$ & $(0,1,0)$  & $\frac{-i}{\sqrt{2}}(e^{i \theta},0,e^{-i \theta})$ & $\frac{1}{\sqrt{2}}(-c_+,0,c_-)$ \\
         \hline \hline
         $\frac{1}{2}$ & $\frac{1}{2}$ & $(1,0)$ & None & None & None\\
         \hline
         \multirow{2}{*}{1} & \multirow{2}{*}{0} & $(0,1,0)$ & $(0,0,1)$ & $(0,1,0)$ & $(0,1,0)$ \\
         & & $(1,0,0)$ & $(1,0,0)$ &  $(-e^{-i\theta},0,e^{i\theta})$ & $(c_-,0,c_+)$ \\
         \hline
         1 & 1 & $(1,0,0)$ & $(0,1,0)$ & $(e^{-i\theta},0,e^{i\theta})$ & $(-c_-,0,c_+)$ \\
         \hline
         \multirow{2}{*}{$\frac{3}{2}$} & \multirow{2}{*}{$\frac{1}{2}$} & $(1,0,0,0)$ & $(1,0,0,0)$ & $(-e^{-i\theta},0,\sqrt{3}e^{i\theta},0)$ &  $(c_-,0,\sqrt{3}c_+,0)$\\
         & & $(0,1,0,0)$ & $(0,0,0,1)$ & $(0,-\sqrt{3}e^{-i\theta},0,e^{i \theta})$ &  $(0,\sqrt{3}c_-,0,c_+)$\\
         \hline
         $\frac{3}{2}$ & $\frac{3}{2}$ & $(1,0,0,0)$ & None & None & None\\
         \hline
         \multirow{2}{*}{$2$} & \multirow{2}{*}{$1$} & $(1,0,0,0,0)$ & $(1,0,0,0,0)$ & $(0,-e^{-i\theta},0,e^{i\theta},0)$ & $(0,c_-,0,c_+,0)$\\
         & & $(0,1,0,0,0)$ & $(0,0,0,0,1)$ & $(e^{-2i\theta},0,-\sqrt{6},0,e^{2i\theta})$ & $(c_-^2,0,\sqrt{6}c_-c_+,0,c_+^2)$\\
         \hline
         2 & 2 & $(1,0,0,0,0)$ & $(0,0,1,0,0)$ & $(\sqrt{3}e^{-2i\theta},0,\sqrt{2},0,\sqrt{3}e^{2i\theta})$ & $(\sqrt{3}c_-^2,0,-\sqrt{2}c_- c_+,0,\sqrt{3}c_+^2)$\\
        \hline \hline
    \end{tabular}
    \caption{Dark states for single ground and excited levels,  with respective angular momenta $F$ and $F'$, for various polarization choices expressed as $(E_{-1},E_0,E_{+1})$. The dark states are expressed in the form $(a_{F,m=-F},...a_{F,m=F})$ and are not normalised. We have used the notation $c_{\pm} = \cos[\theta \pm \tfrac{\phi}{2}]$.}
    \label{tab:dark}
\end{table}

\subsubsection{Destabilizing dark states}
\label{sec:dark_destable}

To maintain a high rate of photon scattering, it is necessary to destabilize the dark states, preferably at a rate that is comparable to the Rabi frequency. There are several ways to do this. One way, already discussed above, is to ensure that the polarization of the light field varies with position. In this case the dark states will also be position dependent, and molecules will not remain in the dark state if they are moving quickly enough. The requirement on the speed is discussed in the context of sub-Doppler cooling in section \ref{sec:Sisyphus1D}. In other cases, a more active method of destabilization is needed. For example, radiation pressure slowing of a molecular beam (see section~\ref{sec:slowing}) usually involves a single counter-propagating laser beam resulting in uniform polarization, and active destabilization of dark states is crucial.

As discussed in detail in \citet{Berkeland2002}, dark states can be made time-dependent either by modulating the polarization of the light, or by applying a magnetic field. Destabilization by polarization modulation is intuitively clear, since the dark state depends on the polarization. For an intuitive picture of destabilization by a magnetic field, take the $z$-axis along the magnetic field and choose the polarization of the light so that the dark states are superpositions of different $m$ sub-levels. The energies of neighbouring sub-levels differ by the Zeeman splitting, $\hbar \omega_{\rm Z}$, so the dark state evolves in time. The system has three characteristic rates - the decay rate of the excited state, $\Gamma$, the Rabi frequency, $\Omega$, and the rate at which the dark state evolves into a bright state, $\gamma_{\text{d}\rightarrow\text{b}}$. The approximate value of $\gamma_{\text{d}\rightarrow\text{b}}$ is either $\omega_{\rm Z}$ or the polarization modulation rate. The photon scattering rate is small when $\gamma_{\text{d}\rightarrow\text{b}} \ll \Omega$ since the molecule adiabatically follows the slowly-varying dark state. It is also small when $\gamma_{\text{d}\rightarrow\text{b}} \gg \Omega$ because then the light is detuned from resonance. For any $\Omega$, the maximum scattering rate is reached when $\gamma_{\text{d}\rightarrow\text{b}} \approx \Omega/2$, and the global maximum is approached when satisfying that condition together with $\Omega \gg \Gamma$.

There are also some cases where dark states are not destabilized by either a magnetic field or polarization modulation. As an example, consider a linear superposition of three states, all with $m=1$, coming from three different hyperfine components of the ground state, and coupled to an excited state with $F'=1$. The state is obviously dark to $\sigma^+$ polarization. There are only two remaining orthogonal polarizations, and the state has two free coefficients, so there must be a particular linear combination that is dark to all three orthogonal polarizations. Polarization modulation will have no influence on this dark state. If, in addition, the ground state has no magnetic moment, a magnetic field will similarly have no influence. Being a linear combination of different hyperfine states, this dark state will evolve at the hyperfine frequency, making it irrelevant if the hyperfine components are resolved. However, if the hyperfine splitting is very small compared to the natural linewidth of the transition, this evolution will be slow, limiting the scattering rate. This situation is exactly the one encountered for TlF driven on the $B ^{3}\Pi_1-X ^{1}\Sigma^+$ transition~\citep{Clayburn2018}. The magnetic moment of the $^{1}\Sigma$ ground state is too small to be useful. The ground state has the same hyperfine structure as shown in figure~\ref{fig:hyperfine}(a), except that the intervals are all very small, spanning only 220~kHz in total. Conversely, the excited state has very large hyperfine structure, so a single laser frequency will only couple to one hyperfine component of the excited state. These features conspire to produce a low scattering rate. This issue can be addressed by coupling to auxiliary states, for example using microwaves to couple to another rotational level. 

\subsubsection{Engineering dark states}

\begin{figure}[!tb]
\centering
\includegraphics[width=0.25\textwidth]{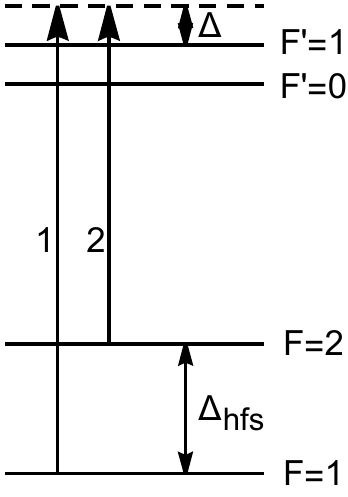}
\caption{Engineering Raman dark states in a level scheme that is typical of many laser-coolable molecules.}
\label{fig:darkRaman}
\end{figure}

It is sometimes desirable to engineer dark states, especially to enhance sub-Doppler cooling and velocity-selective coherent population trapping~\citep{Cohen-Tannoudji1998,Cheuk2018, Caldwell2019}. Dark states that are superpositions of two lower levels (typically different hyperfine components) can be engineered using two laser components with relative frequencies satisfying the Raman condition. 

Figure~\ref{fig:darkRaman} illustrates an example which is pertinent to many of the molecules laser cooled so far. Here, two ground hyperfine levels ($F=1,2$) are coupled by lasers to two excited hyperfine levels ($F'=0,1$). When the frequency difference between the two lasers matches the ground state hyperfine interval, stationary dark states are formed, sometimes known as Raman dark states. When both laser components are polarized to drive $\pi$ transitions there are two Raman dark states in this system, $\pm\sqrt{3/5}\,\eta\ket{1,\pm 1} + \ket{2,\pm 1}$. Here $\eta = d_{2,1}/d_{1,1}$ where $d_{F,F'} = \bra{F'}|d|\ket{F}$, and we have omitted the normalization. The equivalent superposition with $m=0$ is not dark because $\ket{2,0}$ only couples to $F'=1$ while $\ket{1,0}$ only couples to $F'=0$. When both laser components are polarized to drive $\sigma^-$ transitions there is a single Raman dark state, $-\sqrt{1/5}\,\eta\ket{1,0} + \ket{2,0}$. The superposition with $m=1$ is not dark because the $\ket{1,1}$ state can couple to both $F'=1$ and 0. When laser component 1 drives $\sigma^+$ transitions and component 2 drives $\sigma^-$ transitions the Raman dark state is $\sqrt{6/5}\,\eta\ket{1,0} + \ket{2,2}$. It is important to note that in constructing these dark states we have assumed that the ground state hyperfine interval is large compared to the laser detuning, $\Delta_{\rm hfs} \gg \Delta$. When this is not the case, off-resonant couplings (component 1(2) drives transitions from $F=2(1)$) result in residual coupling to the light and there will be some photon scattering.

\subsection{Intermediate electronic states}

One of the desirable properties of a good cycling transition is the absence of any intermediate electronic levels. This simplifies the cooling scheme. Nevertheless, molecules with intermediate states have been laser cooled, and it is interesting to look at these examples.

\begin{figure}[!tb]
    \centering
    \includegraphics[width=\columnwidth]{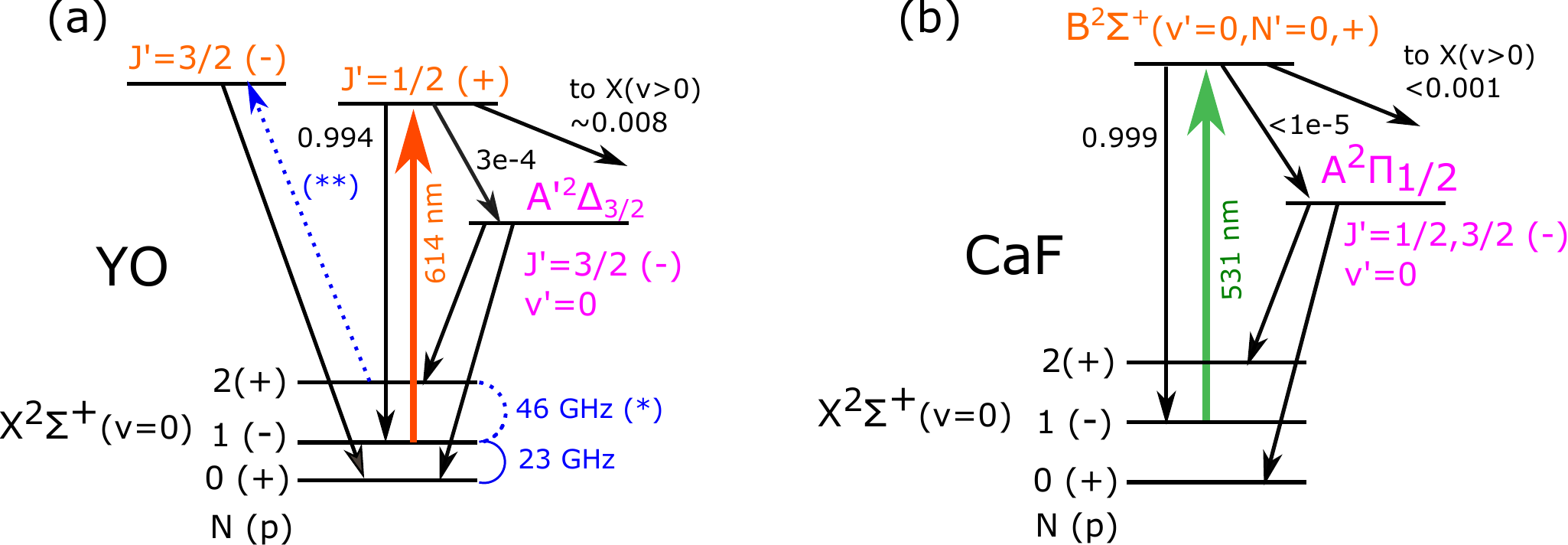}
    \caption{Examples of laser cooling schemes where there are intermediate electronic states. The main cooling transition (solid coloured arrows), and decay pathways from excited states (solid black arrows) are shown. Branching to higher-lying vibrational levels ($v>0$) and hyperfine structure are not shown.  (a) The YO molecule has an intermediate $A'\,^{2}\Delta_{3/2}$ state that disrupts the rotational closure of the main cooling transition.  Microwave remixing and/or additional repump lasers are required to obtain the desired level of cycling closure. (b) The $B^{2}\Sigma^+-X^{2}\Sigma^+$ transition of CaF has highly favourable vibrational branching ratios. The excited state can decay to $A^{2}\Pi_{1/2}$, but the branching ratio for this is small.}
    \label{fig:intermediate_states}
\end{figure}

Molecules such as YO~\citep{Yeo2015} and BaF~\citep{Chen2016,Hao2019,Albrecht2020} have an intermediate $\Delta$ state that can potentially be troublesome. Figure~\ref{fig:intermediate_states}(a) shows the laser cooling scheme used to cool YO molecules~\citep{Collopy2018}.  The main cooling transition is $A^{2}\Pi_{1/2}(v'=0,J'=1/2) - X^{2}\Sigma^{+}(v=0,N=1)$ at 614~nm. Molecules excited to the $A$ state can decay to an intermediate $A'\,^{2}\Delta_{3/2}$ state after scattering about 3000 photons. Molecules that end up in $A'$ can decay to either $N=0$ or $N=2$ of $X ^{2}\Sigma^+$.  Though electric dipole forbidden, this decay occurs with a lifetime of $\approx$1~$\mu$s due to mixing of the $A'$ state with the nearby $A^{2}\Pi_{3/2}$ state (not shown in the figure).  Molecules that decay to $N=0$ are re-introduced into the cooling cycle using resonant microwaves at 23~GHz that mix the $N=0$ and $N=1$ states.  This increases the number of ground states participating in the main cooling cycle and consequently reduces the overall scattering rate, as explained in section~\ref{sec:rate_models}.  Molecules that decay to $N=2$ can be re-introduced into the cooling cycle using resonant microwaves (indicated by (*) in the figure) at 46~GHz that mix $N=2$ and $N=1$, further decreasing the scattering rate by about a factor of two.  Alternatively, they can be optically repumped via $A^{2}\Pi_{1/2}(J'=3/2)$ (indicated by (**)) without affecting the scattering rate.  Yet another option is to optically pump directly on $A(J'=3/2)\leftarrow A'$, though laser light at the required wavelength in the mid infrared is challenging to produce.  We note that although the intermediate state complicates laser cooling, it can also be an advantage because the narrow transition, $A'^{2}\Delta_{3/2}-X^{2}\Sigma^+$, could be used to make a MOT with a very low Doppler cooling limit, around 10$\mu$K in this case~\citep{Collopy2015}. This is similar to narrow line MOTs used for Sr and Yb.

Figure~\ref{fig:intermediate_states}(b) shows the $B^{2}\Sigma^{+}-X^{2}\Sigma^{+}$ cooling transition of CaF, which has excellent vibrational branching ratios for laser cooling. The decay sequence $B\rightarrow A \rightarrow X$ populates $N=0,2$ in $X$, just as for YO and BaF.  However, the branching ratio to the $A$ state appears to be less than $10^{-5}$~\citep{Truppe2017} due to the low frequency and small transition dipole moment. This is small enough that it has little effect on slowing of the beam, though it might be problematic for making a MOT. A similar situation occurs for the polyatomic molecule SrOH, discussed below~\citep{Kozyryev2017}.

Finally, we note that molecules containing a heavy atom, such as YbF, have intermediate electronic states arising from inner-shell excitations. The influence of these states on laser cooling is an important topic for investigation.

\subsection{Polyatomic molecules}
\label{sec:polyatomics}

Laser cooling of polyatomic molecules is more complex than for diatomic molecules due to the proliferation of vibrational modes.  A molecule with $n$ atoms has $\xi = 3n-6$ vibrational modes, or $\xi = 3n-5$ if the molecule is linear. Furthermore, the dense level structure resulting from the many degrees of freedom can lead to a breakdown of the Born-Oppenheimer approximation, and thus to violations of rotational selection rules.  Anharmonic terms in the vibrational Hamiltonian can also give rise to mixing between different vibrational modes. These complications present major challenges to laser cooling of complex molecules. Nevertheless, there has been great recent progress in this direction.

Initial ideas for identifying polyatomic species amenable to laser cooling were presented by \citet{Isaev2016b}.  In general, alkaline earth monohydroxide (MOH) and monoalkoxide (MOR, where R is an organic substituent) molecules with a heavy alkaline earth atom, e.g. M $\in \{$Sr,Ca,Yb,Ba$\}$, have been identified as particularly promising candidates.  As in the diatomic case, vibrational branching is suppressed when the potential energy surfaces of the ground and excited states are very similar.  This is commonly satisfied in alkaline earth monofluoride (MF) and monohydride (MH) diatomic species, where one valence electron of M forms the ionic bond and the other can be excited without significantly affecting that bond. Certain atom complexes or ligands, e.g. OH, can replace F or H with much the same result -- an optically active non-bonding electron having transitions with near diagonal Franck-Condon factors.  Specific ligands, e.g. OCH$_{3}$, are expected to make these factors even closer to unity than in the corresponding diatomic system~\citep{Dickerson2020}.  

Let us consider a linear triatomic monohydroxide molecule, MOH, that has a $X^{2}\Sigma^{+}$ ground state.  Being linear, this is in many ways very similar to the diatomic case, with electronic states associated with excitation of the M-centred electron and rotational states associated with end-over-end rotations.  The most significant difference is the additional modes of vibration. In this instance, there are three fundamental vibrational modes, the M--O stretch, a doubly-degenerate M--O--H bending mode, and the O--H stretch.  The associated vibrational quantum numbers are denoted ($v_{1},v_{2},v_{3}$), respectively.  The degenerate bending mode $v_{2}$ gives rise to a vibrational angular momentum whose projection onto the intermolecular axis is $l \hbar$.  This kind of angular momentum is absent in diatomic molecules.  $l$ takes on $v_{2}+1$ possible integer values in the range $|l| = v_{2},v_{2}-2,v_{2}-4,\ldots,0(1)$ for even (odd) values of $v_{2}$~\citep{Bernath}.  For notation purposes, the value of $l$ appears as a superscript on the bending mode, as in $(v_{1},v_{2}^{l},v_{3})$.  When $v_{2}>0$, our linear molecule becomes similar to a symmetric top with the correlation $l \leftrightarrow K$, where $K$ is the projection of the angular momentum along the symmetry axis~\citep{HerzbergII}.  Just as in that case, we have the constraint that the total angular momentum ignoring spin is $N = R + L + l \geq l$.  States with $|l|>0$ are doubly degenerate due to the clockwise or counter-clockwise motion of the nuclei.  This degeneracy is lifted by Coriolis forces in a splitting known as $l$-doubling, which is akin to $\Lambda$-doubling in diatomic species with $\Lambda>0$, but in polyatomic molecules is present even in a $\Sigma$ state ($\Lambda = 0)$.  A similar parallel can be drawn with $K$-doublets in symmetric tops.  As in those cases, $l$-doubling gives rise to closely spaced levels of opposite parity. As before, $J = N + S$ is the total angular momentum apart from nuclear spin.  We will ignore hyperfine structure in our discussion.

As in a diatomic, rotational closure is obtained by exciting from $N=1$ to $N'=0$ (see figure~\ref{fig:rotationalBranching}). Vibrational branching is governed by the Franck-Condon principle associated with the overlap of vibrational wavefunctions, as in equations (\ref{eqn:FCfactor}) and (\ref{eqn:vibrational_branching}).  Now, however, the integral over the the bond length must be replaced by one over all vibrational coordinates $Q_{i}$.  Specifically, equation (\ref{eqn:d_if}) must be evaluated at the equilibrium separations $Q_{i}^{0}$ and the single overlap integral replaced by the product of $\xi=3n-5$ (or $3n-6$) overlap integrals
\begin{equation}
    \prod\limits_{i=1}^{\xi} 
    \int 
    f_{v_{i}'}^* 
    f_{v_{i}^{\vphantom{-1}}} 
    dQ_{i}
    = 
    \int 
    f_{v_{1}'}^* f_{v_{1}^{\vphantom{-1}}} 
    dQ_{1}
    \int 
    f_{v_{2}'}^* f_{v_{2}^{\vphantom{-1}}} 
    dQ_{2}
    \, \, 
    \cdots
    \,
    \int 
    f_{v_{\xi}'}^* f_{v_{\xi}^{\vphantom{-1}}} dQ_{\xi}.
    \label{eqn:d_if_polyatomic}
\end{equation}
To the same approximation as before, the associated Franck-Condon factors
\begin{equation}
    q_{v_{1}',\ldots,v_{\xi}',v_{1},\ldots,v_{\xi}} =
    \left|
    \prod\limits_{i=1}^{\xi}
    \int 
    f_{v_{i}'}^* 
    f_{v_{i}^{\vphantom{-1}}} 
    dQ_{i}
    \right|^{2}
\end{equation}
determine the vibrational branching ratios.  There are symmetries that cause some of these factors to vanish.  Specifically, for the Franck-Condon factor to be non-zero, the product of vibrational wavefunctions must be totally symmetric with respect to the symmetry operations of the point group to which the molecule belongs.  For the degenerate bending mode vibrations $v_{2}$ with vibrational angular momentum $l$, this leads to the selection rules
\begin{align}
    \Delta l & = 0 \\
    \Delta v_{2} &= 0,\pm2,\pm4,\ldots.
\end{align}
The other two vibrational modes do not alter the molecule's symmetry and are thus governed only by the Franck-Condon principle.  Thus, for laser cooling on a $\Tilde{A}(000)-\Tilde{X}(000)$ transition, one must account for decays to vibrational states of $\Tilde{X}$ with ($v_{1},v_{2},v_{3}$) = ($a,0,0$), ($0,b^{0},0$), and ($0,0,c$) for $a,c \geq 1$ and $b \in \{2,4,\ldots\}$. A good choice of species results in heavily suppressed branching ratios for increasing values of $a,b,c$.

\begin{figure}[!tb]
    \centering
    \includegraphics[width=0.7\columnwidth]{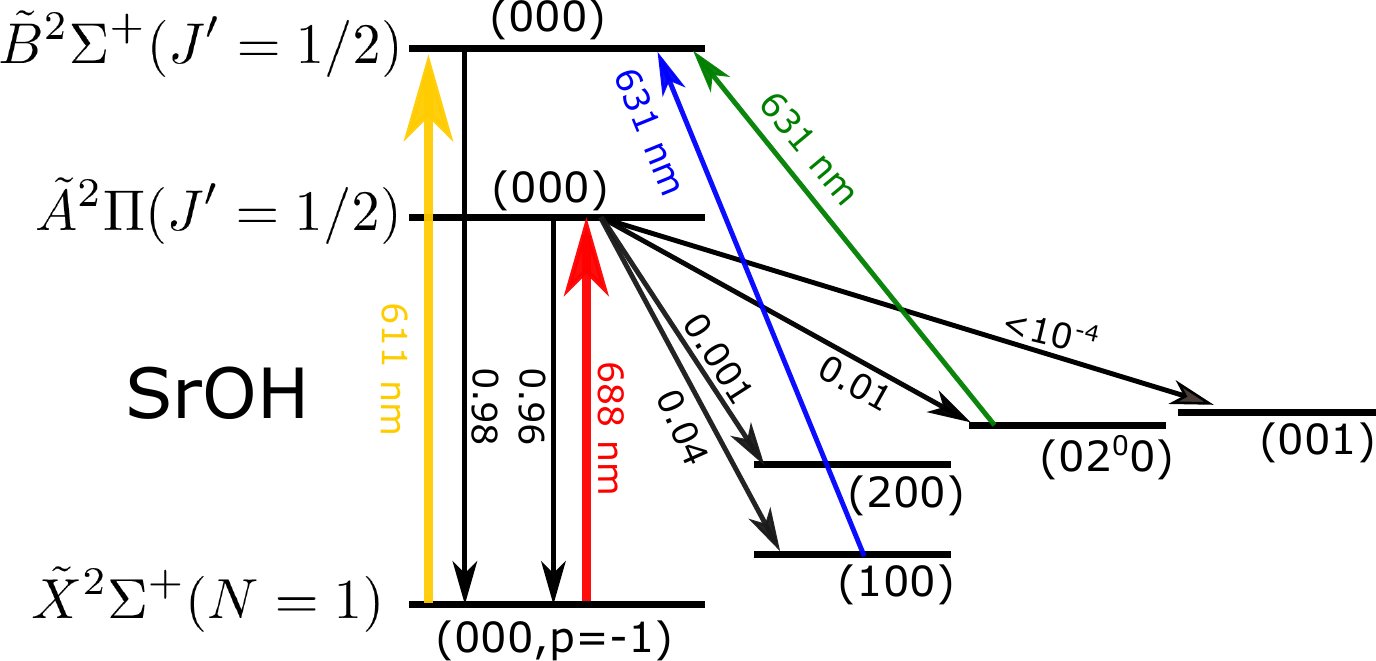}
    \caption{Laser cooling scheme for SrOH, neglecting hyperfine structure.  The main cooling laser drives $\Tilde{A} - \Tilde{X}$ (or, alternatively, $\Tilde{B} - \Tilde{X}$), shown as a thick coloured arrow labeled by its wavelength.  The two most important repump lasers appear as thin coloured arrows.  Significant decay channels (black arrows) are labeled by their respective branching ratios.}
    \label{fig:SrOH}
\end{figure}

To illustrate these ideas, let us consider SrOH, which was the first polyatomic molecule to be laser cooled. The cooling scheme is shown in figure~\ref{fig:SrOH} and discussed in more detail in \citet{Kozyryev2016,KozyryevThesis}.  The first experimental demonstration involved deflection of a molecular beam by radiation pressure~\citep{Kozyryev2016}. Here, scattering of about 100 photons was demonstrated using only two lasers, a main cooling laser operating on the $\Tilde{A}^{2}\Pi_{1/2}(000)(J=1/2,N=0) - \Tilde{X}^{2}\Sigma^{+}(000)(N=1)$ transition at 688~nm and a repump addressing the largest vibrational leak by pumping on $\Tilde{B}^{2}\Sigma^{+}(000) - \Tilde{X}^{2}\Sigma^{+}(100)$ at 631~nm.  The dominant residual vibrational leaks are to $\Tilde{X}^{2}\Sigma^{+}(02^{0}0)$ and (200), with the latter 10 times smaller. Leaks to vibrational states involving a stretch of the O--H bond are below $10^{-4}$.  By closing the leak to the $(02^{0}0)$ level, \citet{Kozyryev2017} demonstrated transverse cooling to below 1~mK using Sisyphus forces. This work also found cooling was improved by using the $\Tilde{B}^{2}\Sigma^{+}(000)-\Tilde{X}^{2}\Sigma^{+}(000)$ transition.  Additional repump lasers addressing the (200) and (01$^1$0) ground vibrational states should allow 10$^{4}$ photons to be scattered. Decay to the latter state (not shown in Fig.~\ref{fig:SrOH}) is dipole forbidden but is known to be significant at this level~\citep{Kozyryev2016}.  In addition to the pioneering work on SrOH, more recent work has extended laser cooling to other triatomic species, including the demonstration of a one-dimensional MOT of CaOH~ \citep{Baum2020} and of both Doppler and Sisyphus forces for YbOH~\citep{Augenbraun2020}.

Ideas and experiments for laser cooling even more complex molecules are also underway.  Following the MOR approach, various ligands (R) can be substituted for the hydrogen atom while maintaining (or even improving) the cycling transition~\citep{Kozyryev2016}.  Increasingly complex ligands are starting to be explored, including CH$_{3}$, (CH$_{2}$)$_{n}$--CH$_{3}$ chains (where $n$ is an integer), and even fullerenes~\citep{Klos2020}.  More extensive exploration of these optical cycling centers, as done by \citet{Li2019}, and the identification of other bonds or structures that behave similarly, will allow for laser cooling to be applied to an increasingly diverse set of complex molecules.  Of particular interest are large organic and chiral molecules for studying fundamental symmetries of nature and ultracold quantum chemistry~\citep{Ivanov2020}. Experimentally, the CH$_{3}$ ligand has been explored in preliminary work on laser cooling of YbOCH$_{3}$~\citep{Augenbraun2020c}.  Remarkably, \citet{Mitra2020} have recently demonstrated laser cooling of the symmetric top CaCOH$_{3}$.

\section{Models of laser cooling}
\label{sec:methods}

We would like to determine how the phase-space distribution of an ensemble of particles evolves under the influence of a force. Generally, the force depends on the spatial coordinates $\vec{x}$ and the velocity $\vec{v}$. It also has a fluctuating part due to the randomness of the photon absorption and emission events, and because the fluctuating dipole moment of the molecule couples to intensity and polarisation gradients to produce a fluctuating dipole force. Most of this section is devoted to the determination of the mean force and its fluctuations, but it is instructive to consider from the outset how to simulate the behaviour of molecules once the force is known. One method is a Monte-Carlo approach where the equations of motion are solved for a large number of particles in order to determine a set of trajectories, incorporating a random walk in order to account for the fluctuations of the force. Another method is to calculate the evolution of the entire probability distribution in phase space, $W(\vec{x},\vec{v},t)$, by solving the Fokker-Planck-Kramers (FPK) equation. In three dimensions, a suitable equation is ~\citep{Molmer1994, Marksteiner1996}
\begin{equation}
    \frac{\partial W}{\partial t}+\sum_{i}v_i\frac{\partial W}{\partial x_i}=\sum_{i}\frac{\partial }{\partial v_i}\left(\frac{-F_i(\vec{x},\vec{v})}{m}W+\frac{D(\vec{x},\vec{v})}{m^2}\frac{\partial W}{\partial v_i}\right),
\label{eqn:fpe}
\end{equation}
where $i\in \{x,y,z\}$, $\vec{F}(\vec{x},\vec{v})$ is the mean force and $D$ is the momentum diffusion constant which describes the fluctuations of the force. We have taken the diffusion constant to be independent of the direction of motion. 

In general, $\vec{F}$ and $D$ vary on the scale of a wavelength, $\lambda$, but we are often interested in the behaviour on a much larger scale. In that case, it is appropriate to average $\vec{F}$ and $D$ over a region of size $\lambda$. In the special case where the applied fields are uniform on the scale of the molecular distribution, the position dependence vanishes and the force is always in the direction of motion. In that case, the equation for the probability density reduces to
\begin{equation}
\frac{\partial}{\partial t}v^2W(v,t)=\frac{\partial }{\partial v}\left(-\frac{F(v)}{m}v^2W(v,t)+\frac{v^2D(v)}{m^2}\frac{\partial W(v,t)}{\partial v}\right),
\label{eqn:fpe3}
\end{equation}
where $F$ and $D$ are now the wavelength-averaged values of the force and the diffusion constant. Once $F(v)$ and $D(v)$ are known, this equation can be solved to determine the evolution of the velocity distribution over time. 

The steady-state solution of equation (\ref{eqn:fpe3}) is
\begin{equation}
W(v) = W_0 \exp\left[m \int_0^{v}\frac{F(v')}{D(v')}\textrm{d}v'\right],
\label{eqn:fpeSteady}
\end{equation}
where $W_0$ is a constant defined by the normalisation condition $\int W(v) 4\pi v^2 \textrm{d}v = 1$. As we will see, for low velocities we often have a damping force which is linear in velocity, $F(v) \approx - \alpha v$ where $\alpha$ is the damping constant, and a diffusion constant which is independent of velocity, $D(v) \approx D$. In this case, the velocity distribution is a Gaussian function
\begin{equation}
    W(v) = W_0 \exp\left(-\frac{m \alpha v^2}{2 D}\right) = W_0 \exp\left(-\frac{m v^2}{2 k_{\rm B} T}\right)
\end{equation}
where we identify the temperature as
\begin{equation}
    T = \frac{D}{k_{\rm B}\alpha}.
\label{eqn:T_D_alpha_relation}
\end{equation}

In some cases, it is sufficient to use a rate model to estimate $\vec{F}$, $D$, and other useful quantities. In other cases, it is necessary to solve generalised optical Bloch equations for the multi-level molecule interacting with multiple frequencies of light. Both approaches are described below.

\subsection{Rate model}
\label{sec:rate_models}

A great deal can be learned about laser cooling and magneto-optical trapping by neglecting all coherences and using rate equations to determine the populations of the lower and upper levels of the molecule. Following~\citet{Tarbutt2015}, we consider the case where $N_{\rm g}$ levels of the ground electronic state are coupled by laser light to $N_{\rm e}$ levels of an excited electronic state. The populations are $n^{\rm g}_j$ and $n^{\rm e}_k$ where $j$ and $k$ are indices labelling the ground and excited states. The transition angular frequencies are $\omega_{j,k}$. The light field has several components, labelled by an index $p$, each described by an angular frequency $\omega_{p}$, wavevector $\vec{k}_p$ and polarization $\vec{\epsilon}_{p}$. In general, we should allow every transition to be driven by every component of the light, each with its own Rabi frequency $\Omega_{j,k,p}$ and detuning $\delta_{j,k,p} = \omega_p - \omega_{j,k} - \vec{k}_p\cdot\vec{v} = \Delta_{j,k,p} - \vec{k}_p\cdot\vec{v}$. Here, we have included the Doppler shift $-\vec{k}_p\cdot\vec{v}$ for a molecule moving at velocity $\vec{v}$. The rate equations for the populations in this general case are

\begin{subequations}
\begin{align}
    \frac{d}{d t}n^{\rm g}_j&=\sum_{k,p} R_{j,k,p} (n^{\rm e}_k - n^{\rm g}_j) + \Gamma \sum_{k} r_{j,k} n^{\rm e}_k,\label{eqn:popg}\\
   \frac{d}{d t} n^{\rm e}_k&=\sum_{j,p} R_{j,k,p} (n^{\rm g}_j - n^{\rm e}_k) - \Gamma n^{\rm e}_k,\label{eqn:pope}
\end{align}
\label{eqn:pops}
\end{subequations}
where $\Gamma$ is the decay rate of the excited states which we take to be the same for all $k$, $r_{j,k}$ is the branching ratio for excited state $k$ to decay to ground state $j$, and $R_{j,k,p}$ is the excitation rate between $j$ and $k$ due to laser component $p$. This excitation rate is
\begin{equation}
    \label{eqn:excitation_rate}
    R_{j,k,p} = \Gamma \frac{\Omega_{j,k,p}^{2} / \Gamma^{2}}{1+4\delta_{j,k,p}^{2}/\Gamma^{2}}.
\end{equation}
We can also write an equation for the mean force acting on the molecule by considering the rate of change of momentum due to absorption and stimulated emission,
\begin{equation}
   \vec{F} = m \frac{d}{d t}\vec{v} = \hbar \sum_{j,k,p} \vec{k}_p R_{j,k,p}(n^{\rm g}_j - n^{\rm e}_k).
    \label{eqn:rateModelAccel}
\end{equation}
Numerical simulations based on these equations have proven to be useful for simulating magneto-optical traps of molecules and laser deceleration of molecular beams (see sections \ref{sec:slowing} and \ref{sec:mots}). 

\subsubsection{Scattering rate}

The photon scattering rate is $R_{\rm sc} = \Gamma n^{\rm e}$, where $n^{\rm e}=\sum_k n^{\rm e}_k$. This is easily calculated for any set of parameters by solving equations (\ref{eqn:pops}) in the steady state.

In order to obtain more insight, and some simple and useful results, let us consider the case where there is only one excited state, making the index $k$ redundant. We also suppose that there is just one laser beam (one value of $\vec{k}_p$) and that each transition is driven by only a single component of the light. This makes $p$ redundant, since each component can instead be labelled by the transition it drives. The intensity of each component in the polarization state needed to drive the intended transition is $I_{j}$. In this case equations (\ref{eqn:popg}) reduce to
\begin{equation}
    R_j (n^{\rm e} - n^{\rm g}_j) +\Gamma r_j n^{\rm e}  = 0.
\label{eqn:popeSimple}
\end{equation}
We define a saturation intensity $I_{\rm s} = \pi h c \Gamma / (3\lambda^3)$, where $\lambda$ is the wavelength, approximated equal for all transitions. We also define a saturation parameter
$$
s_j = \frac{2 \Omega_j^2}{r_j \Gamma^2} = \frac{I_{j}}{I_{\rm s}},
$$
and a modified saturation parameter
$$
s_j'= s_j \frac{1}{1+4\delta_j^2/\Gamma^2}.
$$
Then, equations (\ref{eqn:popeSimple}) become
\begin{equation}
    \frac{1}{2} s_j'(n^{\rm e} - n^{\rm g}_j) + n^{\rm e} = 0.
\end{equation}
From this set of $N_{\rm g}$ equations we find the photon scattering rate to be
\begin{equation}
    R_{\rm sc}  = \frac{\Gamma_{\rm eff}}{2}\frac{s_{\rm eff}'}{1+s_{\rm eff}'}
    \label{eqn:Rsc}
\end{equation}
where
\begin{equation}
    \Gamma_{\rm eff} = \frac{2}{N_{\rm g} + 1}\Gamma
    \label{eqn:Gamma_eff1}
\end{equation}
and
\begin{equation}
    s_{\rm eff}' = \frac{N_{\rm g} + 1}{2}\left(\sum_j (s_j')^{-1}\right)^{-1}.
    \label{eqn:s_eff1}
\end{equation}
In the special case where all the detunings are equal and all the intensities are equal and sum to $I_{\rm tot}$, we arrive at
\begin{equation}
    R_{\rm sc} = \frac{\Gamma_{\rm eff}}{2}\frac{s_{\rm eff}}{1+s_{\rm eff}+4\delta^{2}/\Gamma^{2}}
    \label{eqn:RscSimple}
\end{equation}
where
\begin{equation}
    s_{\rm eff} = \frac{I_{\rm tot}}{I_{\rm s,eff}} = \frac{N_{\rm g}+1}{2 N_{\rm g}^2}\frac{I_{\rm tot}}{I_{\rm s}}.
\end{equation}

\begin{figure}[!tb]
\centering
\includegraphics[width=0.55\textwidth]{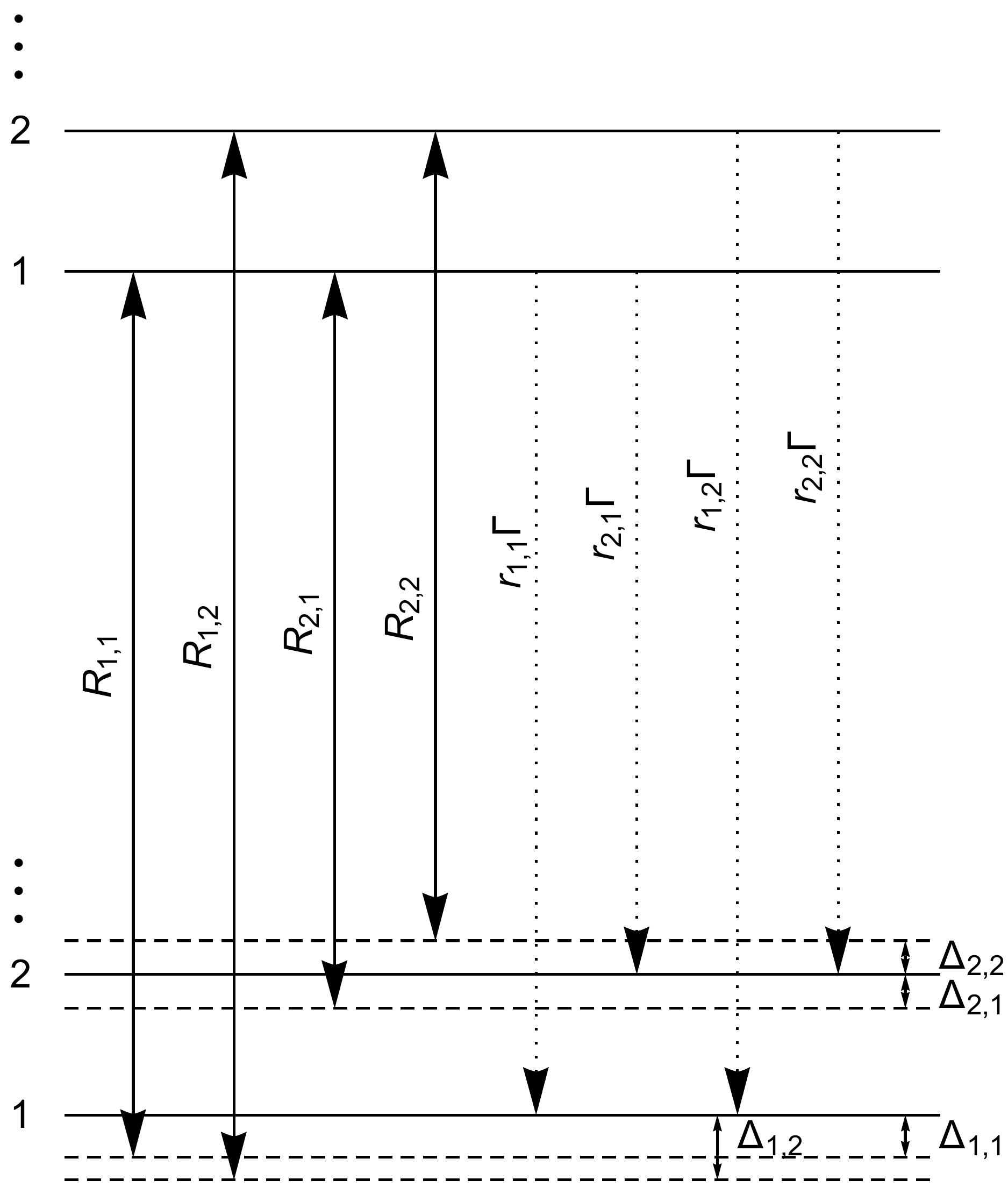}
\caption{Multi-level rate model. The transition between ground level $j$ and excited level $k$ is driven at rate $R_{j,k}$ by light of intensity $I_{j,k}$ and detuning $\Delta_{j,k}$. The spontaneous decay rate from $k$ to $j$ is $r_{j,k}\Gamma$. }
\label{fig:rateModel}
\end{figure}

Figure~\ref{fig:rateModel} illustrates the more general case where there are several excited states. Here the results are more complicated, but we can recover some simple expressions by arranging the excitation rates so that the excited states have equal populations. This is an important case because it maximizes the scattering rate.  To handle this case, it is helpful to define quantities averaged over excited states, 
\begin{align}
\bar{R}_j &= \frac{1}{N_{\rm e}}\sum_k R_{j,k},\nonumber\\
\bar{r}_j &= \frac{1}{N_{\rm e}}\sum_k r_{j,k}, \nonumber\\
\bar{s}_j\,'&= \frac{1}{N_{\rm e}}\sum_k \frac{I_{j,k}}{I_{\rm s}} \frac{r_{j,k}}{\bar{r}_j}\frac{1}{1+4\delta_{j,k}^2/\Gamma^2},\nonumber
\end{align}
where $I_{j,k}$ is the intensity driving the transition from $j$ to $k$. The expression for the scattering rate in this case is identical to equation (\ref{eqn:Rsc}), but with
\begin{equation}
    \Gamma_{\rm eff} = \frac{2 N_{\rm e}}{N_{\rm g} + N_{\rm e}}\Gamma
    \label{eqn:Gamma_eff}
\end{equation}
and
\begin{equation}
    s_{\rm eff}' = \frac{N_{\rm g} + N_{\rm e}}{2}\left(\sum_j (\bar{s}_j\,')^{-1}\right)^{-1}.
    \label{eqn:s_eff}
\end{equation}

Equation (\ref{eqn:Rsc}) has the same form as the familiar photon scattering rate for a two level atom. Although it is strictly only valid under the conditions outlined above, in practice it has been found to be a good approximation to the results obtained from simulations based on the full rate equations, and is found to give a reasonable estimate of the scattering rate measured in several experiments (see section \ref{sec:RateModelApplications}).  We see that the scattering rate has a maximum value of $R_{\rm sc}^{\rm max}=\Gamma_{\rm eff}/2$, which depends on the ratio of the number of excited states to the total number of states, as given by equation (\ref{eqn:Gamma_eff}). This result can be understood by realizing that when all transitions are strongly saturated the population is distributed equally amongst all levels, making $n_{\rm e} = N_{\rm e}/(N_{\rm g}+N_{\rm e})$~\citep{Shuman2009}. For the rotationally closed transition discussed in section~\ref{sec:rotationalTransitions}, the degeneracy of the ground state is three times that of the excited state, implying $R_{\rm sc}^{\rm max} = \Gamma/4$ in the best case. This shows that the scattering rate, and associated force, will always be at least two times smaller for a molecule than for a two-level atom with an identical linewidth. In practice, it is common for the $v=1$ repump laser to couple to the same excited state as the main cooling laser, doubling $N_{\rm g}$ and reducing $R_{\rm sc}^{\rm max}$ to $\Gamma/7$. In cases where there are small leaks to other rotational states, requiring microwave remixing~\citep{Yeo2015,Norrgard2016}, $N_{\rm g}$ is even larger and the maximum scattering rate is reduced even further. The use of more than one excited electronic state in the laser cooling scheme can be a useful way to increase the scattering force. This method has been used for laser slowing of CaF where both the $A^{2}\Pi_{1/2}$ and $B^{2}\Sigma^{+}$ excited states were used~\citep{Truppe2017}.

The scattering rate reaches half its maximum value when  $s_{\rm eff}' = 1$. In the case of a single excited state, the laser intensity needed to reach this condition is $I_{\rm tot} = 2N_{\rm g}^2/(N_{\rm g}+1) I_{\rm s}$. Since there are always several ground states, this intensity is considerably higher than the equivalent atomic case. We see from equation (\ref{eqn:s_eff}) that the individual $s_j'$ are summed in parallel in $s_{\rm eff}'$. If one of the set of $s_{j}$ is significantly smaller than all the rest, either because the $I_{j,k}$ are too small or the $\delta_{j,k}$ are too large, that level becomes a bottleneck, limiting the value of $s_{\rm eff}$ and therefore $R_{\rm sc}$. This shows the importance of ensuring all ground states are coupled to at least one of the excited states at an adequate rate. 

\begin{figure}[!tb]
\centering
\includegraphics[width=0.7\textwidth]{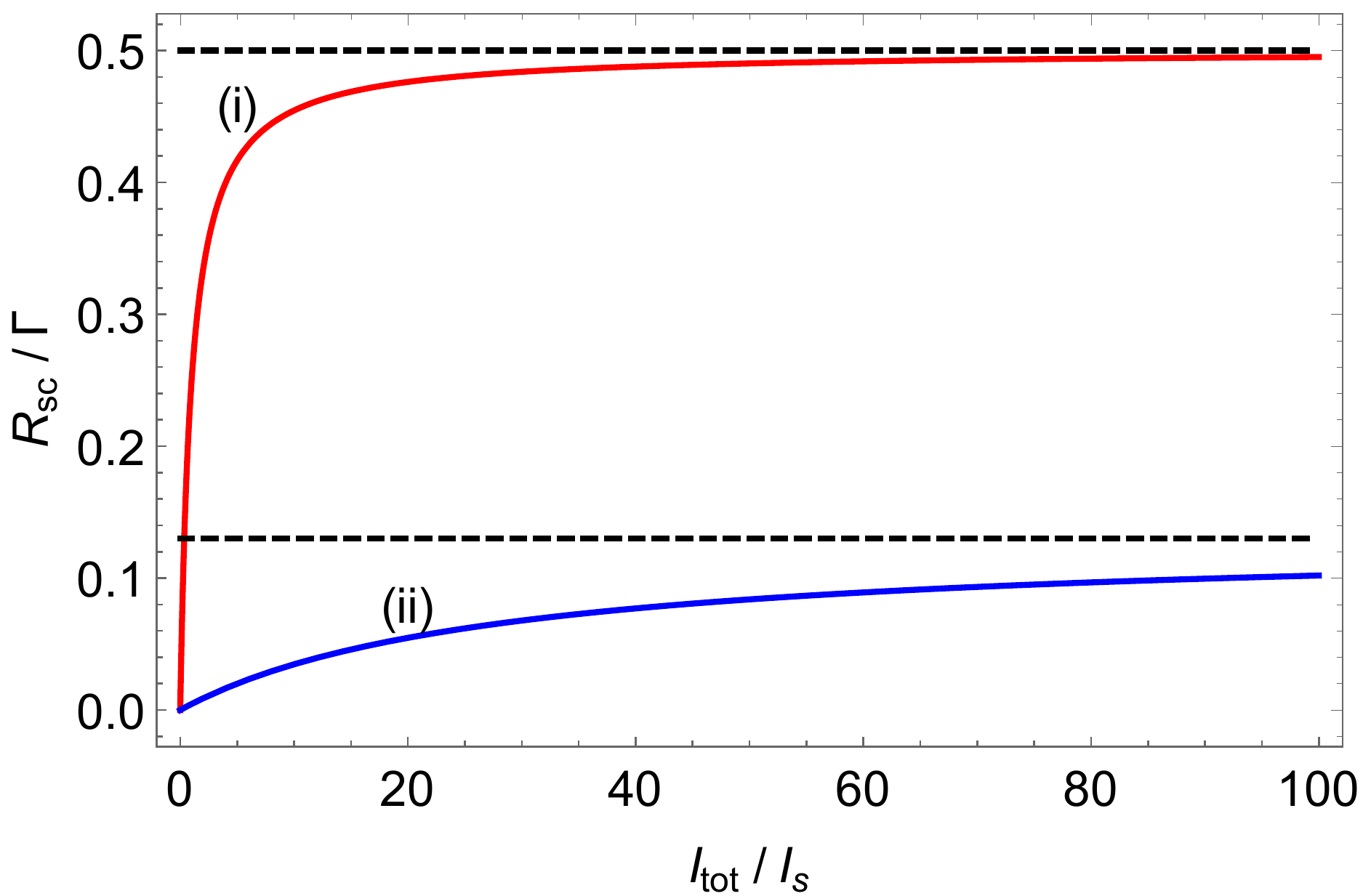}
\caption{Photon scattering rate as a function of total intensity for two situations. (i) A two-level atom driven on resonance. (ii) A CaF molecule with all levels of $X^{2}\Sigma^{+}(v=0,1,N=1)$ driven on resonance to $A^{2}\Pi_{1/2}(v'=0,J'=1/2)$. Dashed lines are the high intensity limits.}
\label{fig:scatteringRateComparison}
\end{figure}

To summarize, we see that in many cases of practical interest the maximum photon scattering rate is reduced compared to the case of an ideal two-level atom, that high laser intensity is needed to approach that maximum rate, and that a single level excited at a smaller rate than all the others can greatly reduce the overall scattering rate. Figure~\ref{fig:scatteringRateComparison} compares the resonant scattering rate achievable in typical atom cooling and molecule cooling experiments. Atomic experiments often approximate the ideal two-level system where $R_{\rm sc}$ tends towards $\Gamma/2$ and the total intensity required is only a few times $I_{\rm s}$. For the molecular case,  we use the rate equations to model a CaF molecule in a six-beam MOT where the levels of $X^{2}\Sigma^{+}(v=0,1,N=1)$ are driven to $A^{2}\Pi_{1/2}(v'=0,J'=1/2)$. The total intensity is divided equally between 8 frequency components which address the 8 hyperfine levels, all at zero detuning. The maximum rate is $R_{\rm sc}^{\rm max}=0.13\Gamma$, about 4 times smaller than the atomic case, and the total intensity needed to reach a certain fraction of $R_{\rm sc}^{\rm max}$ is 28 times higher for the molecule than for the atom.

\subsubsection{Force, damping constant and spring constant}

In the context of a rate model, the motion of a molecule interacting with a set of laser beams is found by solving equation (\ref{eqn:rateModelAccel}) together with equations (\ref{eqn:pops}). Approximate expressions for the scattering force for a set of beams can also be obtained from the approximate expressions for the scattering rate, most straightforwardly from equation (\ref{eqn:RscSimple}). Because $R_{\rm sc}$ is not linear in intensity, the scattering rate due to one beam is altered by the presence of other beams. An ad hoc approach to handling this is to argue that the scattering rate due to a particular beam should have the intensity of that beam in the numerator of equation (\ref{eqn:RscSimple}), but have the total intensity of all beams in the denominator so that the rate saturates in the right way~\citep{Lett1989}. In this spirit, when there are $N_{\rm b}$ beams of equal intensity, the scattering force due to beam $i$ whose wavevector is $\vec{k}_i$ can be written as
\begin{equation}
    \vec{F}_{i} \approx  \frac{\hbar \vec{k}_{i}\Gamma_{\rm eff}}{2} \frac{s_{\rm eff}/N_{\rm b}}{1+ s_{\rm eff}+4(\Delta-\vec{k}_i\cdot \vec{v})^2/\Gamma^2 },
\label{eqn:F_i}    
\end{equation}
where $s_{\rm eff}$ is the effective saturation parameter corresponding to the total intensity of all $N_{\rm b}$ beams.

If two of the beams counter-propagate along $x$, so that $\vec{k}_1 = - \vec{k}_2 = k \hat{x}$, and all the others are orthogonal, the total force in the $x$-direction is $F_{x} = \vec{F}_1 + \vec{F}_2$. Using a Taylor expansion for small $v$, this is $F_x = -\alpha v_x$, where
\begin{equation}
    \alpha = -\frac{\Gamma_{\rm eff}}{\Gamma} \frac{8 \hbar k^2 (\Delta/\Gamma) (s_{\rm eff}/N_{\rm b})}{(1+ s_{\rm eff}+4\Delta^2/\Gamma^2 )^2}.
\label{eqn:dampingConstant}
\end{equation}
is the damping coefficient.

A similar approach can be used to estimate the restoring force in a MOT, where the counter-propagating beams have opposite handedness and the magnetic field at $x$ is $A x$ for some constant $A$ . In a simple picture of the MOT, the transitions driven by the beams with wavevectors $\pm k \hat{x}$ are Zeeman shifted by the angular frequencies $\pm \Delta\omega_{\rm Z} = \pm \mu_{\rm eff} A x / \hbar$, where $\mu_{\rm eff}$ is a characteristic magnetic moment. This changes $\Delta$ to $\Delta \mp \Delta\omega_{\rm Z}$ in the expression for the force due to each beam, equation (\ref{eqn:F_i}). Using a Taylor expansion for small $x$, the restoring force along $x$ is $F_x = -\kappa x$ where
\begin{equation}
    \kappa = -\frac{\Gamma_{\rm eff}}{\Gamma} \frac{8 k \mu_{\rm eff} A (\Delta/\Gamma)(s_{\rm eff}/N_{\rm b})}{(1+ s_{\rm eff}+4\Delta^2/\Gamma^2 )^2}
\label{eqn:springConstant}
\end{equation}
is the spring constant.

\subsubsection{Temperature}
\label{sec:temperature}

We can find an expression for the temperature following a standard argument. We take the typical three-dimensional case where there are $N_{\rm b}=6$ beams and consider the low velocity limit. Each photon scattering event involves two randomly directed momentum kicks of size $\hbar k$, one due to absorption and the other due to spontaneous emission. This is a random walk process, resulting in the mean square momentum increasing by $\delta \langle p^2 \rangle  = 2 R_{\sc} \delta t (\hbar k)^2$ in a time $\delta t$. The rate of increase of energy for a molecule of mass $m$ is 
$$
P_{\rm heat}=\frac{1}{2m}\frac{\delta\langle p^2 \rangle}{\delta t}=\frac{R_{\rm sc}\hbar^2 k^2}{m}.
$$
The damping force is $\vec{F}=-\alpha \vec{v}$ and the corresponding cooling power is 
$$
P_{\rm cool} = \vec{F}\cdot\vec{v} = -\alpha v^2 = -\frac{2 \alpha E}{m},
$$
where $E$ is the kinetic energy. At equilibrium, $P_{\rm heat}+P_{\rm cool}=0$. We equate the equilibrium energy to $3/2 k_{\rm B} T_{\rm D}$, giving the Doppler temperature,
\begin{equation}
    T_{\rm D} = \frac{\hbar^2 k^2}{3} \frac{R_{\rm sc}}{k_{\rm B}\alpha}.
    \label{eqn:T_D1}
\end{equation}
Comparing equations (\ref{eqn:T_D_alpha_relation}) and (\ref{eqn:T_D1}) we see that, within this model, the diffusion constant is
\begin{equation}
    D = \frac{\hbar^2 k^2 R_{\rm sc}}{3}.
    \label{eqn:D_Simple}
\end{equation}

Equation (\ref{eqn:T_D1}) can be used to find the Doppler temperature along with any method for determining the scattering rate and damping constant, either analytical or numerical. In the simplest approach, we can use equations (\ref{eqn:RscSimple}) and (\ref{eqn:dampingConstant}) for $R_{\rm sc}$ and $\alpha$, giving the result
\begin{equation}
    T_{\rm D} = - \frac{\hbar \Gamma^2}{8 k_{\rm B}\Delta} (1+s_{\rm eff}+4\Delta^2/\Gamma^2)
    \label{eqn:T_D2}
\end{equation}
for negative $\Delta$. This is the same result as for a two-level atom. The minimum temperature is obtained when $\Delta = - \tfrac{1}{2} \Gamma \sqrt{1+s_{\rm eff}}$ and is
\begin{equation}
    T_{\rm D, min}=\frac{\hbar \Gamma}{2 k_{\rm B}}\sqrt{1+s_{\rm eff}}.
\label{eqn:T_D,min}
\end{equation}

\subsubsection{Applications of the rate model}
\label{sec:RateModelApplications}

The rate model gives a good estimate of the photon scattering rate in cases where dark states are destabilised at a rate close to or exceeding the scattering rate (see section \ref{sec:dark_destable}). This is the typical situation for laser slowing and magneto-optical trapping. For SrF molecules, the scattering rates measured in an rf MOT and a dc MOT (see section \ref{sec:mots}) agree reasonably well with the predictions of the complete rate model and with the approximate result of equation (\ref{eqn:RscSimple})~\citep{Norrgard2016}. For CaF in a dc MOT, the measured scattering rate was found to follow equation (\ref{eqn:RscSimple}) but with $\Gamma_{\rm eff}$ about half the value given by equation (\ref{eqn:Gamma_eff}), while the approximate analytical results for $R_{\rm sc}$ were found to agree well with the full numerical results over a wide range of intensities~\citep{Williams2017}.

The trapping forces in a MOT are also described well by the rate model. For example, measurements of the spring constant as a function of scattering rate in an rf MOT of SrF were found to fit well to the form of equation (\ref{eqn:springConstant}) re-expressed in terms of $R_{\rm sc}$~\citep{Norrgard2016}. 
For a dc MOT of CaF, the measured spring constant was found to be in good agreement with the numerical results of the rate model, differing by less that 40\% for measurements spanning two orders of magnitude in intensity~\citep{Williams2017}. Those results also fit well to equation (\ref{eqn:springConstant}) with $I_{\rm s,eff}$ and $\mu_{\rm eff}\Gamma_{\rm eff}$ treated as free parameters.

\subsubsection{Limitations of the rate model}

In most situations, the rate model fails to describe either the damping constant or the temperature. In the molecule MOTs studied so far, damping constants are typically an order of magnitude smaller than predicted by either equation (\ref{eqn:dampingConstant}) or numerical results of the rate model. Temperatures in a MOT are often an order of magnitude higher than given by equation (\ref{eqn:T_D1}) or (\ref{eqn:T_D2}) and show a much stronger dependence on intensity than suggested by those equations. In an optical molasses, with suitably chosen parameters, temperatures far below $T_{\rm D}$ are obtained. These departures from the rate model results are all due to strong Sisyphus-type forces which often overwhelm the Doppler cooling forces, especially in the low velocity regime where damping constants and temperatures are typically measured. The proper description of these forces requires a model based on the optical Bloch equations.

\subsection{Optical Bloch equations}
\label{sec:optical_bloch_equations}

The use of optical Bloch equations to model the motion of atoms in laser fields was first developed by \citet{Gordon1980} and later extended to multi-level atoms by \citet{Ungar1989}. These methods were recently extended to molecules and to three-dimensional light fields by \citet{Devlin2016, Devlin2018}.

\subsubsection{The model}
\label{sec:OBE_the_model}

The Hamiltonian of the system is
\begin{equation}
    H = H_{\rm field} + H_{\rm mol} + H_{\rm int}
\end{equation}
where 
\begin{align}
    H_{\rm field} &= \frac{1}{2} \int \left( \epsilon_0 \vec{E}^2 + \frac{1}{\mu_0}\vec{B}^2\right) dV,\\
    H_{\rm mol} &= \frac{\vec{P}^2}{2m} + \sum_{n} \hbar \omega_{n} \ket{n}\bra{n},\\
    H_{\rm int} &= -\vec{\mu}\cdot \vec{B} - \vec{d}\cdot\vec{E}.
\end{align}
Here, $\vec{E}$ and $\vec{B}$ are the electric and magnetic field operators, $\vec{d}$ and $\vec{\mu}$ are the electric and magnetic dipole moment operators, $\vec{P}$ is the momentum operator, $\hbar \omega_n$ is the energy of state $n$ and the sum runs over all relevant internal states of the molecule. It is convenient to use the notation $\ket{g,j}$ and $\ket{e,k}$ to distinguish levels of the electronic ground and excited states, with $j$ and $k$ representing all quantum numbers needed to identify each state. We also introduce the classical electric field of the light $\vec{E}'$.

The optical Bloch equations (OBEs) are obtained from the Heisenberg equation of motion
\begin{equation}
    \frac{d A}{d t} = -\frac{i}{\hbar}[A,H]
    \label{eqn:Heisenberg}
\end{equation}
applied to the molecular operators $\ket{g,j}\bra{e,k}$, $\ket{e,k}\bra{e,k'}$ and $\ket{g,j}\bra{g,j'}$. The commutator with $H_{\rm int}$ is evaluated by expanding $\vec{d}$ as
\begin{equation}
\vec{d} = \sum_{j,k}  \vec{d}_{j,k}  \ket{g,j}\bra{e,k} + {\rm h.c.}
\end{equation}
where
\begin{equation}
    \vec{d}_{j,k} = \bra{g,j} \vec{d} \ket{e,k} 
\end{equation}
and h.c. stands for the hermitian conjugate. The magnetic moment can be expanded in a similar way. Spontaneous emission is introduced using the radiation reaction approximation in which the total electric field is the sum of the applied electric field of the light and a reaction field $\vec{E}_{\rm RR} = \frac{1}{6\pi\epsilon_0 c^3}\frac{d^3}{d t^3} \vec{d}$~\citep{Ungar1989}. The resulting equations are given in \citet{Devlin2016} for the case where the ground and excited states each have a single hyperfine level, and in \citet{Devlin2018} where there are multiple hyperfine levels. 

The force is
\begin{equation}
    \vec{F} = \frac{d \vec{P}}{d t} = -\frac{i}{\hbar}[\vec{P},H] = -\vec{\nabla}H.
\end{equation}
In the second step we have used equation (\ref{eqn:Heisenberg}), and in the third step we have used the fact that $H$ can be written in powers of the coordinates $x_i$ and momenta $p_i$, and the general result that $[p_i,g(\vec{x})]=-i\hbar\, \partial g/\partial x_i$ for any $g(\vec{x})$ that can be expanded in powers of $x_i$. The force due to the direct interaction with a small applied magnetic field can often be neglected in comparison to the much larger force exerted by the light field. In that case the expectation value of the force can be written in terms of $\vec{\nabla}(\vec{d}_{j,k}\cdot \vec{E}')$ and  expectation values of the molecular operators, $\langle\ket{g,j}\bra{e,k}\rangle$. The latter are found by solving the OBEs for a molecule dragged at constant velocity through the light field. In 1D, depending on the polarization configuration, the solutions either reach a steady state or a quasi steady state where they become periodic with the same periodicity as the Hamiltonian. In 3D, the Hamiltonian is periodic along certain directions and it is convenient to solve for motion along those directions. After initial transients have decayed away, the expectation value of the force becomes periodic and averaging over this period, and over a set of initial positions, gives the force along that direction. Repeating this procedure for many different velocities produces a velocity-dependent force curve. 

We also need to determine the diffusion coefficient, which is~\citep{Cohen-Tannoudji1992}
\begin{equation}
    D_i = {\rm Re} \int_{-\infty}^{t} {\rm d}\tau \left( \langle F_i(\tau)F_i(t) \rangle - \langle F_i(\tau)\rangle\langle F_i(t) \rangle \right),
\end{equation}
where $i$ denotes the direction and $\vec{F}(t)$ is the force at time $t$. This involves quantities of the form $\langle (\ket{g,j}\bra{e,k}(\tau))(\ket{g,j}\bra{e,k}(t))\rangle$, the expectation value of the product of molecular operators at two different times. This is not given directly by the OBEs, but for simple systems at rest can be related to the solutions of the OBEs using the quantum regression method described by \citet{Gordon1980} and \citet{Molmer1991b}. Calculations of $D(v)$ for a molecule moving through a 3D light field have not yet been attempted. However, numerical results for $D(v)$ for various $F \rightarrow F'$ systems in 1D light fields can be found in \citet{Chang2002} and could potentially be used to approximate molecules in 3D light fields. Alternatively, and more simply, equation (\ref{eqn:D_Simple}) can be used as a lower limit to the diffusion constant.

\subsubsection{Sisyphus forces in 1D}\label{sec:Sisyphus1D}

There are two variations of the Sisyphus force for molecules, both utilizing dark states. They differ according to the mechanism that takes the molecule out of the dark state - either motion through a changing polarization, or a magnetic field. 

\begin{figure}[!tb]
\centering
\includegraphics[width=0.8\textwidth]{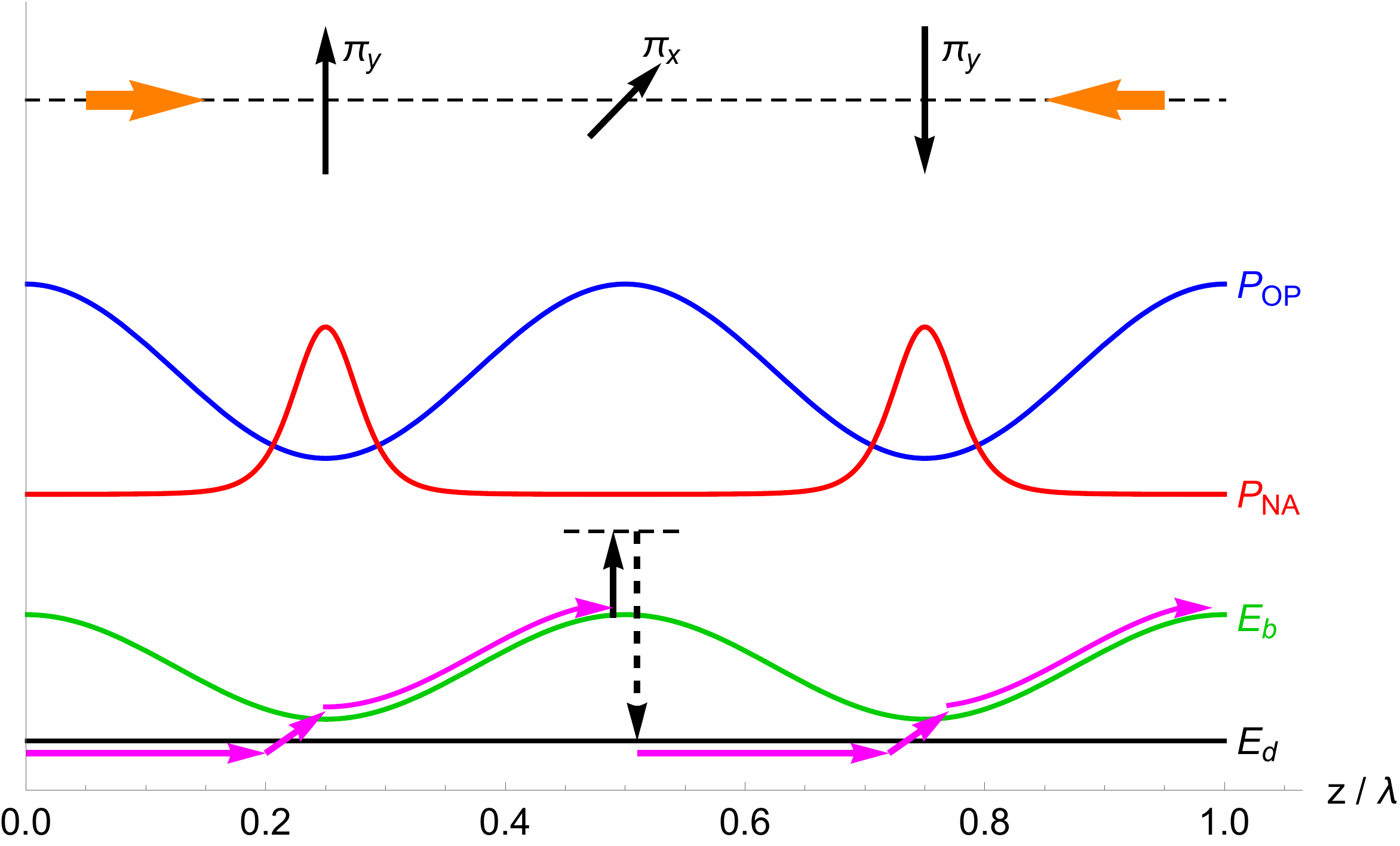}
\caption{Illustration of the non-adiabatic Sisyphus mechanism for an $F=1$ to $F'=1$ transition, positive detuning, and the lin-$\phi$-lin polarization configuration with $\phi=\pi/4$. The polarization changes from linear along $y$ at $z=\lambda/4$ to circular at $z=3\lambda/16$ to linear along $x$ at $z=\lambda/2$. $E_{\rm d}$ and $E_{\rm b}$ are the energies of the dark and bright states. $P_{\rm OP}$ is the optical pumping probability which is proportional to the intensity and peaks at $z=n\lambda/2$ (integer $n$) where $E_{\rm b}$ is largest. $P_{\rm NA}$ is the non-adiabatic transition probability from the dark to the bright state which peaks at $z=(2n+1)\lambda/4$ where $E_{\rm b}$ is smallest.}
\label{fig:subDopplerMechanism}
\end{figure}

Figure~\ref{fig:subDopplerMechanism} illustrates the first mechanism, in which the motion of the molecule through a changing polarization takes it from a dark to a bright state. The method was described by \citet{Weidemuller1994} in the context of cooling atoms below the recoil limit, and our description closely follows the one given there. We consider a molecule with an $F=1$ ground state and an $F'=1$ excited state. The transition dipole moment between $m_F=1$ and $m_F'=0$ is $d_{\rm ge}$. The molecule interacts with a pair of laser beams of wavelength $\lambda=2\pi/k$, counter propagating along $z$ and linearly polarized at an angle $\pm\phi/2$ to the $x$-axis. We call this the lin-$\phi$-lin configuration. It results in an intensity standing wave and a polarization which changes every $\lambda/4$ from linear along $y$ to linear along $x$, with elliptical in between. The electric field of the light is
\begin{align}
    \vec{E}'\! &= \!E_0\!\left[\!\left(\!\cos(\frac{\phi}{2}) \hat{x}\! +\! \sin(\frac{\phi}{2}) \hat{y}\!\right)\!\cos(k z\! -\! \omega t)\! +\! \left(\!\cos(\frac{\phi}{2}) \hat{x} \!-\! \sin(\frac{\phi}{2}) \hat{y}\!\right)\!\cos(k z\! +\! \omega t)\!\right] \nonumber \\ &= \frac{E_0}{\sqrt{2}} \left[ -\cos(k z + \frac{\phi}{2}) \hat{e}_{+1} +  \cos(k z - \frac{\phi}{2}) \hat{e}_{-1}\right] e^{i \omega t} + {\rm c.c.}
\end{align}
where $\hat{e}_{\pm 1}=\mp(\hat{x}\pm i\hat{y})/\sqrt{2}$.

Using the method described in section \ref{sec:dark}, we find the position-dependent dark state of the molecule to be
\begin{equation}
    \ket{d} = A\left[ \cos(k z-\phi/2)\ket{-1} - \cos(k z + \phi/2)\ket{1} \right],
    \label{eqn:darkState}
\end{equation}
where $A$ is the normalization and we use $\ket{m_F}$ to label the ground states. There are two bright states, $\ket{0}$ which plays no role here, and
\begin{equation}
    \ket{b} = A\left[ \cos(k z+\phi/2)\ket{-1} + \cos(k z - \phi/2)\ket{1} \right].
    \label{eqn:brightState}
\end{equation}
The energy of the dark state is $E_{\rm d}=0$ and the energy of the bright state, which is its ac Stark shift, is
\begin{equation}
    E_{\rm b} = \frac{\hbar\Delta}{2}\left(-1+\sqrt{1+\frac{2\Omega^2}{\Delta^2}(1+\cos(2k z)\cos(\phi))}\right) \approx \frac{\hbar \Omega^2}{2\Delta}(1+\cos(2k z)\cos(\phi)),
    \label{eqn:brightEnergy}
\end{equation}
where the approximation holds when $\Omega \ll \Delta$. Here, $\Delta$ is the detuning and $\Omega = d_{\rm ge} E_0/\hbar$ is the Rabi frequency. $E_{\rm b}$ is positive when $\Delta$ is positive, which is the case illustrated in figure~\ref{fig:subDopplerMechanism} and described below.

A molecule in a bright state will be optically pumped to the dark state with a probability proportional to the intensity of the light field. This probability is plotted in figure~\ref{fig:subDopplerMechanism} and we see that it is largest when $E_{\rm b}$ is largest. A stationary molecule will remain in the dark state, but a molecule with speed $v$ can make a non-adiabatic transition to the bright state with a probability $P_{\rm NA} = |\hbar v \bra{d} \tfrac{d}{d z}\ket{b} / E_{\rm b}|^2$~\citep{Messiah2014}. Note that there is no transition probability from $\ket{d}$ to $\ket{0}$, which is why $\ket{0}$ plays no role. Using equations (\ref{eqn:brightState}) and (\ref{eqn:brightEnergy}) we obtain
\begin{equation}
    P_{\rm NA} = \frac{\sin^2\phi}{(1+\cos\phi\cos(2k z))^4} \left(\frac{2\Delta}{\Omega}\right)^2\left(\frac{k v}{\Omega}\right)^2.
    \label{eqn:P_NA}
\end{equation}
Figure~\ref{fig:subDopplerMechanism} illustrates how $P_{\rm NA}$ varies with $z$ when $\phi=\pi/4$, showing that it is strongly peaked near the regions where $E_{\rm b}$ is smallest.  Since transitions to $\ket{b}$ mainly happen at the bottom of the potential hill and transitions to $\ket{d}$ mainly happen at the top, molecules repeatedly climb potential hills and lose energy. This is the non-adiabatic Sisyphus mechanism which cools molecules when $\Delta > 0$ and heats them when $\Delta < 0$.

The second Sisyphus mechanism is similar to the first but does not require polarization gradients. Instead, a magnetic field couples the dark and bright states. We consider the same 1D situation described above with both beams polarized along $\hat{x}$ (i.e. $\phi=0$), producing a standing wave of intensity with uniform polarization. The dark and bright states are simply $\ket{d}=\tfrac{1}{\sqrt{2}}(\ket{-1}-\ket{1})$ and $\ket{b}=\tfrac{1}{\sqrt{2}}(\ket{-1}+\ket{1})$. The energy of the bright state, $E_{\rm b} = \hbar \omega_{\rm b}$ is given by equation (\ref{eqn:brightEnergy}). In a magnetic field $\vec{B}=B\hat{z}$, the $\ket{\pm 1}$ states have a Zeeman shift of $\pm \hbar \omega_{\rm Z} = \pm g_{F}\mu_{\rm B} B$. In the basis of $\ket{d}$ and $\ket{b}$ the Hamiltonian is
\begin{equation}
H = \hbar\left(
\begin{array}{cc}
 0 & -\omega_{\rm Z} \\
 -\omega_{\rm Z} & \omega_{\rm b} \\
\end{array}
\right).
\label{eqn:magneticSisyphusHam}
\end{equation}

\begin{figure}[!tb]
\centering
\includegraphics[width=\textwidth]{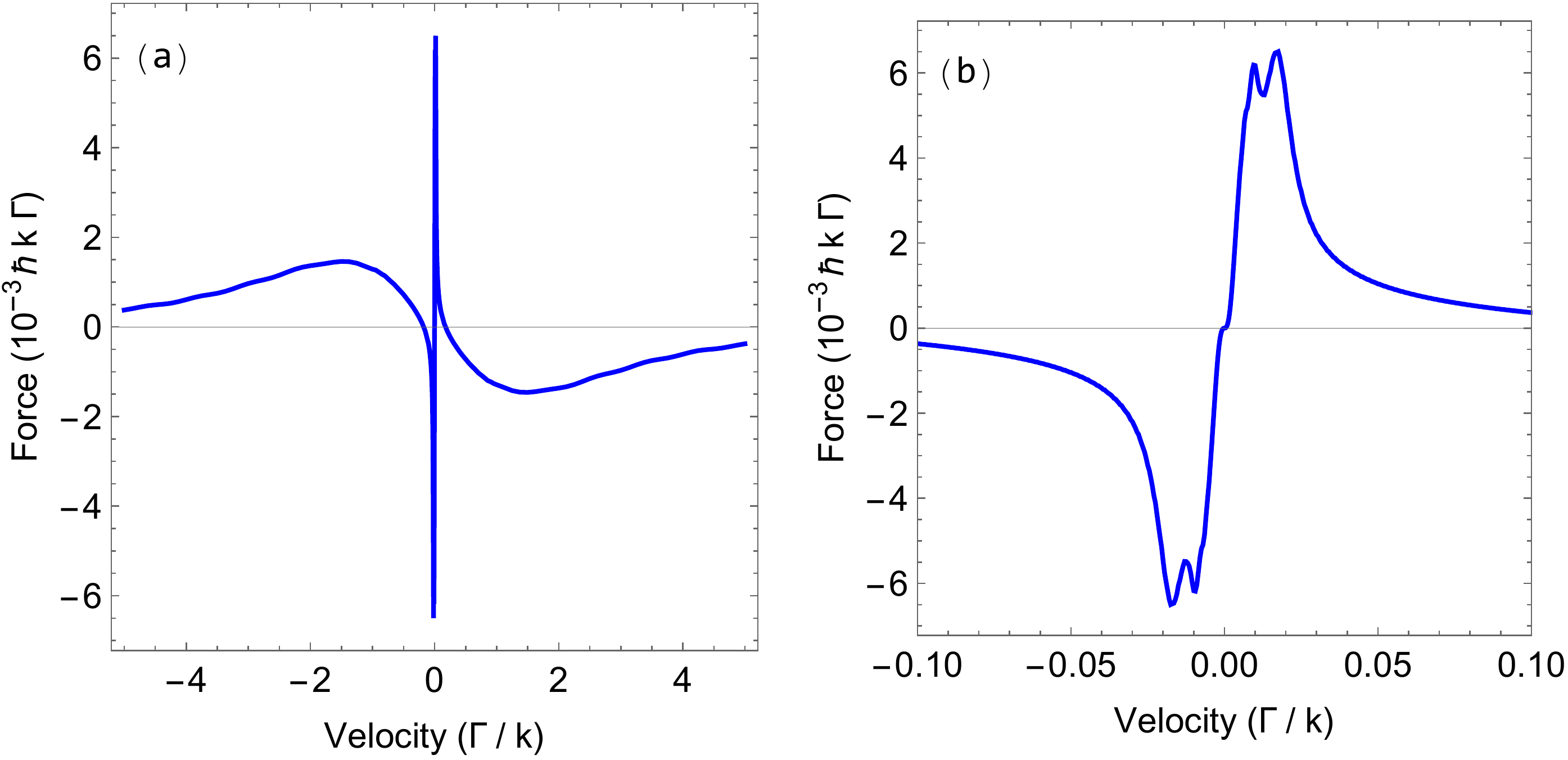}
\caption{Force versus velocity for a molecule with ground state $F=1$ and excited state $F'=1$ interacting with light in the 1D lin-$\phi$-lin polarization configuration discussed in the text. Parameters are $\phi = \pi/4$, $\Delta = -2.5\Gamma$, $\Omega = \Gamma/2$, $B=0$. (a) shows a wide velocity range and (b) a narrow one. Modified from \citet{Devlin2016}.}
\label{fig:OBE1DExample}
\end{figure}

Consider a stationary molecule that is in the dark state at time $t=0$. The probability of being in the bright state at time $t$ is
\begin{equation}
    P_{\rm b}(t) = \frac{1}{1+\eta^2}\sin^2(\sqrt{1+\eta^2}\, \omega_{\rm Z} t)
\end{equation}
where $\eta = \omega_{\rm b}/(2\omega_{\rm Z})$. When $\Omega \ll \Delta$ we can use the approximate form of $E_{\rm b}$ in equation (\ref{eqn:brightEnergy}), giving $\eta \approx \eta_{\rm max} \cos^2(k z)$ where $\eta_{\rm max} = \Omega^2/(2\Delta g_{F} \mu_{\rm B} B)$. We focus on the case where $\eta_{\rm max} \gg 1$. Away from the nodes of the standing wave, the probability of being in the bright state is always very small and oscillates at high frequency, while at the nodes the probability oscillates between 0 and 1 at angular frequency $\omega_{\rm Z}$. For a moving molecule the Hamiltonian in equation (\ref{eqn:magneticSisyphusHam}) is time-dependent and the Schr\"odinger equation has to be solved numerically, but for a slowly-moving molecule the picture is the same as already described -- the molecule is highly likely to remain in the dark state except when it passes through the nodes where it has a high probability of transferring to the bright state. This results in exactly the same picture of Sisyphus cooling as in figure~\ref{fig:subDopplerMechanism}, and is sometimes called the magnetically-assisted Sisyphus effect. It was first elucidated in the context of sub-Doppler cooling of atoms \citep{Emile1993,Sheehy1990}, and was described in the very first work on laser cooling of molecules where it was found to be an effective cooling mechanism~\citep{Shuman2010}.

Figure~\ref{fig:OBE1DExample} shows an example of the force obtained from the solutions to the optical Bloch equations for an $F=1$ to $F'=1$ system. In this example, $\phi = \pi/4$ and $B=0$, so the force at low velocity is due to the non-adiabatic Sisyphus effect. Because the detuning is negative, this force results in strong heating. Reversing the detuning reverses the sign of the force. The maximum Sisyphus force in this system is similar in magnitude to the sub-Doppler cooling forces obtained in type-I systems (i.e. when $F'>F$) at these values of $\Omega$ and $\Delta$. At higher speeds, $v > 0.2 \Gamma/k$, there is a damping force in this configuration, but this is about 100 times weaker than the Doppler cooling force in a type-I system. The equivalent force curves for the magnetically-assisted Sisyphus effect are very similar to the ones shown in figure~\ref{fig:OBE1DExample}, and can be found in \citet{Emile1993}.

\subsubsection{Sisyphus forces in 3D}

In a three-dimensional light field there are always intensity gradients and polarization gradients. As a result, the non-adiabatic Sisyphus force outlined above plays a central role in laser cooling of type-II systems. The magnetically-assisted Sisyphus effect can also be important in 3D if a magnetic field of a suitable size is applied, typically in the range 0.1~G to a few G.

\begin{figure}[!tb]
\centering
\includegraphics[width=\textwidth]{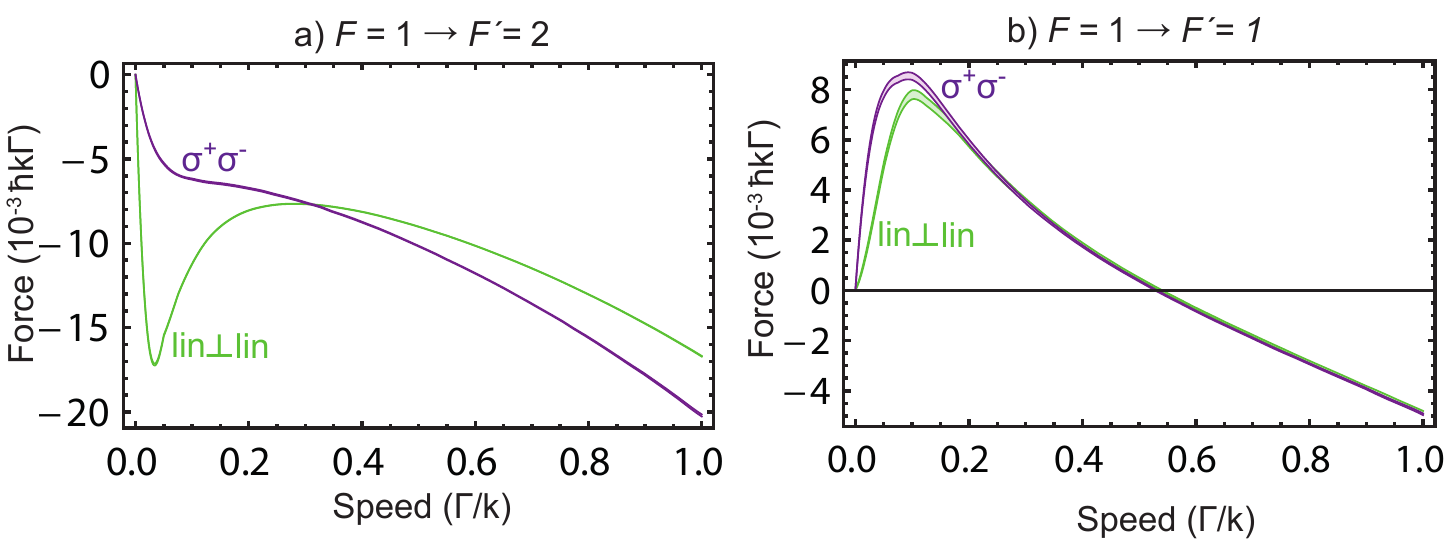}
\caption{Force in the direction of the velocity, plotted as a function of velocity, for two different 3D polarization configurations and two different systems. a) $F=1 \rightarrow F'=2$, b) $F=1 \rightarrow F'=1$. Parameters are $\Delta = -2.5\Gamma$, $\Omega = \Gamma/2$, $B=0$. The widths of the lines show the 68\% confidence intervals on the mean force. Modified from \citet{Devlin2016}.}
\label{fig:OBE3DExample}
\end{figure}

In 3D, the force has components both parallel and perpendicular to the velocity. Although the perpendicular components can be significant~\citep{Devlin2016}, especially at low speeds, we focus here on the parallel component, which has been explored in much more detail. Figure~\ref{fig:OBE3DExample} shows these parallel forces for two standard 3D polarization configurations, the $\sigma^{+}\sigma^{-}$ configuration and the lin-$\perp$-lin configuration, and compares the results for a type-II system ($F=1 \rightarrow F'=1$) to those for a type-I system ($F=1 \rightarrow F'=2$). Here, the detuning is negative. The results in 3D for the type-I system are very similar to their 1D counterparts and can be interpreted in terms of well known 1D models of sub-Doppler cooling - the Sisyphus effect for the lin-$\perp$-lin configuration and orientational cooling for $\sigma^{+}\sigma^{-}$~\citep{Dalibard1989, Ungar1989}. These polarization gradient forces have the same sign as the Doppler cooling force and provide much stronger damping at low velocities, leading to temperatures below the Doppler limit. The results in 3D for the $F=1 \rightarrow F'=1$ system are quite different to the 1D case. In fact, for these two polarization configurations, the force is zero at all velocities in 1D, while in 3D the force is large, similar in magnitude to the type-I case. In type-II systems the Sisyphus-type force, which dominates at low velocity, has the opposite sign to the Doppler cooling force which dominates at high velocity and the dominance of the Sisyphus force extends to a higher velocity than in the type-I systems.  This leads to a zero crossing of the force at a critical velocity, $v_{\rm c}$, whose value is far greater than the mean velocity at the Doppler limit, even at modest intensities. For negative detuning, the force drives the molecules towards $v_{\rm c}$ and this high equilibrium speed leads to the high temperatures found in magneto-optical traps of type-II systems. Unlike type-I systems where polarization gradient forces are inhibited by a magnetic field, the Sisyphus forces in type-II systems persist over the whole range of magnetic fields typically seen by molecules in a MOT~\citep{Devlin2018}. The critical velocity is found to be independent of detuning and proportional to the square root of the laser intensity~\citep{Devlin2016}. This is consistent with the observation that the temperature in molecular MOTs increases with increasing laser intensity~\citep{Norrgard2016, Truppe2017b, Williams2017, Anderegg2017}.

For positive detuning, molecules with speeds above $v_{\rm c}$ are accelerated to high speed, while those below $v_{\rm c}$ are strongly damped towards zero velocity. As can be seen from the illustration in figure~\ref{fig:subDopplerMechanism}, slow-moving molecules spend much of their time in dark states, so the photon scattering rate is low. This combination of a low heating rate and strong damping can lead to very low temperatures, as described in section~\ref{sec:sub-Dopp}. 

\subsubsection{Applications of the OBE model}

\begin{figure}[!tb]
\centering
\includegraphics[width=1.0\textwidth]{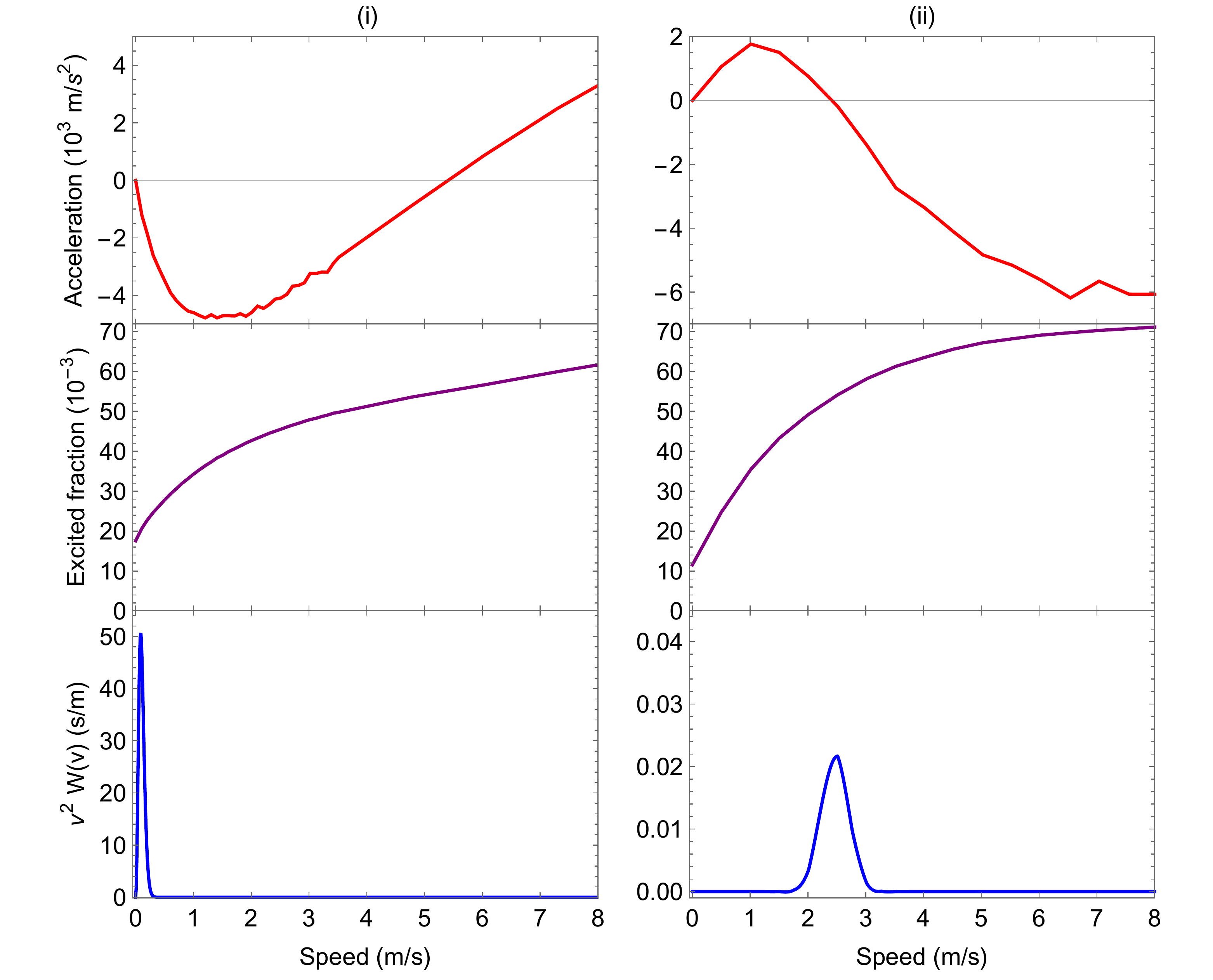}
\caption{Acceleration, excited-state fraction and probability density versus speed for CaF molecules in two situations. (i) Optical molasses. Intensity of cooling light is 456~mW/cm$^2$ and detuning is $2.61\Gamma$. (ii) Centre of MOT. Intensity of cooling light is 234~mW/cm$^2$ and detuning is $-0.64\Gamma$. Modified from~\citet{Devlin2018}.
}
\label{fig:CaFSims}
\end{figure}

OBE models have been used to simulate experiments demonstrating one-dimensional transverse cooling of YbF~\citep{Lim2018} and YbOH~\citep{Augenbraun2020} beams and to estimate the scattering rate for transverse cooling of BaH~\citep{McNally2020}. For YbF, the spatial distributions determined using the FPK equation are in fairly good agreement with measurements. For YbOH, the force curves generated by solving the OBEs were used for trajectory simulations of a set of molecules. These simulations include random decays to dark vibrational levels and the random momentum diffusion due to photon recoil. The simulated beam profiles generated this way agree well with the measured ones, and the simulations are accurate in predicting how the beam width and fraction of ultracold molecules depend on the detuning and intensity. For both YbF and YbOH, the simulations predict temperatures in the range 1-10~$\mu$K, which is below the temperature resolution of the experiments.

In 3D, generalised optical Bloch equations have been used to calculate the velocity-dependent force for CaF in an optical molasses, and the position dependence and velocity dependence of the force in a MOT~\citep{Devlin2018}. The OBEs also give the scattering rate as a function of velocity, which was used in equation (\ref{eqn:D_Simple}) to estimate the diffusion constant. The resulting $F(v)$ and $D(v)$ were then used in the FPK equation in order to determine velocity distributions, and to estimate how temperatures and other parameters depend on the various experimental parameters. Figure~\ref{fig:CaFSims} summarizes some key information obtained from these simulations. We see that the force curves for the full molecular system are found to be similar to those for the model $F=1 \rightarrow F'=1$ system presented in figure~\ref{fig:OBE3DExample}b). The fraction of molecules in the excited state is smaller than predicted by a rate model at all speeds, and is in good agreement with measured values. The excited-state population drops at low speed due to optical pumping into transient dark states, though does not go to zero at zero velocity. This shows that motion through the light field is an important factor in de-stabilizing dark states, though not the only factor. 

In the molasses, where the light is blue-detuned (figure~\ref{fig:CaFSims}(i)), the velocity distribution is close to a thermal distribution. The modelling predicts a damping time in the molasses of about 100~$\mu$s, close to the measured value, but the predicted temperature is 3-6 times lower than measured over a wide range of parameters, indicating that the fluctuating dipole force, which is neglected in the model, is a major source of heating in the optical molasses. The predicted temperature dependence on intensity, detuning and magnetic field were all found to be similar to the measured trends. In the MOT, where the light is red-detuned (figure~\ref{fig:CaFSims}(ii)), the velocity distribution peaks near the velocity where the force crosses zero. The distribution is determined by a balance between Doppler cooling and Sisyphus heating. Although the distribution is far from thermal, the ballistic expansion of molecules with this distribution looks similar to that of a thermal distribution. The temperatures determined this way depend on the intensity, and are similar to the measured values.

OBE simulations have also proven useful in designing deep laser cooling schemes that cool molecules towards the recoil limit~\citep{Cheuk2018, Caldwell2019}. These methods typically rely on the combination of strong Sisyphus cooling to damp the velocity towards zero, and optical pumping into robust dark states so that stationary molecules do not scatter photons. In~\citet{Caldwell2019} two deep laser cooling schemes are modelled and compared to experiment. For one scheme, the model suggests that it is feasible to cool below the recoil limit, although this has yet to be achieved and the simulation method breaks down in that limit.

\subsubsection{Limitations of the OBE model}

As described in section~\ref{sec:OBE_the_model}, the force is obtained by finding the quasi steady state solutions of the OBEs for a molecule dragged at constant velocity through the light field. The accuracy of this method is questionable since it ignores the transient response of the molecule. It takes many photon scattering events for the transient to diminish, and in that time the force may change the speed appreciably, while the momentum diffusion may randomise it. Nevertheless, for speeds $v \gtrsim 0.2 \sqrt{\hbar \Gamma/m}$, the method is found to agree quite well with the force obtained from Monte Carlo simulations that include a random walk to simulate the momentum diffusion~\citep{Ungar1989}. This means that we might expect the method to work well for temperatures down to about $0.2 T_{\rm D,min}$, but to be inaccurate at lower velocities. The method also treats the motion of the molecule classically, so must break down once the temperature approaches the recoil limit.

Another significant limitation of the model is the difficulty of accurately determining the diffusion constant. While methods exist to evaluate $D(v)$ in some cases~\citep{Gordon1980, Molmer1991b, Agarwal1993}, they have not yet been extended to the more complex situation of a molecule moving through a 3D light field. Without a proper evaluation of $D(v)$, the temperatures predicted by the model will not be accurate. The development of methods to address this shortcoming would be particularly valuable.

 \section{Laser slowing}
\label{sec:slowing}

\subsection{Molecular beams and radiation-pressure slowing}

Radiation-pressure slowing is crucial for loading a magneto-optical trap (MOT), discussed in section~\ref{sec:mots}, and is even more important for molecules than it is for atoms. The capture velocity of an atomic MOT (the maximum atom velocity that the MOT can capture) is typically a few tens of meters per second. For atoms with high enough vapour pressure near room temperature, the vapour contains enough of these low velocity atoms to load the MOT directly~\citep{Monroe1990}. Alternatively, to protect the high vacuum region of the MOT from the background vapour, a 3D MOT can be loaded from a slow beam extracted from a 2D MOT~\citep{Weyers1997}. When the vapour pressure is not sufficient for these methods, the MOT is loaded from a continuous atomic beam that is slowed to below the MOT capture velocity using the radiation pressure of counter-propagating laser light. As the atoms slow down their Doppler shift changes, and this can be compensated by tailoring the Zeeman shift using a carefully designed magnetic field profile~\citep{Phillips1982}, applying a frequency chirp to the slowing light~\citep{Ertmer1985}, or frequency broadening the slowing light to address all molecules regardless of their forward velocity~\citep{Zhu1991}. These atomic beams are almost always continuous, so, within limits, more trapped atoms can be obtained by loading the MOT for longer. MOT load times of 1--10~s are typical.   

The situation is quite different for molecules. First, because the photon scattering rate is reduced (see figure~\ref{fig:scatteringRateComparison}), the MOT capture velocity is smaller than for atoms, typically about 10~m/s~\citep{Williams2017}. Second, the molecules are often chemically unstable free radicals that must be created in-situ, typically as a \emph{pulsed} molecular beam. So far, all molecular MOTs have started with pulsed molecular beams produced by a cryogenic buffer-gas source~\citep{Hutzler2012}.  These sources generate short pulses, typically 0.1--10~ms in duration, with typical forward velocities of 100--200~m/s. The low initial velocities make these beam sources particularly attractive for laser cooling experiments. Because their repetition rates are only a few Hz, it is normal to only load a single pulse of molecules.  For this reason, and because the capture velocity is low, the number of particles in a molecular MOT is several orders of magnitude smaller than most atomic MOTs. This makes it especially important to develop efficient methods of decelerating the molecules.   

To date, all molecular MOT experiments have used frequency-chirped or frequency-broadened laser slowing methods.  For typical speeds of 100--200~m/s, and a cycling transition at a visible wavelength, the required frequency range of the chirp or broadening is a few hundred MHz. With scattering rates of around $10^{6}$ photons/second, the acceleration is about 10$^{4}$~m/s$^{2}$ and the molecules will come to rest in about 10~ms after scattering around $10^{4}$ photons.  Rapid destabilization of dark states is important for efficient slowing and can be done using a magnetic field or polarization modulation (see section~\ref{sec:dark}). The first radiation pressure slowing of a molecular beam was demonstrated by \citet{Barry2012}. Starting from a SrF beam with a mean speed of 140~m/s, about 6\% of the molecules were decelerated to speeds below 50~m/s. The frequency spectrum of the laser was broadened to a few hundred MHz to implement the frequency broadened slowing method. Using this method, the final velocity is determined by the frequency spread and average detuning of the light, and the duration of the slowing. In frequency-chirped slowing~\citep{Zhelyazkova2014, Truppe2017}, the final beam velocity is determined by the initial detuning, the chirp rate, and the duration of the slowing period. We note that some experiments use a combination of frequency-broadened and frequency-chirped slowing~\citep{Yeo2015}.

The slowing method should bring as many molecules as possible into the capture volume and capture velocity of the MOT. This is challenging, because the initial distribution from the molecular source is often broad in velocity, position, and time. Without any transverse cooling, slow molecules diverge rapidly, so ideally all molecules would reach the target velocity just as they reach the MOT. In practice, laser slowing tends to bring molecules to a particular forward velocity at a particular time, and the beam often has a wide spread of axial positions at that time. Those molecules that are still far from the MOT will diverge too much and are unlikely to be captured.  More molecules could be captured if the slowing procedure naturally brought them to the desired forward velocity at a specific position.  That happens in the Zeeman slowers commonly employed in cold atom experiments~\citep{Phillips1982}, and the development of Zeeman slowing for molecules is ongoing~\citep{Petzold2018,Liang2019}. Quasi-continuous beams based on high repetition rate molecular beams sources are also currently being developed~\citep{Shaw2020} and are likely to substantially increase the number of molecules loaded, especially in combination with a Zeeman slower. Other ways to improve the efficiency of beam-to-MOT transfer are discussed below.

\subsection{Simulating the slowing sequence}

Simulations are useful to evaluate the performance of a chosen slowing method, and to maximize the number of slow molecules loaded into the MOT as a function of the many variable parameters, e.g amount of frequency broadening, chirp amplitude, chirp rate and functional form. Simulating individual molecule trajectories using a semi-classical Monte-Carlo approach based on a rate-model has been found to accurately predict most observed slowing results~\citep{Truppe2017,Williams2017}. Coherences appear to play little role in beam slowing, so the rate model is adequate.  

Once the frequency spectrum and polarization of the light is fixed, the steady-state scattering rate, $R_{\rm sc}= \Gamma n^e$, depends on only two parameters, a detuning $\delta$ and the total laser intensity $I$.  In this reduced parameter space, it is computationally efficient to calculate $R_{\rm sc} $ for a large set of $\delta$ and $I$ values, interpolate over these values to make a scattering rate map $R_{\rm sc}(\delta, I)$, and then use this map to simulate individual molecule trajectories. After calculating the transition amplitudes between all magnetic sublevels of the ground and excited states for the chosen magnetic field and laser polarization, $R_{\rm sc}$ is calculated by finding the steady-state solutions to equations~(\ref{eqn:pops}). The steady-state solutions are sufficient because the populations reach the steady state on a timescale much shorter than any changes in the slowing parameters. To account for population remixing amongst the magnetic sublevels by an applied magnetic field, an additional ad hoc term
\begin{equation}
    \label{eqn:magnetic_remixing}
    \sum\limits_{j'}\frac{\omega_{j} - \omega_{j'}}{2\pi}(n^{g}_{j} - n^{g}_{j'})\delta_{F F'}
\end{equation}
is added to the right hand side of equation~(\ref{eqn:popg}).  Here,
$\hbar \omega_{j}$ is the Zeeman shift of level $j$. 

Simulating molecule trajectories begins by choosing a set of molecules with initial positions and velocities reflecting the characteristics of the molecular beam. Their trajectories are calculated as follows. At each time step $dt$ the probability of scattering a photon is $P = R_{\rm sc} dt$. The time step is chosen so that $P \ll 1$. A random number, $r$, between 0 and 1 is chosen, and if $P > r$ a photon is scattered. In that case, the momentum in the direction of the laser's $k$ vector is increased by $\hbar k$ to account for the absorption, and the momentum in a direction chosen at random from an isotropic distribution is increased by $\hbar k$ to account for spontaneous emission. The state distribution of the molecules in the beam can be tracked by choosing the excited and subsequent ground state according to a quasi-random sampling of the set of available paths weighted by the the relative transition dipole strengths in excitation and the corresponding branching ratios in decay. In this way, slowing, stochastic heating, and the evolution of the internal degrees of freedom are taken into account. The spatial profile of the slowing laser, overall intensity, initial detuning, and the time-dependence of any chirp, are all parameters that can be optimized through these simulations and compared to experiments.  Many of these parameters can be explored and optimized without re-calculating the scattering rate map.

\subsection{Frequency chirped versus frequency broadened slowing}

\begin{figure}[!tb]
    \centering
    \includegraphics[width=\linewidth]{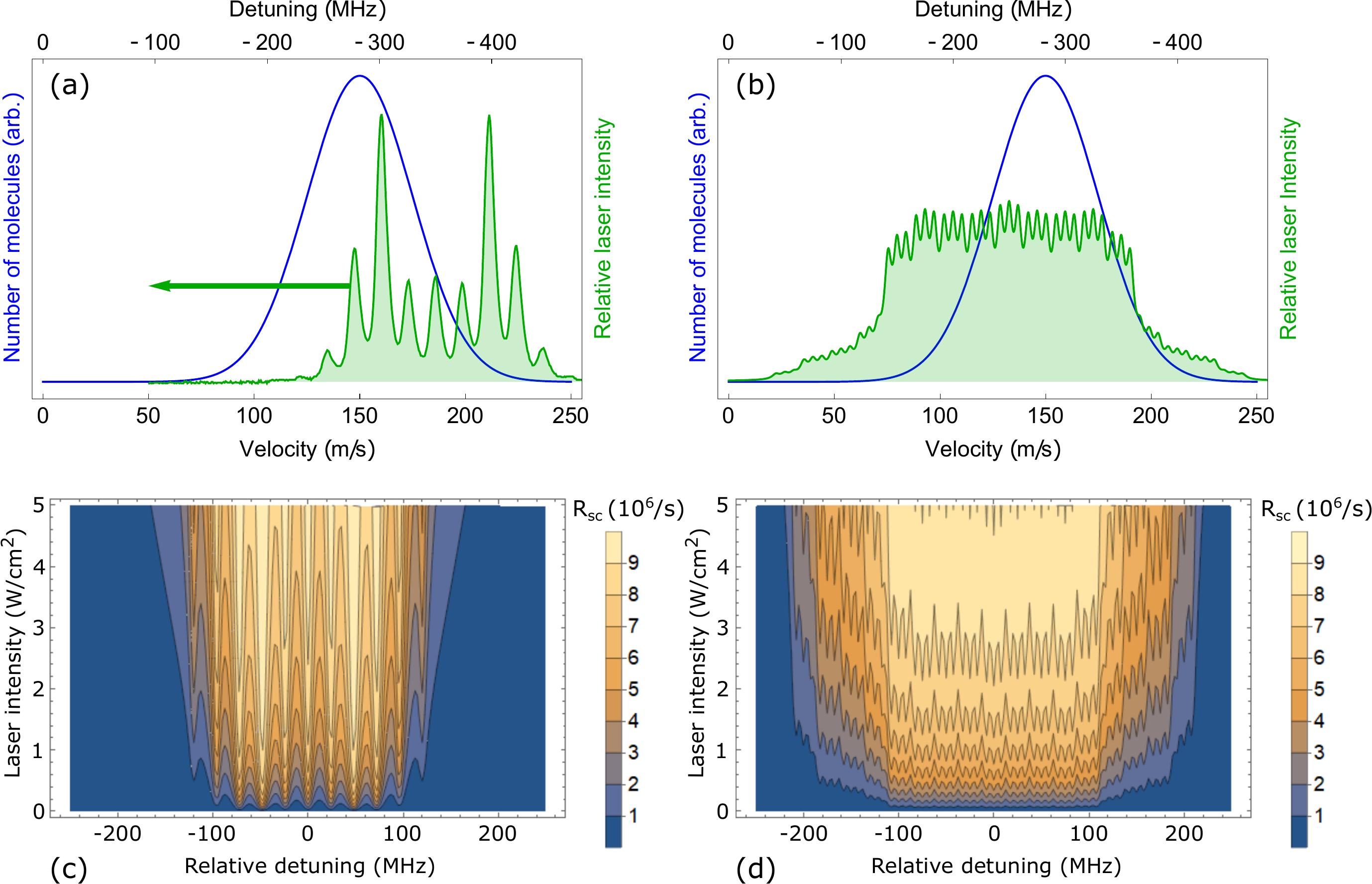}
    \caption{Two general approaches for laser slowing.  The blue curve(s) indicate a typical initial velocity distribution for a buffer-gas molecular beam.  (a) In chirped-laser slowing, frequency sidebands are added to the slowing laser to address the hyperfine components of the transition.  As the molecules slow down, the light is chirped towards resonance, indicated by the green arrow.  (b) In frequency-broadened slowing, a fixed frequency spectrum is broadened more significantly and detuned such that molecules stop scattering photons once the target velocity has been reached.  (c,d) Examples of scattering rate maps for laser slowing of CaF molecules on the $B^{2}\Sigma^{+}-X^{2}\Sigma^{+}$ transition using the laser frequency spectra shown in (a,b).}
    \label{fig:slowing}
\end{figure}

Which is most effective, frequency chirped slowing or frequency broadened slowing? This is an important question that simulations can help answer. Both approaches have been studied for CaF molecules, slowed using the $B^{2}\Sigma^{+}(v'=0) - X^{2}\Sigma^{+} (v=0)$ cycling transition~\citep{Truppe2017}. Some results are illustrated in figure~\ref{fig:slowing}.  For frequency-chirped slowing, shown in figure~\ref{fig:slowing}(a), frequency sidebands are added to the slowing light in order to address all hyperfine components rather than addressing molecules with many different speeds. In this case, an EOM was used to generate the frequency sidebands shown. The initial overall detuning is chosen so that molecules near the peak of the velocity distribution (solid blue line) are Doppler shifted into resonance. The overall chirp of the entire sideband structure is indicated by the green arrow.  In frequency-broadened slowing, shown in figure~\ref{fig:slowing}(b), the spectrum of the laser light is broadened over a wider range.  The structure shown was generated by sequentially passing the slowing light through EOMs driven at 74, 24, and 8~MHz.  Example scattering rate maps corresponding to these two sideband structures are shown in figure~\ref{fig:slowing}(c) and (d).

\begin{figure}[p]
    \centering
    \includegraphics[width=0.9\linewidth]{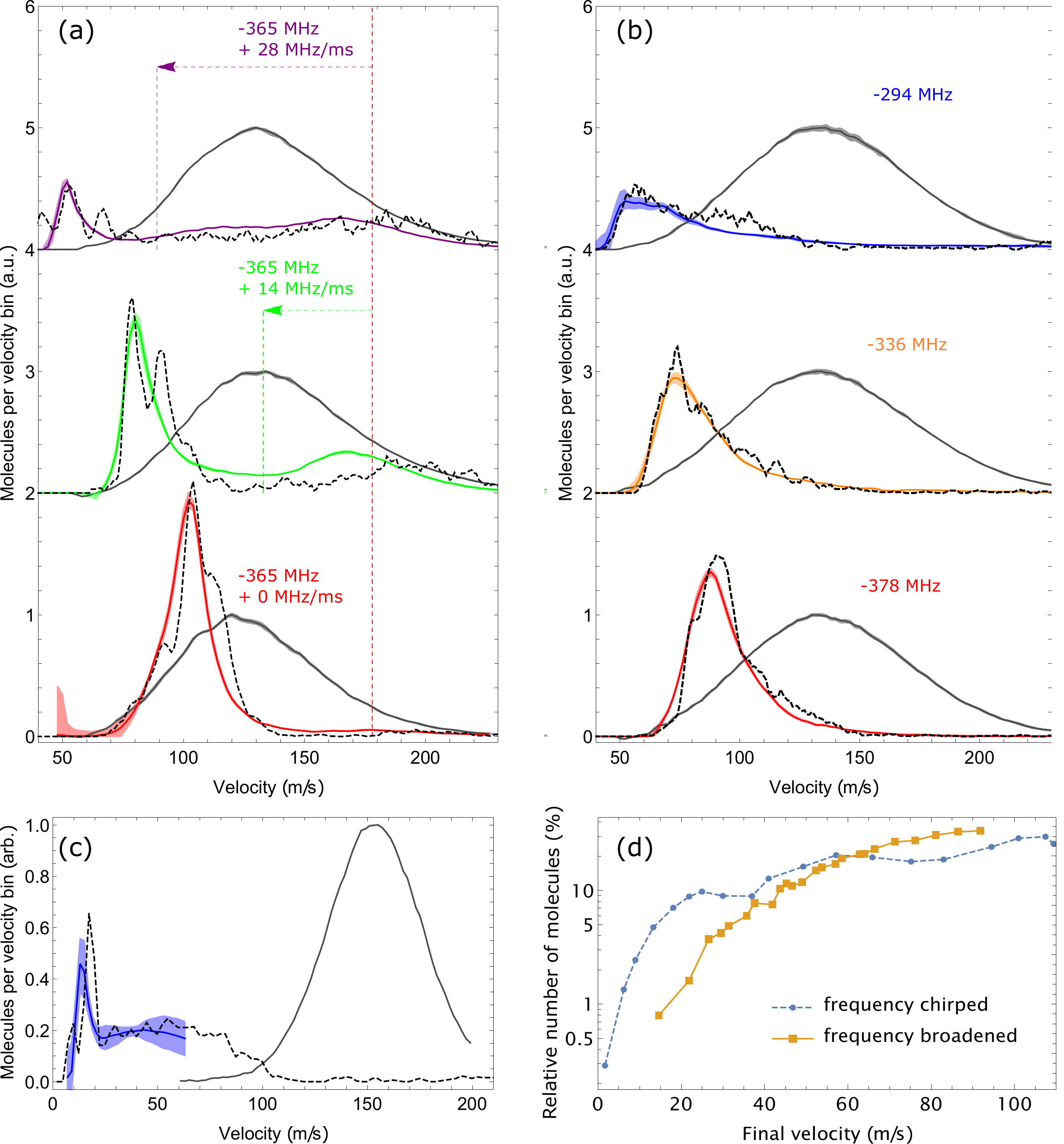}
    \caption{A comparison of frequency-chirped and frequency-broadened laser slowing approaches for CaF molecules.  The initial velocity distribution present in the molecular beam appears as a solid black line.  (a) Velocity distributions resulting from frequency-chirped laser slowing experiments (solid coloured lines) and simulations (dashed lines) for various chirp rates over a 7~ms slowing time, as measured 1.3~m from the beam source.  Coloured bands indicate 68\% confidence limits. (b) Similar velocity distributions resulting from frequency-broadened slowing for various overall detunings.  (c) Results from a longer and faster chirp suitable for producing molecules at low speeds appropriate for loading a MOT.  (d) Simulated relative number of molecules within a 10~m/s wide range centered at different mean forward velocities.  Values are expressed as a percentage of the total number of molecules arriving at the same detector when no slowing is applied.  Adapted from \citet{Truppe2017} with permissions.}
    \label{fig:slowing_sims}
\end{figure}

Figure~\ref{fig:slowing_sims} compares the results of frequency chirped and frequency broadened slowing experiments for CaF molecules, with corresponding simulations~\citep{Truppe2017}. We see that frequency chirped slowing provides significantly more axial cooling of the molecular beam. This is because chirped slowing slows the faster molecules before the slower ones, which compresses the axial velocity distribution.  By contrast, frequency-broadened slowing addresses most forward velocities present in the molecular beam at all times and does not produce significant cooling until the molecules are slow enough to fall out of resonance with the light. The higher level of control of the beam forward velocity, together with increased axial cooling, makes frequency-chirped slowing more efficient at bringing molecules to the very low velocities needed for loading a MOT, as shown in figure~\ref{fig:slowing_sims}(c) and figure~\ref{fig:slowing_sims}(d).  Chirped slowing is also a more efficient way to use the available laser power since, at all times, most of the power is at frequencies that are close to resonance. On the other hand, frequency-broadened slowing is significantly simpler, with fewer control parameters and less sensitivity to the exact characteristics of the molecular beam. It may be the preferred method when the distribution of forward velocities is large, or when the source-to-trap distance is too short for a chirp to be useful, or when the molecular beam properties tend to vary in time.

\subsection{Reducing losses during slowing}

In current experiments, only a small fraction\footnote{The trajectory simulations discussed above for frequency-chirped or frequency-broadened slowing estimate that typically $<0.01$\% of the molecules that leave the buffer-gas source are captured in the MOT.}  of the molecules produced by the source are captured in the MOT because the beam diverges as it is slowed. This divergence is exacerbated by the stochastic heating associated with spontaneous emission (see section \ref{sec:temperature}). Typically, the transverse temperature increases by a few mK, corresponding to an increase in the transverse velocity spread of $\lesssim$ 1~m/s. One way to reduce the transverse divergence slightly is to use a converging slowing laser, which has been shown to increase the number of molecules in the MOT by a factor of 2--3~\citep{Steinecker2016}.  A number of approaches are being pursued to more significantly reduce losses associated with beam divergence during slowing. Ideas include: (i) decelerate the beam so rapidly that it does not have time to diverge; (ii) focus or guide the beam while it is being slowed; (iii) apply transverse cooling either before or during slowing to minimize beam divergence. All have the potential to increase the number of molecules in a MOT by a large factor.

The first option can work well for a light molecule that has a very strong transition at a short wavelength, so that the scattering rate is high and the number of photons that have to be scattered is relatively low. The AlF molecule is an interesting example, having a cycling transition at 228~nm with a decay rate of 84~MHz (see table \ref{tab:laser-cooled-molecules}).  Rapid deceleration can also be achieved using \emph{stimulated} scattering processes~\citep{Chieda2011,Ilinova2015,Kozyryev2018,Galica2018,Long2019,Wenz2020}. These approaches can transfer momentum at a high rate and can even work for molecules that lack good cycling transitions since they largely bypass spontaneous emission.  However, the intensities available from current continuous wave lasers are probably not sufficient  to produce these strong forces over the large spatial area needed to slow typical molecular beams.  Current work by \citet{Beyer2020} is seeking to fill this technological gap.

Option (ii), guiding while slowing, is an attractive option that could be achieved in several ways. For example, \citet{Fitch2016} have proposed a magnet geometry where focusing stages alternate with strong-field stages, and optical pumping transfers molecules between weak- and strong-field seeking states, resulting in both guiding and deceleration. Since the magnetic field is used to remove kinetic energy, rather than laser photons, far fewer photons need to be scattered in order to slow the molecules down.  As such, this technique may be particularly useful for molecules that lack highly-closed cycling transitions.  Another option, proposed by \citet{DeMille2017}, is to use a microwave field blue detuned from a rotational transition in the molecule to provide transverse confinement.  Alternatively, laser-coolable molecules are usually amenable to more established slowing methods such as Stark~\citep{Bethlem1999,Osterwalder2010}, Zeeman~\citep{Vanhaecke2007b, Narevicius2008, Lavert-Ofir2011}, or centrifuge~\citep{Chervenkov2014} deceleration.  All of these provide transverse guiding during slowing but typically select only a small fraction of the axial positions and velocities in the molecular beam for deceleration to the target velocity. The combination of these slowing methods with laser cooling techniques~\citep{Wall2011} is an interesting avenue which, at present, remains largely unexplored. 

Option (iii), transverse cooling, is potentially very powerful since it produces a highly collimated molecular beam that can then be slowed efficiently. Additionally, an electric or magnetic lens can be used to focus molecules into the laser cooling region in order to capture a larger fraction of the initial transverse velocity distribution, resulting in an intense, slow, collimated beam~\citep{Fitch2020b}.     

\section{Magneto-optical trapping}
\label{sec:mots}

The magneto-optical trap (MOT)~\citep{Raab1987} is the starting point of almost all experiments that use ultracold atoms. The MOT cools and traps atoms using pairs of counter-propagating laser beams, typically along three orthogonal axes, along with a magnetic field gradient to distinguish between the beams. In recent years, MOTs of SrF, CaF and YO molecules have been made~\citep{Barry2014,Truppe2017b,Anderegg2017,Collopy2018}, and MOTs of several other molecular species are being developed. Magneto-optical traps can be classified according to the total angular momenta of the ground and excited states, $F$ and $F'$. In type-I MOTs, $F' > F$ so there are no dark states, whereas in type-II MOTs $F' \le F$ so there are dark states in the system that play an important role. Type-II atomic MOTs~\citep{Raab1987, Prentiss1988, Shang1994, Flemming1997, Nasyrov2001, Atutov2001, Tiwari2008} result in lower densities and higher temperatures than type-I MOTs, so are very rarely used. The closed cooling transitions in molecules are type-II transitions (see section \ref{sec:rotationalTransitions}), so molecular MOTs are type-II MOTs. In the following, we consider the nature of the cooling and trapping forces in these MOTs and review the progress in producing them. As a reminder, we use $F$ and $F'$ to refer to the total angular momentum of the ground and excited states, $g$ and $g'$ to refer to their magnetic moments, and $\Delta$ to refer to the detuning of the light from resonance. We use $\sigma^{\pm}$ to refer to polarizations that drive $\Delta m = \pm 1$ transitions, and the term `restoring beam' to refer to the beam that pushes a displaced molecule back towards the centre. Unless stated otherwise, the $z$-axis is in the direction of the magnetic field.

\begin{figure}[!tb]
    \centering
    \includegraphics[width=\linewidth]{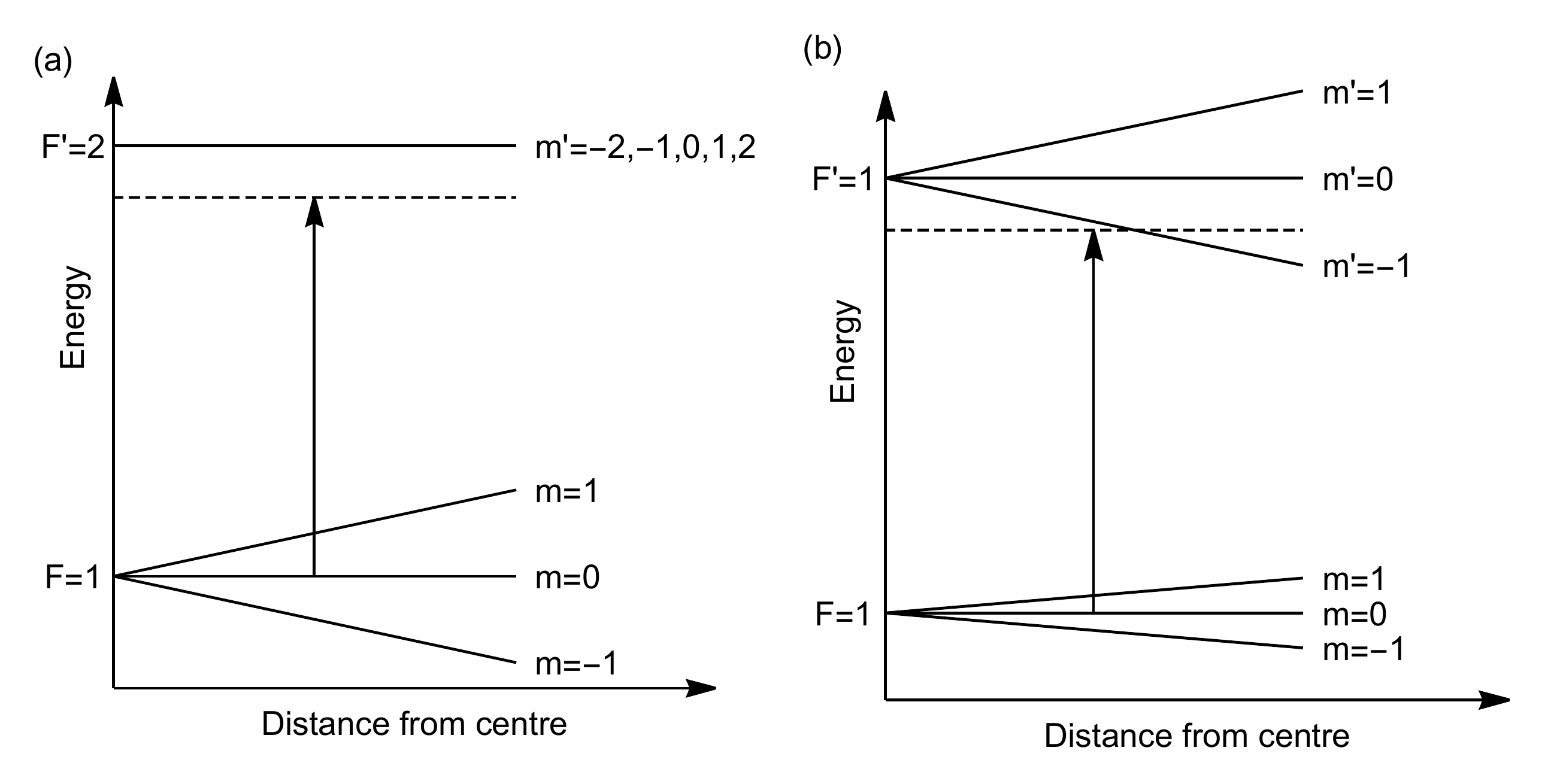}
    \caption{Level structures where MOTs fail. (a) No Zeeman splitting in the excited state. (b) $F=1 \rightarrow F'=1$ in 1D.}
    \label{fig:badMOT}
\end{figure}

An analysis of MOTs for atoms or molecules with a single transition $F \rightarrow F'$ driven by a single frequency of light brings out several useful insights~\citep{Tarbutt2015}. (1) If there is no Zeeman splitting in the excited state ($g'=0$) there is no confining force. The confining forces in a MOT arise because the Zeeman splitting brings the atom closer to resonance with the restoring beam. When $g'=0$, and for any $m$, the $\Delta m = \pm 1$ transitions have the same detuning, so there is no net force. This is illustrated in figure~\ref{fig:badMOT}(a) and is true for any choice of $F$ and $F'$. (2) When $g' > 0$ and $\Delta < 0$, the restoring beams should be polarized to drive $\sigma^{-}$ transitions when $F' \ge F$, and to drive $\sigma^+$ transitions when $F' < F$. These polarization requirements are reversed when $g'<0$ or when $\Delta > 0$. The sign of $g$ has no influence on the required polarization. (3) In a one-dimensional MOT there is no confining force when $F=F'=1$. This case is illustrated in figure~\ref{fig:badMOT}(b). When the atom is in $m=1$ it is only bright to the beam that drives $\Delta m = -1$ transitions, and when it is in $m=-1$ is is only bright to the opposing beam that drives $\Delta m = +1$ transitions. In both cases the excited state is $m'=0$ which decays to $m=\pm 1$ with equal probability. It follows that equal numbers of photons are scattered from the counter-propagating beams, so there is no net force. Although this result is special to this case, it illustrates the general problem of making MOTs when there are dark states - the atom quickly goes dark to the beam that would otherwise push it towards the centre. (4) In three dimensions the orthogonal beams play an important role in pumping molecules out of dark states, and a confining force can be recovered. Nevertheless, the presence of dark states results in substantially smaller forces than for type-I MOTs.

The first issue discussed above is particularly important for molecules with $^{2}\Pi_{1/2}-{}^{2}\Sigma $ cooling transitions. A pure $^{2}\Pi_{1/2}$ state has $g'\approx 0$ because the magnetic moments associated with the orbital and spin angular momenta are very nearly equal but opposite. Mixing of the $^{2}\Pi_{1/2}$ state with nearby $\Sigma$ states results in a non-zero value of $g'$, but this is usually small. Without mitigation, the small value of $g'$ and the presence of dark states would result in very weak confining forces. Fortunately, there are two good ways to avoid these problems.

\subsection{Dual-frequency MOT}

\begin{figure}[!tb]
    \centering
    \includegraphics[width=\linewidth]{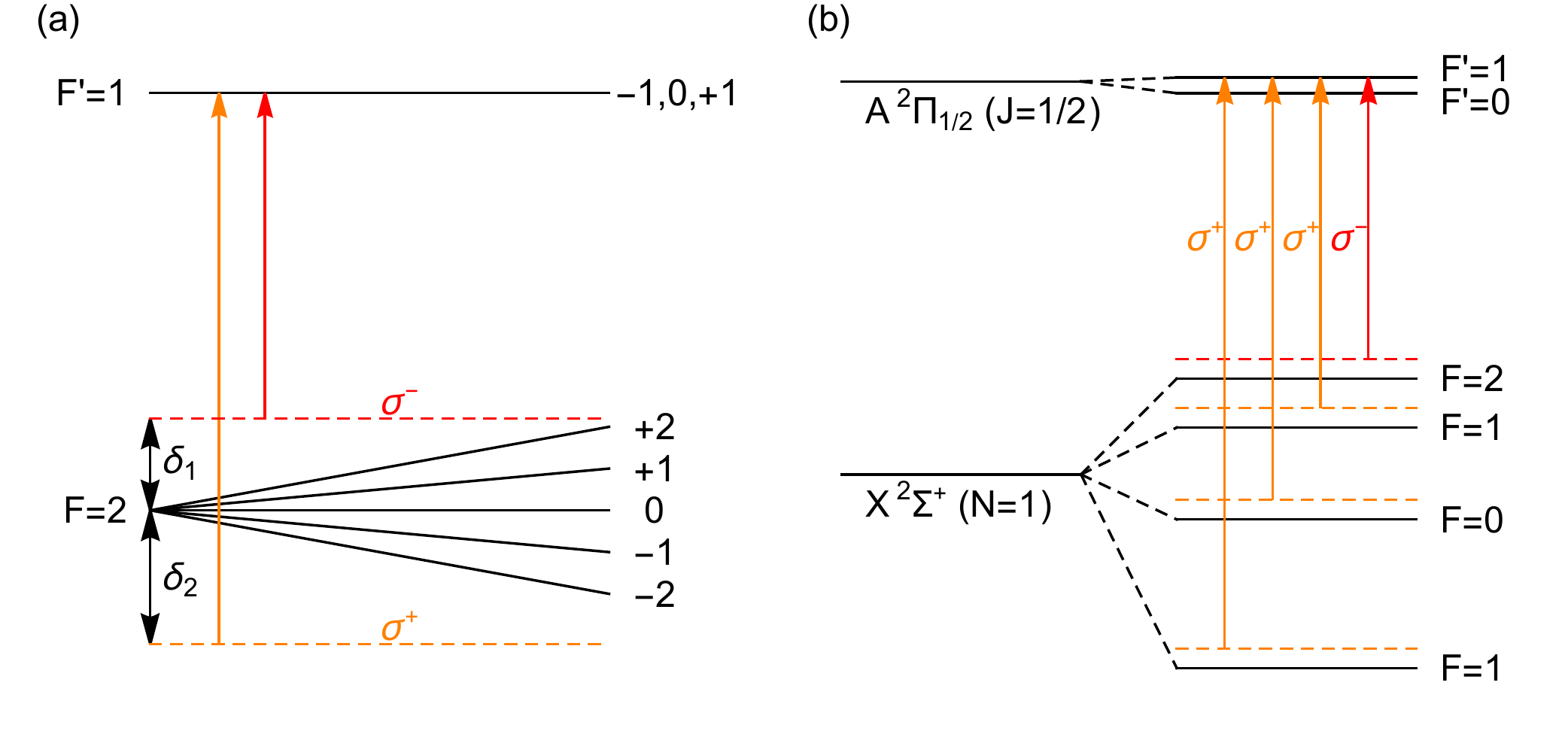}
    \caption{(a) Scheme for dual frequency MOT, illustrated for a transition with $F=2$ and $F'=1$. The transition is driven by two frequency components of opposite detuning and polarization handedness. (b) Implementation of this scheme for CaF.}
    \label{fig:dcMOT}
\end{figure}

The difficulties outlined above apply when there are isolated transitions driven by a single frequency of light. They can be avoided by using two frequencies of opposite polarization. This scheme, introduced in \citet{Tarbutt2015b}, is known as a dual-frequency MOT or a dc MOT, and is illustrated in figure~\ref{fig:dcMOT}(a) for the particular case of an $F=2$ ground state and an $F'=1$ excited state. Each MOT beam contains two frequency components of opposite handedness, one red detuned and the other blue detuned. The red detuned restoring beam drives $\sigma^-$ transitions, and the blue detuned one drives $\sigma^+$ transitions. As always, the opposing beams drive the opposite transitions. A molecule in a state of positive $m$ can only interact with a $\sigma^{-}$ component. The one from the restoring beam is closer to resonance so the molecule preferentially scatters photons from this beam. Similarly, when the molecule is in a negative $m$ state it can only interact with a $\sigma^+$ component. Once again, it is the component from the restoring beam which is closest to resonance, so the molecule still scatters more photons from this beam and there is a net trapping force. The presence of both red and blue detuned components leads to a competition between Doppler cooling and heating, but the cooling dominates for the arrangement shown in figure~\ref{fig:dcMOT}(a) where the detuning is larger for the blue component ($|\delta_2/\delta_1| > 1$).

The dual frequency scheme can often be implemented in an easy and natural way because the ground-state hyperfine structure already requires that rf sidebands be applied to the cooling light. Figure~\ref{fig:dcMOT}(b) illustrates the example of CaF cooled using the $A^{2}\Pi_{1/2} (J'=1/2)-X^{2}\Sigma^{+} (N=1)$ transition. The ground state hyperfine structure is the same as shown in figure~\ref{fig:hyperfine}(a), and we have also included the small (unresolved) hyperfine splitting of the excited state. Four frequency components are used, each red detuned from the hyperfine component it is designed to address. Choosing the polarizations in the way shown implements the dual frequency scheme shown in figure~\ref{fig:dcMOT}(a).

\begin{figure}[!tb]
    \centering
    \includegraphics[width=\linewidth]{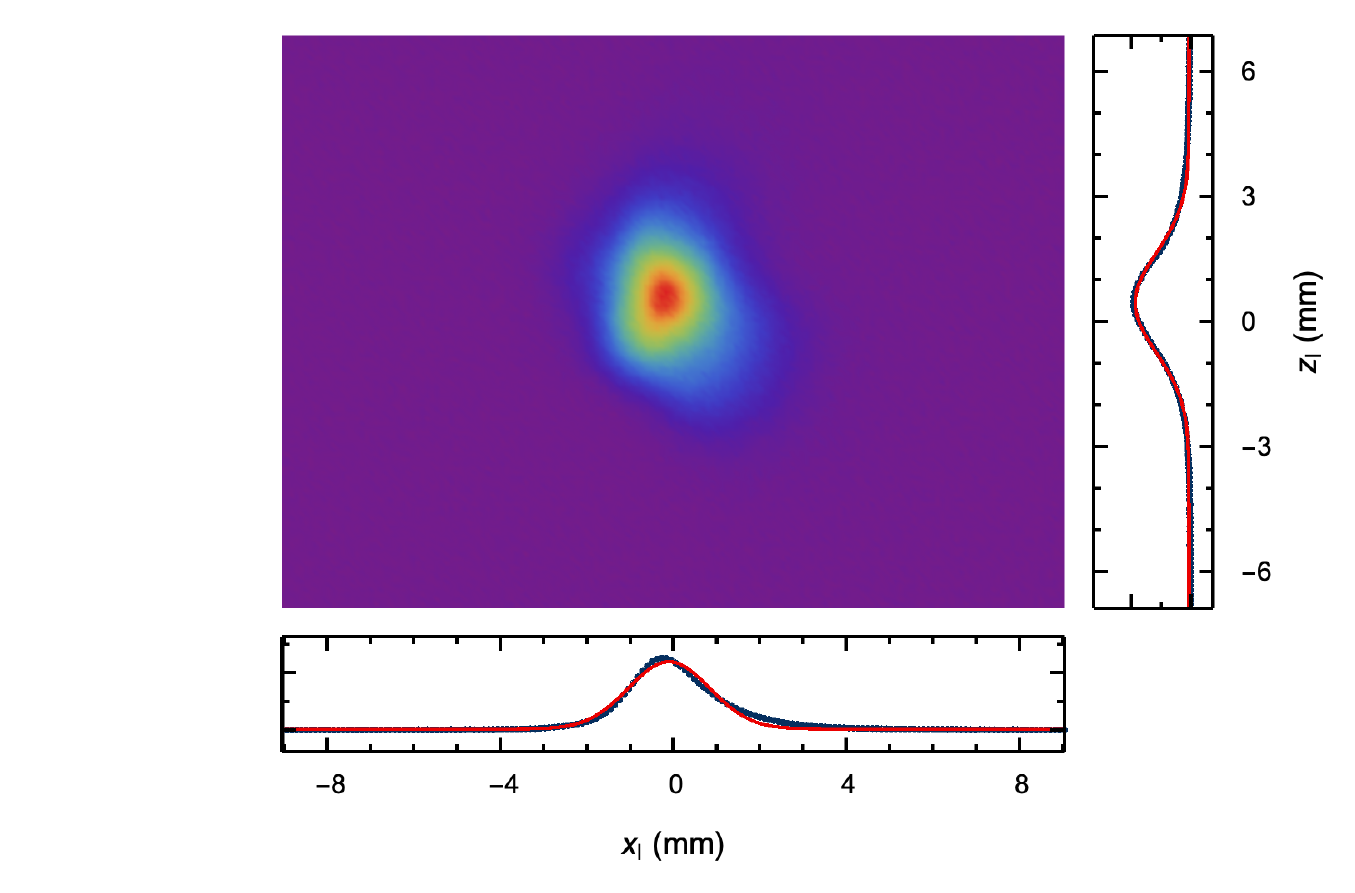}
    \caption{Single shot fluorescence image of a MOT of CaF, using an exposure time of 10~ms. Also shown are the profiles in the radial ($x_{\rm l}$) and axial ($z_{\rm l}$) directions, together with Gaussian fits to these profiles. Figure from \citet{CaldwellThesis}.}
    \label{fig:dcMOTImage}
\end{figure}

The first 3D MOT of molecules was reported by \citet{Barry2014}. They captured a few hundred SrF molecules and measured a density of 600~cm$^{-3}$, a temperature of 2.3(4)~mK, a radial trap frequency of 17.2(6)~Hz, a damping constant of 140(10)~s$^{-1}$, and a lifetime of 56~ms. By changing the polarizations of some of the frequency components, following a suggestion by \citet{Tarbutt2015}, an improved SrF MOT was obtained, yielding higher density, higher trap frequency and longer lifetime, though at the expense of a higher temperature~\citep{McCarron2015}. The dual-frequency mechanism provides most of the confinement in these MOTs. In 2017, dc MOTs of a second species, CaF, were reported by \citet{Truppe2017b} and by \citet{Anderegg2017}, followed by detailed characterisation work~\citep{Williams2017}. A dc MOT of YO molecules has also recently been studied~\citep{Ding2020}. In these MOTs the polarizations of the various frequency components were deliberately chosen to implement the dual-frequency method. Figure~\ref{fig:dcMOTImage} shows a dc MOT of CaF. The picture is obtained by imaging the MOT fluorescence onto a camera. This MOT contains about $2 \times 10^4$ molecules with a peak density of around $1.2 \times 10^6$~cm$^{-3}$.

\subsection{Radio-frequency MOT}

\begin{figure}[!tb]
    \centering
    \includegraphics[width=\linewidth]{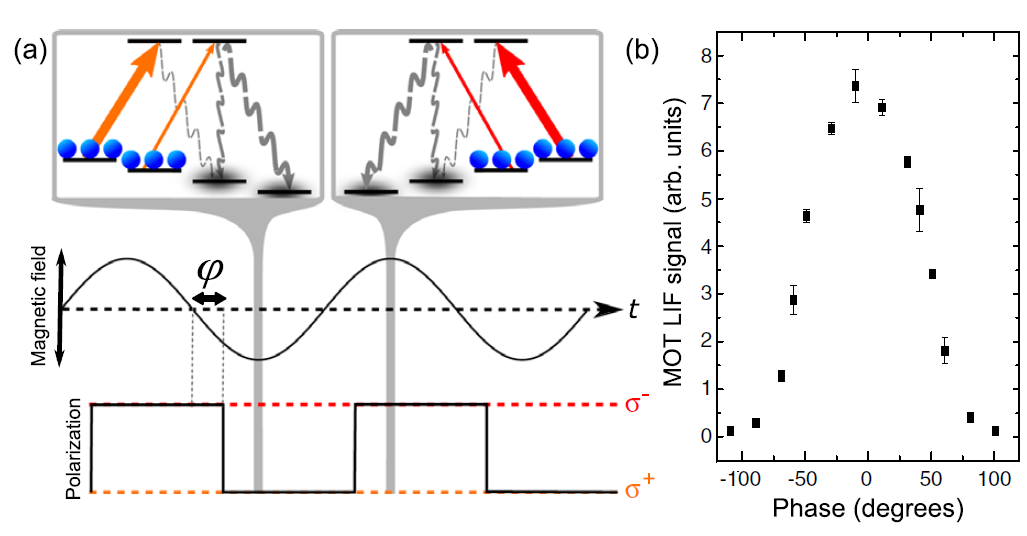}
    \caption{(a) Principle of the rf MOT. The magnetic field and laser polarizations alternate at the same frequency. Population pumped to a dark state during one half-cycle becomes bright during the next half-cycle. When the magnetic field switches in phase with the polarization a net confining force is obtained. (b) Laser-induced fluorescence signal from an rf MOT of SrF as a function of the phase difference $\varphi$, showing that the trapping is most effective near $\varphi=0$. Reprinted from \citet{Norrgard2016} with permissions.  Copyright (2016) by the
American Physical Society.}
    \label{fig:rfMOT}
\end{figure}

In an rf MOT, the quadrupole magnetic field and laser polarization are synchronously switched at a frequency that is comparable to the optical pumping time, so that no molecule remains in a dark state for long. Figure~\ref{fig:rfMOT}(a) illustrates this scheme using a picture where the $z$-direction is fixed. We imagine a molecule displaced towards positive $z$. The polarization handedness of the laser beams alternates from left to right so that a beam driving $\sigma^-$ transitions in one half-cycle drives $\sigma^+$ transitions in the next half-cycle. The figure indicates the transition driven by the restoring beam (i.e. the beam propagating towards $-z$). In the half-cycle where the magnetic field is negative, the negative $m$ states are Zeeman shifted into resonance with the restoring beam which is polarized to drive $\sigma^+$ transitions. Molecules rapidly scatter photons from this beam and are optically pumped to the positive $m$ states. Although they can scatter photons from the opposing beam, this process is slow because the light is further from resonance. In the next half-cycle the magnetic field is positive and the polarization of the restoring beam is switched so that scattering from the restoring beam is fast once again. In this way, a substantial net restoring force is applied. Note that adiabatic following of the magnetic field would ruin this scheme. The field must switch rapidly enough that (in a picture with a fixed $z$-axis) a molecule stays in the same $m$ state as the field switches its direction.\footnote{Pictures with the $z$-axis in the direction of the magnetic field are also common. In this picture, the population is transferred diabatically from $m$ to $-m$ as the field passes through zero.}

The rf MOT was first introduced by \citet{Hummon2013} who used the method to demonstrate transverse magneto-optical compression of a beam of YO molecules. More recently, \citet{Baum2020} applied the same technique to a beam of CaOH, the first time magneto-optical forces have been demonstrated for polyatomic molecules. The first 3D rf MOT of molecules was demonstrated by \citet{Norrgard2016} using SrF. Figure~\ref{fig:rfMOT}(b) shows how the laser-induced fluorescence signal from the MOT varies with the phase difference ($\varphi$) between the magnetic field and polarization switching cycles. As expected, molecules are only trapped when $|\varphi| < 90^{\circ}$, and the signal is greatest close to $\varphi = 0$ where the field and polarization switch in phase. The rf MOT of SrF was found to perform better than the earlier dc MOTs of the same molecule. More molecules could be loaded, the lifetime was much longer, and molecules remained trapped at much lower laser intensity. It was found that the temperature could be lowered to 400~$\mu$K by ramping down the intensity. Soon after, 3D rf MOTs of CaF~\citep{Anderegg2017} and YO~\citep{Collopy2018} were also demonstrated.

\subsection{Features of molecular MOTs}

Despite their increased complexity, molecular MOTs behave much like atomic MOTs in many ways. In particular, the scattering rate and spring constant are reasonably well described by simple equations of the same form as for atomic MOTs, equations (\ref{eqn:Rsc}) and (\ref{eqn:springConstant}). One strikingly different feature of molecular MOTs is the relatively high temperature of the molecules, and the increase in temperature with laser intensity. Temperatures of a few mK are typical, and temperatures as high as 15~mK are observed in some cases. These temperatures are in strong contrast to type-I atomic MOTs of alkali atoms which are usually at a few hundred $\mu$K, but are similar to the elevated temperatures observed in type-II atomic MOTs. The high temperatures can be understood in the context of the Sisyphus forces discussed in section~\ref{sec:Sisyphus1D} and the force curves shown in figure~\ref{fig:OBE3DExample}. For type-II transitions, Sisyphus forces dominate at low velocity and heat the molecules when the light is red-detuned. Doppler cooling dominates at higher velocities, so there is a critical velocity, $v_{\rm c}$, where the force crosses zero. Molecules in the MOT are driven towards $v_{\rm c}$, as can be seen in figure~\ref{fig:CaFSims}(ii). Simulations~\citep{Devlin2016,Devlin2018} predict MOT temperatures similar to the measured values and show that $v_{\rm c}^2$ is proportional to the laser intensity, which explains the observed increase of temperature with intensity.  

The lifetimes of molecular MOTs are much shorter than most atomic MOTs. In most cases lifetimes are near 100~ms, though for the SrF rf MOT lifetimes up to 500~ms have been observed at low laser intensity. The lifetimes are likely to be limited by optical pumping into states that are not part of the cooling cycle. As evidence for this, we note that lifetimes are usually found to be shorter at higher scattering rates. In addition, since trap depths may only be a few times the typical temperature, and laser-cooling rapidly establishes thermal equilibrium, molecules can be lost continuously from the high velocity tail of the velocity distribution. This loss mechanism was thought to be responsible for the short lifetimes observed in the first molecular MOT~\citep{Barry2014} and may play a significant role in all molecular MOTs produced so far.

Comparisons of dc and rf MOTs suggests that, for SrF and CaF, rf MOTs tend to perform better~\citep{Norrgard2016, Anderegg2017}, yielding more molecules at a higher density and a lower temperature. In the case of YO however, the dc MOT yields more molecules and somewhat lower temperatures~\citep{Ding2020}. A dc MOT is also easier to implement. An analysis of the advantages of each MOT type is likely to be needed for each new species investigated.

\section{Sub-Doppler cooling}
\label{sec:sub-Dopp}

The main mechanisms of sub-Doppler cooling were described in section \ref{sec:optical_bloch_equations}, together with results of simulations based on multi-level optical Bloch equations. Here, we review the experimental work.

\subsection{Cooling in one or two dimensions}

In the first demonstration of laser cooling applied to molecules, \citet{Shuman2010} cooled a beam of SrF molecules in one transverse direction. The molecular beam passed through a 15~cm long sheet of light formed by reflecting a laser beam back and forth at a slight angle. The transverse density distribution of the laser-cooled molecular beam was then measured and compared to the uncooled beam. All the ingredients needed to cool molecules to the ultracold regime were demonstrated in these remarkable first experiments. When the detuning was negative, and a 5~G magnetic field was applied to destabilize dark states, Doppler cooling was observed. With a positive detuning, and the magnetic field reduced to 0.6~G, Sisyphus cooling was observed. The cooling light was linearly polarized in these experiments, and the retro-reflected beam formed standing waves of intensity resulting in magnetically-assisted Sisyphus cooling (see section ~\ref{sec:Sisyphus1D}). With the help of simulations, the temperature obtained by Sisyphus cooling was estimated to be about 300~$\mu$K, with a conservative upper limit of 5~mK. Transverse laser cooling using Sisyphus forces has since been used to cool several other molecular species including alkaline-earth monohydrides~\citep{McNally2020}, polyatomic molecules~\citep{Kozyryev2017, Mitra2020} and the heavy polar molecules of interest for EDM measurements~\citep{Lim2018, Augenbraun2020}. These experiments show good agreement with the results of OBE simulations. At the typical laser intensities used, the capture velocity for Sisyphus cooling is found to be around 1~m/s, and temperatures as low as 100~$\mu$K have been measured. A comparison of the magnetically-assisted and non-adiabatic Sisyphus mechanisms suggests that the former tends to be most effective, at least in one dimension~\citep{Lim2018}.

\subsection{Cooling in three dimensions}

Sub-Doppler cooling was first demonstrated in three dimensions by \citet{Truppe2017b}. In these experiments, CaF molecules were first prepared in a magneto-optical trap at a temperature near 1~mK. Then, the magnetic field was turned off and the detuning of the light switched from negative to positive to form a blue-detuned optical molasses, often known as a gray molasses because of the important role of the dark states in the sub-Doppler cooling. In the molasses, the molecules cooled below the minimum Doppler temperature of $T_{\rm D,min} = 200$~$\mu$K on a timescale of less than 1~ms. After zeroing the magnetic field and optimizing the intensity, a temperature of $55(2)$~$\mu$K was reached. No molecules were lost in this molasses cooling step. The same gray molasses cooling method has been used to cool CaF molecules into an optical dipole trap~\citep{Anderegg2018}, showing that the cooling is still effective in the presence of the ac Stark shifts of the trap, and has been used to cool SrF to a similar temperature~\citep{McCarron2018}.

An analysis of gray molasses cooling of CaF shows that the cooling is limited by the presence of multiple frequency components of light with different polarizations. A state that is dark to one component is not dark to another component designed to address a different hyperfine level. Because the hyperfine splittings are quite small, the dark states are transient and the residual photon scattering is a limitation to cooling. Two approaches have been designed to avoid this limitation. One approach is to turn off three of the frequency components shown in figure~\ref{fig:dcMOT}(b), leaving just a single frequency of light~\citep{Caldwell2019}. In this case, for any polarization there are two dark states that are superpositions of the $F=2$ Zeeman sub-levels. This configuration was found to yield low temperatures ($T \lessapprox 10$~$\mu$K) provided the light was detuned to the blue of all hyperfine components of the cooling transition and the magnetic field was tuned to within 50~mG of zero. The cooling was found to be remarkably insensitive to laser detuning, working well even with the laser detuned by 350~MHz ($\Delta = 43\Gamma)$ from the $F=2$ component. The lowest temperature obtained was 5.4(7)~$\mu$K. The second approach is to use two frequency components, the uppermost and lowermost ones in figure~\ref{fig:dcMOT}(b), with the relative frequencies tuned to match the interval between the $F=2$ and the lower $F=1$ hyperfine levels~\citep{Cheuk2018}. This sets up Raman dark states which can enhance gray molasses cooling, as discussed in section \ref{sec:dark} and illustrated in figure~\ref{fig:darkRaman}. This configuration, sometimes called $\Lambda$-enhanced cooling, has been used to cool alkali atoms~\citep{Grier2013} and was shown to work well for CaF molecules. The lowest temperature of $5.0(5)$~$\mu$K was obtained when the relative laser frequency was tuned within 100~kHz of the two-photon Raman resonance condition, but the temperature was not very sensitive to the single-photon detuning from the excited state. It was shown that the cooling remains effective for molecules in an optical dipole trap, and that the residual photon scattering from the molasses can be used to image single molecules in optical tweezer traps~\citep{Anderegg2019}.

\begin{figure}[!tb]
    \centering
    \includegraphics[width=\linewidth]{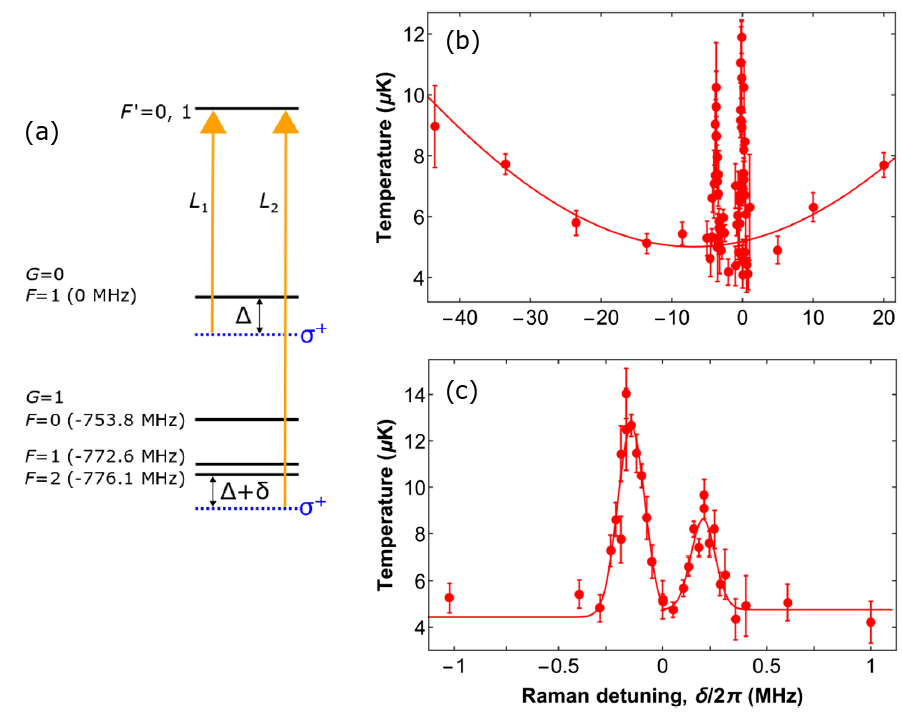}
    \caption{Sub-Doppler cooling of YO molecules. (a) Level scheme used for gray molasses cooling. Note the large frequency interval between the two hyperfine manifolds labelled $G=0$ and $G=1$. (b) Temperature versus Raman detuning $\delta$ over a wide range. (c) Temperature versus Raman detuning near the Raman resonance $\delta=0$. Adapted from \citet{Ding2020} with permissions.}
    \label{fig:YOCool}
\end{figure}

Gray molasses cooling has also recently been applied to YO molecules~\citep{Ding2020}, reaching a temperature of 4~$\mu$K. Here, the ground state hyperfine structure consists of three closely spaced levels $F=0,1,2$ and a single $F=1$ component which is very well separated from the others. This structure is interesting because it allows an exploration of various sub-Doppler mechanisms (and MOT mechanisms) for some simple, well-isolated, level schemes. The gray molasses cooling scheme used by \citet{Ding2020} is shown in figure~\ref{fig:YOCool}(a). Away from the two-photon resonance ($\delta \ne 0$) there are two approximately independent systems, each addressed by a single frequency of light that is blue detuned. This is analogous to two independent single-frequency molasses and we may expect results similar to the single-frequency molasses investigated for CaF molecules~\citep{Caldwell2019}. For example, the cooling should be robust to the two detunings $\Delta$ and $\Delta + \delta$. That is indeed what is observed, as can be seen from figure~\ref{fig:YOCool}(b) which shows that the temperature is below 10~$\mu$K over a wide range of $\delta$ away from $\delta=0$. Conversely, close to the two-photon Raman resonance ($\delta = 0$), the system is strongly influenced by the Raman dark state and behaves in a similar way to the $\Lambda$-enhanced cooling studied in \citet{Cheuk2018}. Figure~\ref{fig:YOCool}(c) shows the temperature in this region, showing features that are characteristic of $\Lambda$-enhanced cooling~\citep{Grier2013,Nath2013}. There is a minimum in the temperature at $\delta =0$ where the Raman dark state forms, and a pair of narrow peaks when $|\delta| \simeq 200$~kHz, which the authors attribute to de-stabilization of Zeeman dark states within the two separate systems due to cross-coupling near the Raman resonance. While the experiments with CaF illustrate that Raman dark states can enhance gray molasses cooling by setting up a dark state that is more robust than would otherwise exist, the experiments with YO show that adding a new one does not result in further cooling when the dark states are already stable.

The preparation of laser-cooled molecules has come a long way since the earliest demonstrations. For example, in the recent work with YO molecules~\citep{Ding2020}, $5 \times 10^4$ molecules were prepared at a peak density of $5.4 \times 10^{7}$~cm$^{-3}$ and a temperature of $4$~$\mu$K. The corresponding peak phase-space density is $3.3 \times 10^{-8}$, 9 orders of magnitude higher than in the very first magneto-optical trap of molecules~\citep{Barry2014}. As we will see in section~\ref{sec:optical_dipole}, the density and phase-space density can be increased further by cooling the molecules into an optical dipole trap.

\section{Magnetic trapping}

The ultracold molecules prepared using the methods described above need to be transferred into a conservative trap for most applications. In a conservative trap the lifetime can be much longer, the quantum state can be selected and preserved, and the phase-space density can be increased by sympathetic, evaporative or Raman-sideband cooling. Molecules with $^{2}\Sigma$ ground states can be trapped magnetically as demonstrated for ultracold SrF and CaF~\citep{McCarron2018,Williams2018}. For these molecules, a magnetic trap can have large depth and large volume and is well suited to sympathetic or evaporative cooling. We note that magnetic traps have also been used to confine warmer molecules produced by other methods, including buffer gas cooling, Stark deceleration and Zeeman deceleration~\citep{Weinstein1998,Sawyer2007, Hogan2008, Tsikata2010,Riedel2011,Lu2014,Akerman2017,Liu2017,Heazlewood2021}.

\subsection{Zeeman effect}

The complete effective Hamiltonian describing the Zeeman effect of a diatomic molecule is given in section 7.6 of \citet{BrownCarrington2003}. Here, we focus on the Zeeman effect of $^{2}\Sigma$ states since this is most relevant to our discussion of magnetic trapping of laser-cooled molecules. The effective Zeeman Hamiltonian is
\begin{equation}
    H_{\rm Z} = g_{S} \mu_{\rm B} \vec{S}\cdot\vec{B} + g_{l}\mu_{\rm B}\left[\vec{S}\cdot\vec{B} - (\vec{S}\cdot\hat{z}')(\vec{B}\cdot\hat{z}') \right] - g_{\rm r}\mu_{\rm B}\vec{N}\cdot\vec{B} - \sum_i g_{\rm N}^{i}\mu_{\rm N}\vec{I}_i \cdot \vec{B},
    \label{eq:HZ}
\end{equation}
where $\vec{B}$ is the magnetic field and $\hat{z}'$ is a unit vector in the direction of the internuclear axis. The first term is the electron spin interaction and has coefficient $g_{S}$, the second term is an anisotropic correction to the electron spin interaction and has coefficient $g_l$, the third term is the rotational Zeeman interaction characterised by $g_{\rm r}$, and the last is the nuclear Zeeman interaction characterised by the nuclear $g$-factors $g_{\rm N}^i$ and summed over the nuclei. The first term is usually about 1000 times larger than the other terms, but the small terms play an important role in determining the residual magnetic sensitivity of rotational transitions that would otherwise be magnetically insensitive.

\begin{figure}[!tb]
    \centering
    \includegraphics[width=\linewidth]{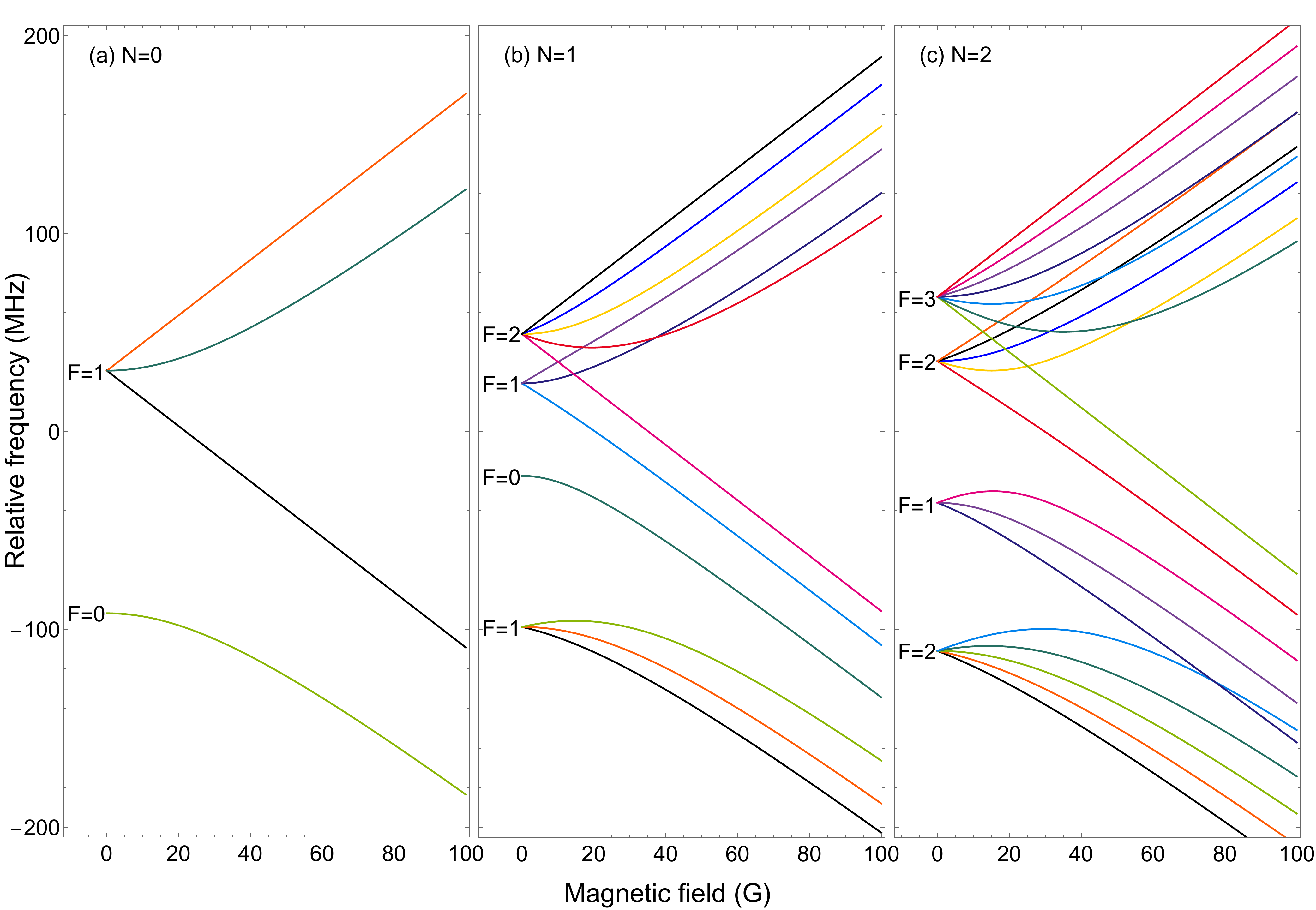}
    \caption{Zeeman shifts for the three lowest-lying rotational levels of CaF.}
    \label{fig:CaFZeeman}
\end{figure}

Figure~\ref{fig:CaFZeeman} shows the Zeeman shifts for the first three rotational states of CaF up to $B=100$~G. These are calculated by finding the eigenvalues of $H_{\rm hyp} + H_{\rm Z}$ for various values of $B$, where $H_{\rm hyp}$ is given by equation (\ref{eq:hypDoubletSigma}). When $B$ is small, so that the Zeeman shifts are much smaller than the hyperfine splittings, the Zeeman shifts are linear and the levels can be labelled by the quantum numbers $(N,F,m_{F})$. When the Zeeman shifts are much larger than the hyperfine splittings, they are again linear in $B$ and the levels can be labelled by the projection quantum numbers $(m_S,m_N,m_I)$. In this strong field regime, the levels form two well separated manifolds of opposite $m_S$, with much smaller splittings within these manifolds corresponding to the various values of $m_N$ and $m_I$.

\subsection{State preparation and trapping}
\label{sec:state_prep_trap}

After cooling in the MOT and optical molasses, the molecules are distributed amongst the Zeeman and hyperfine components of the $N=1$ rotational state, only some of which are weak field seeking states suitable for magnetic trapping. In the example shown in figure~\ref{fig:CaFZeeman}, 4 of the 12 magnetic sublevels have positive Zeeman shifts in the weak field regime. Ideally, all the molecules should be transferred to a single one of these states. One way to do this is to apply optical pumping in an arrangement where one (and only one) of the states is a dark state. This was demonstrated in~\citet{McCarron2018} for SrF molecules, which has a similar structure to the one shown in figure~\ref{fig:CaFZeeman}. By driving the closed rotational transition from $N=1$ using one beam polarized to drive $\pi$ transitions and other to drive $\sigma^+$ transitions, the only dark state within $N=1$ is $\ket{N,F,m_F}=\ket{1,2,2}$, which has the largest positive Zeeman shift. Using this approach, about 40\% of the molecules present in the MOT were confined in the magnetic trap in the target state. Molecules were also present in one other state, but they could be removed from the magnetic trap by driving a microwave transition to $N=0$, leaving a pure sample. The temperature of these SrF molecules in the magnetic trap was about 80~$\mu$K. 

\begin{figure}[!tb]
    \centering
    \includegraphics[width=\linewidth]{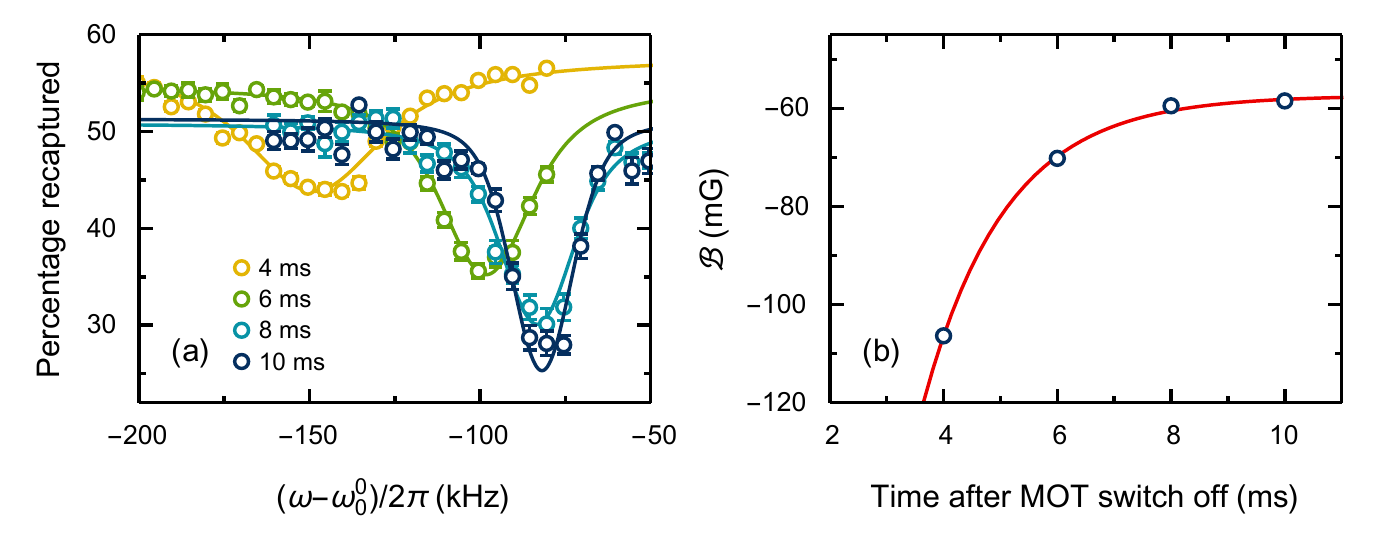}
    \caption{(a) Preparing CaF molecules in the ground rotational state by driving the microwave transition from $\ket{N,F, m_F}=\ket{1,0,0}$ to $\ket{0,1,1}$. The transition is driven using a microwave pulse, $40$ $\mu$s in duration, applied at various times after switching off the MOT coils. The line centre and shape change over time due to eddy currents induced when the MOT coils are turned off. Points and error bars are mean and standard error of nine experiments. Lines are fits to a Voigt profile. (b) Magnetic field at the position of the molecules determined from the data in (a), showing the decay of the eddy currents over time. The offset of $-60$~mG is the field deliberately applied in the experiment. The line is an exponential fit. Figure from \citet{CaldwellThesis}. }
    \label{fig:Eddy}
\end{figure}

A second approach to producing a pure sample was demonstrated by \citet{Williams2018} for CaF. The molecules were optically pumped into $\ket{1,0,0}$ by turning off the frequency component that addresses the $F=0$ level and lowering the intensity of the other components. This pumps about 60\% of the population to $F=0$, limited by off-resonant excitation. The molecules were then transferred to the state $\ket{0,1,1}$ by driving a microwave $\pi$-pulse. Figure~\ref{fig:Eddy}(a) shows an example where this transition is driven at various times after turning off the MOT coils. The molecules are held in an optical molasses during this wait time. The line is shifted and broadened at early times due to the magnetic field resulting from eddy currents produced when the MOT coils are turned off. The field measured by this microwave spectroscopy is shown in figure \ref{fig:Eddy}(b). Fortunately, the cloud expands very slowly in the molasses so can be held there for 10~ms without much loss of density, to let the eddy currents decay away. These data illustrate that coherent control of the molecules requires very precise control over magnetic fields. After preparing molecules in $N=0$, all molecules remaining in $N=1$ were removed by using a pulse of resonant light from a single beam (the slowing laser beam) to push them away, leaving a pure sample. These $N=0$ molecules were trapped directly, and by using a second microwave pulse the population could be transferred back to a selected sub-level of $N=1$. In these experiments, the molecules in the magnetic trap were at a density of $1.2 \times 10^{5}$~cm$^{-3}$ and a temperature of 70(8)~$\mu$K. No heating was observed on a 500~ms timescale.

Lifetimes in the magnetic trap were initially limited to about 1~s by collisions with the helium gas from the cryogenic buffer gas sources, but later improved by installing mechanical shutters to stop the flow of gas into the trap regions. After collisional loss is reduced to a low enough rate, the next most significant source of trap loss for many relevant molecules is rotational and vibrational excitation by blackbody radiation~\citep{Vanhaecke2007,Buhmann2008}. At room temperature, the vibrational excitation rate is the dominant one for most molecules, though rotational excitation is important for molecules with large rotational constants, especially the hydrides. For the vibrational ground state, the vibrational excitation rate is
\begin{equation}
    \Gamma_{\rm vib} = \frac{\omega_{01}^3 d_{01}^2}{3\pi\epsilon_0\hbar c^{3}} \frac{1}{e^{\hbar\omega_{01}/(k_{\rm B}T)}-1},
\end{equation}
where $T$ is the temperature, $\omega_{01}$ is the frequency of the vibrational transition and 
\begin{equation}
    d_{01} = \sqrt{\frac{\hbar}{4\pi m \omega_{01}}}  \left.\frac{d \mu^{\rm e}}{d R_N}\right|_{R_0}
\end{equation}
is the transition dipole moment expressed in terms of the mass $m$ and the gradient of the dipole moment function at the equilibrium separation. We have neglected the small dependence of the transition frequency on the rotational state. In this approximation, the result is independent of rotational state. For the alkaline earth monofluorides the lifetime, $1/\Gamma_{\rm vib}$, is typically a few seconds at room temperature, rising to thousands of seconds at $77$~K. For CaF, a magnetic trap lifetime of $4.5(4)$~s has been measured~\citep{CaldwellThesis}, consistent with the lifetime limit due to blackbody heating at room temperature. We note here that, because of the lifetime limit imposed by blackbody heating, sympathetic or evaporative cooling of molecules is likely to require a cryogenically-cooled environment.

\subsection{Rotational coherences in magnetic traps}
\label{sec:mag_coherence}

\begin{figure}[!tb]
    \centering
    \includegraphics[width=\linewidth]{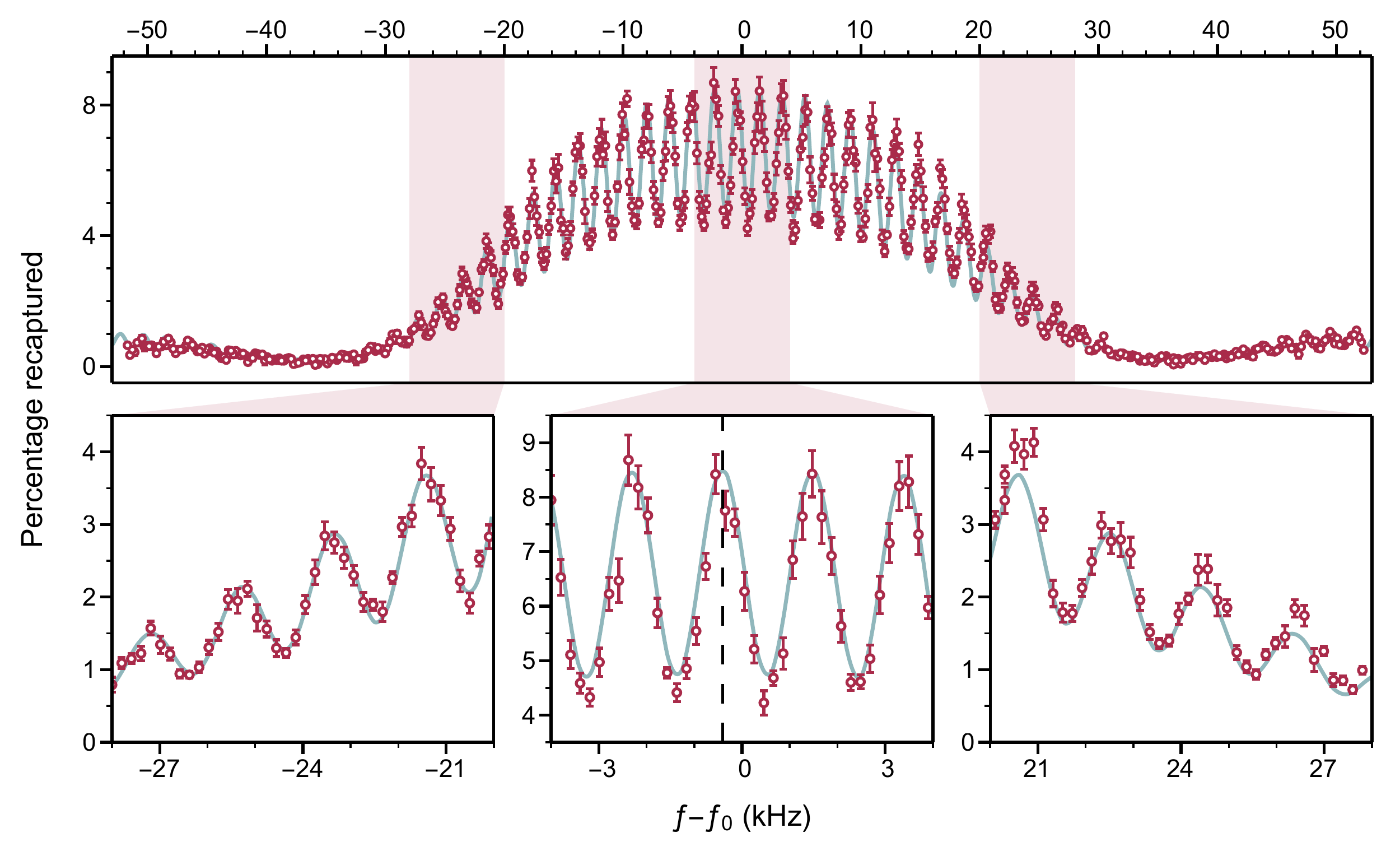}
    \caption{Ramsey interferometry for magnetically-trapped CaF molecules prepared in a superposition of the two states $\ket{N,F,m_F}=\ket{0,1,1}$ and $\ket{1,2,2}$, whose frequency spacing is $f_0 \approx 20.55$~GHz. Two short $\pi/2$ pulses of frequency $f$ are separated by a free evolution time of 493~$\mu$s. The data show the number of molecules in $N=1$ after the two-pulse sequence, as a function of the detuning $f-f_0$. From \citet{Blackmore2018}.}
    \label{fig:Ramsey}
\end{figure}

Most applications of ultracold molecules call for coherent control of the internal states and often rely on long coherence times. Rotational coherences are particularly important because rotational superpositions have large oscillating electric dipole moments, providing the long-range dipole-dipole interactions needed for quantum simulation and information processing (see section \ref{sec:applications}). The strong coupling of rotational states to microwave photons could also be used to interface molecules with solid-state systems (see section \ref{sec:information}). 

It is challenging to engineer long coherence times for trapped molecules because the trap potential is typically different for the different states involved. This leads to decoherence. In the classical regime, the decoherence can be described in terms of the spatial variation of the transition frequency due to the inhomogeneous field of the trap. In the quantum regime, the decoherence arises when there is a distribution over motional states and when the transition frequency depends on the motional state. In both regimes, temporal fluctuations of the trapping field also cause decoherence.  Taking the example of magnetically trapped molecules in thermal equilibrium at temperature $T$, where the two states of interest have magnetic moments $\mu$ and $\mu + \Delta \mu$, the decoherence rate will be at least $(k_{\rm B} T/h)(\Delta \mu/\mu)$. When $T = 10$~$\mu$K and $(\Delta \mu/\mu)=10^{-2}$, this rate is about 2~kHz. 

The coherent control of magnetically-trapped molecules has been studied in \citet{Williams2018, Blackmore2018, Caldwell2020}. Figure~\ref{fig:Ramsey} shows an example of Ramsey interferometry with CaF molecules in a magnetic quadrupole trap. Here, the molecules are prepared in a superposition of two different rotational states, $\ket{N,F,m_F}=\ket{0,1,1}$ and $\ket{1,2,2}$. For $^{2}\Sigma$ states, the Zeeman shifts of the stretched states\footnote{The stretched states are the ones with maximal (or minimal) values of all projection quantum numbers, here $m_N$, $m_S$ and $m_I$.} are almost independent of the rotational quantum number $N$. This can be seen in figure~\ref{fig:CaFZeeman} - notice the nearly identical shifts of the uppermost level of each $N$. Consequently, transitions between these stretched states are relatively insensitive to magnetic fields. The residual sensitivity is determined by the middle two terms in equation (\ref{eq:HZ}), whose coefficients are $g_l$ and $g_{\rm r}$. For many laser-coolable molecules, there exist rotational transitions between stretched states where these two terms almost cancel, resulting in transitions that are exceptionally insensitive to magnetic fields~\citep{Caldwell2020}. In these cases, long rotational coherence times are feasible for magnetically-trapped molecules. For CaF, the pair of states $\ket{0,1,1}$ and $\ket{1,2,2}$ have $\Delta \mu/\mu = -7.4 \times 10^{-5}$, while the pair $\ket{1,2,2}$ and $\ket{2,3,3}$ have $\Delta \mu/\mu = -3.4 \times 10^{-6}$. For the latter pair, the magnetic sensitivity is so small that rotational coherence times exceeding 1~s seem feasible for molecules cooled to $5$~$\mu$K or below. So far, a coherence time of 6.4(8)~ms has been demonstrated, limited by the large size of the trapped cloud, an inhomogeneous change in phase resulting from the change in position of the molecules between microwave pulses, and a geometric phase which depends on the trajectory of each molecule between microwave pulses. These effects can be reduced by using smaller clouds and a magnetic trap geometry with a large bias field.

\section{Optical traps}
\label{sec:optical_dipole}

Optical dipole traps~\citep{Grimm2000} are commonly used to trap ultracold atoms and have recently been used to trap laser-cooled molecules~\citep{Anderegg2018, Cheuk2018}. They have smaller volumes and lower depths than magnetic traps, but they can be used to trap molecules in any internal state and they can provide much tighter confinement. Importantly, the molecules can be cooled inside the trap, which can significantly enhance both the spatial and phase-space densities. There are many possible configurations of optical dipole traps, but the most common uses light that is red detuned from all electronic transitions. The interaction of the induced electric dipole moment with the electric field of the light -- the ac Stark shift -- produces a potential proportional to the intensity, which attracts the molecule to the location of highest intensity.  Here, we first outline the theory of the ac Stark effect and then describe recent work on optical trapping of laser-cooled molecules. 

\subsection{AC Stark effect}

The theory of the ac Stark effect for a diatomic molecule is given in the appendix of \citet{Caldwell2020b}. Here we highlight some salient features. We consider a molecule interacting with light which has electric field amplitude $\mathcal{E}_0$, angular frequency $\omega_{\rm L}$, and unit polarization vector $\hat{\epsilon}$: $\vec{E}=\frac{1}{2}\mathcal{E}_0( \hat{\epsilon} e^{-i \omega_{\rm L} t} + \hat{\epsilon}^*e^{i \omega_{\rm L} t}) $. It is convenient to write an effective operator for the Stark effect that acts within a small subspace, often a single rotational state. This operator can be written in the form
\begin{equation}
H_{\rm S}=\sum _{K=0}^2 H_{\rm S}^{(K)} = -\frac{\mathcal{E}_0^2}{4}\sum _{K=0}^2 \sum _{P=-K}^K  (-1)^P {\mathcal A}^{(K)}_P \mathcal{P}_{-P}^{(K)}.
\label{eq:HStarkAC}
\end{equation}
It is a sum of three parts, each of which is the scalar product of two rank-$K$ spherical tensors, a polarizability tensor ${\mathcal A}^{(K)}$ that depends only on properties of the molecule, and a polarization tensor ${\mathcal P}^{(K)}$ that depends only on the polarization of the light. The matrix elements of $H_{\rm S}^{(K)}$ can be expressed as
\begin{equation}
    \bra{\gamma'} H_{\rm S}^{(K)} \ket{\gamma} = - \frac{{\cal E}_0^2}{4} \alpha^{(K)} \sum _{P=-K}^K  (-1)^P h(\gamma,\gamma',K,P) \mathcal{P}_{-P}^{(K)},
\end{equation}
where $\gamma$ and $\gamma'$ stand for the set of quantum numbers needed to define states within the subspace. The parameters $\alpha^{(K)}$ depend on the frequency of the light and the frequencies of transitions to states outside the subspace, while the factor $h$ contains all of the dependence on the quantum numbers. 
The scalar part produces the same shift for all states within the subspace, $\langle H_{\rm S}^{(0)} \rangle = - \alpha^{(0)} {\cal E}_0^2/4$, where $\alpha^{(0)}$ is the scalar polarizability. The vector ($\alpha^{(1)}$) and tensor ($\alpha^{(2)}$) polarizabilities produce different shifts for different states. The vector part is zero for a dc electric field. For an ac field it is only non-zero when the molecule has a spin and the light has a component of circular polarization. The effect of the vector part is the same as that of a magnetic field applied along the axis of circular handedness. Expressions for $\alpha^{(K)}$ and $h$ for molecules in $^{1}\Sigma$ or $^{2}\Sigma$ states are given in \citet{Caldwell2020b}.

\begin{figure}[!tb]
    \centering
    \includegraphics[width=\linewidth]{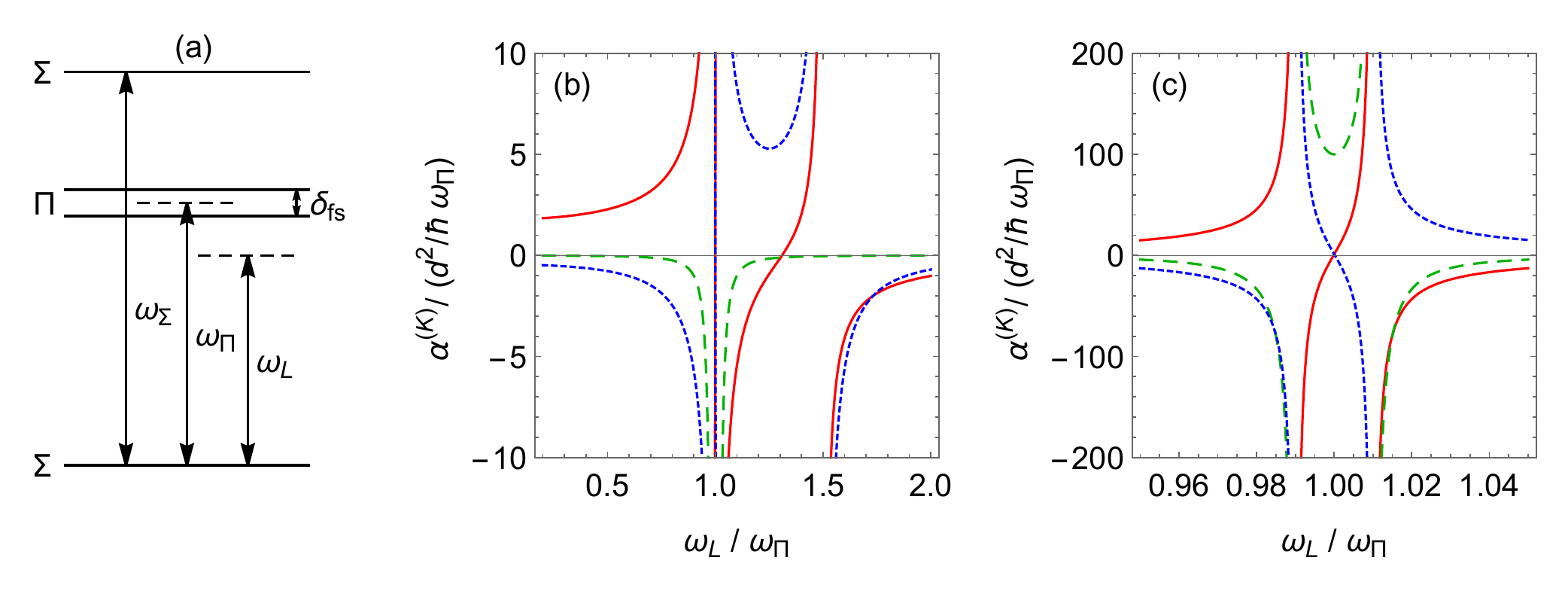}
    \caption{Polarizability for a model molecule. (a) Illustration of the model. (b) Polarizability components versus laser frequency in scaled units. We have chosen $\omega_{\Sigma}=1.5\omega_{\Pi}$, $\delta_{\rm fs} = \omega_{\Pi}/50$ and $d_{\Sigma}=d_{\Pi}=d$. (c) Same as (b), plotted over a narrow range around the $\Sigma-\Pi$ transition. Solid red: $\alpha^{(0)}$. Dashed green: $\alpha^{(1)}$. Dotted blue: $\alpha^{(2)}$.}
    \label{fig:Polarizability}
\end{figure}

To illustrate the dependence of the polarizability components on the frequency of the light, $\omega_{\rm L}$, we consider a model molecule with an electronic energy structure as depicted in figure~\ref{fig:Polarizability}(a). This structure is typical for many molecules with $^{1}\Sigma$ or $^{2}\Sigma$ ground states. The model has two electronic excited states, a $\Pi$ state and a $\Sigma$ state, with energies $\hbar \omega_{\Pi}$ and $\hbar \omega_{\Sigma}$ above the ground state. Due to the spin-orbit interaction, the $\Pi$ state is split into two fine-structure components shifted by $\pm\delta_{\rm fs}/2$ (with $\delta_{\rm fs} = 0$ for a $^{1}\Pi$ state). The transition dipole moments for transitions from the ground state to the $\Pi$ and $\Sigma$ states are $d_{\Pi}$ and $d_{\Sigma}$. For convenience, we introduce the scaled quantities $l=\omega_{\rm L}/\omega_{\Pi}$, $s=\omega_{\Sigma}/\omega_{\Pi}$, and $f=\delta_{\rm fs}/\omega_{\Pi}$. In this model, the polarizability components are
\begin{subequations}
\begin{align}
    \alpha^{(0)} &= \frac{1}{3}\left( \alpha_{||} + \alpha_{\perp,-} + \alpha_{\perp,+} \right), \\
    \alpha^{(1)} &= \frac{1}{2}\left( \beta_{\perp,-} - \beta_{\perp,+}\right),\\
    \alpha^{(2)} &=  \frac{1}{3}\left( 2\alpha_{||}-\alpha_{\perp,-}-\alpha_{\perp,+}\right),
\end{align}
\label{eq:pol_components}
\end{subequations}
where
\begin{subequations}
\begin{align}
    \alpha_{||} &= \frac{d_{\Sigma}^2}{\hbar \omega_{\Pi}} \left( \frac{1}{s+l}+\frac{1}{s-l}\right), \\
    \alpha_{\perp,\pm} &= \frac{d_{\Pi}^2}{\hbar \omega_{\Pi}}\left( \frac{1}{1 \pm f/2 + l} + \frac{1}{1 \pm f/2 - l}  \right), \\
    \beta_{\perp,\pm} &= \frac{d_{\Pi}^2}{\hbar \omega_{\Pi}}\left( \frac{1}{1 \pm f/2 + l} - \frac{1}{1 \pm f/2 - l}  \right).
\end{align}
\label{eq:pol_molframe}
\end{subequations}
Note that these results are relevant for frequencies far above rotational transition frequencies. At microwave frequencies the polarizability is dominated by the rotational structure, which is absent in our model.

Figure~\ref{fig:Polarizability}(b) plots these polarizability components for the case where $s=3/2$, $f=1/50$ and $d_{\Sigma}=d_{\Pi}=d$. The scalar polarizability, $\alpha^{(0)}$, is positive when the light is red detuned from all resonances ($\omega_{\rm L} < \omega_{\Pi}$), and negative when blue detuned ($\omega_{\rm L} > \omega_{\Sigma}$). It tends towards a constant in the low frequency limit ($\omega_{\rm L} \ll \omega_{\Pi}$) and towards zero in the high frequency limit ($\omega_{\rm L} \gg \omega_{\Sigma}$). It has a large magnitude near the resonances and changes sign through them. The vector polarizability, $\alpha^{(1)}$, is very small for all frequencies other than $\omega_{\rm L}=\omega_{\Pi}$. Its behaviour in this region is best seen in the expanded view shown in figure~\ref{fig:Polarizability}(c). Halfway between the fine-structure components, $\alpha^{(1)}$ is large whereas $\alpha^{(0)}$ and $\alpha^{(2)}$ both pass through zero. Away from resonances, $|\alpha^{(2)}|$ is smaller than $|\alpha^{(0)}|$ but much larger than  $|\alpha^{(1)}|$. It therefore dominates the state-dependence of the polarizability in these regions. There is a range of frequencies between the $\Sigma-\Pi$ and $\Sigma-\Sigma$ transition frequencies where the polarizability is dominated by $\alpha^{(2)}$ because $\alpha^{(0)},\alpha^{(1)}$ are near zero whereas $\alpha^{(2)}$ is not. The behaviours of the polarizability components between resonances can be exploited to make state-dependent tweezer traps as proposed by \citet{Caldwell2020c, Caldwell2020d}. 
 
\subsection{Optical dipole traps}

When the ac Stark shift is either much smaller or much larger than other relevant interactions (e.g. hyperfine and Zeeman), the effect of $H_{\rm S}$ can be considered in isolation leading to a shift of the form $U=- \alpha {\cal E}_0^2/4 = - \alpha I / (2c\epsilon_0)$, where $I$ is the intensity of the light field and $U$ is the potential in which the molecules move. Here, unlike above, we have not separated the polarizability into its components. Instead, $\alpha$ is a parameter that depends on the state of the molecule and the polarization of the light. Its value can be determined from the theory outlined above. In the case where the light is detuned far below all electronic resonances, $\alpha$ can be approximated by $\alpha^{(0)}$, which in turn can be roughly estimated using equations (\ref{eq:pol_components}) and (\ref{eq:pol_molframe}).

A simple optical dipole trap is formed using a single, red detuned, focussed laser beam. For a Gaussian beam with waist $w_0$ and wavelength $\lambda$, the intensity distribution as a function of the radial and axial coordinates $r$ and $z$ is
\begin{equation}
    I(r,z) = \frac{I_0}{1+z^{2}/z_0^2} e^{-2r^2/w^2},
    \label{eq:gaussianBeam}
\end{equation}
where $w = w_0 \sqrt{1+z^2/z_0^2} $ and $z_0 = \pi w_0^2/\lambda$. The intensity at the centre of the focus is related to the power $P$ by $I_0 = 2P/(\pi w_0^2)$. For a molecule close to the centre of the trap, we can expand equation (\ref{eq:gaussianBeam}) around $r=z=0$, giving a trap potential
\begin{equation}
    U(r,z) = -U_0 + \frac{1}{2}m \omega_r^2 + \frac{1}{2}m \omega_z^2,
\end{equation}
where $\omega_r = \sqrt{4U_0/(m w_0^2)}$ and $\omega_z = \sqrt{2U_0/(m z_0^2)}$ are the angular oscillation frequencies in the radial and axial directions, and $U_0 = \alpha I_0/(2c\epsilon_0)$ is the trap depth.

Let us consider a CaF molecule in the state $\ket{N,F,m_F}=\ket{1,2,m_F}$ interacting with light linearly polarized along the $z$-axis. For this case, $\alpha = \alpha^{(0)} + \tfrac{1}{10}(2-m_F^2)\alpha^{(2)}$. Using the transition frequencies and dipole moments for the $A^{2}\Pi-X^{2}\Sigma^+$ and $B^{2}\Sigma^+-X^{2}\Sigma^+$ transitions, and taking a wavelength of $\lambda = 1064$~nm, equations (\ref{eq:pol_components}) and (\ref{eq:pol_molframe}) give $\{\alpha^{(0)},\alpha^{(2)}\} = \{2.97,-1.47\} \times 10^{-39}~\text{J}~(\text{V}~\text{m}^{-1})^{-2}$. Taking $P=15$~W, $w_0=45$~$\mu$m and $m_F=2$, we find $U_0 / k_{\rm B}= 210$~$\mu$K, $\omega_r = 2\pi\times 1.2$~kHz and $\omega_z = 2\pi\times 6.5$~Hz. We should remember that all these values are approximate, since the real molecule has more excited states than the model molecule. Nevertheless, they give reasonable estimates of the trap depths and oscillation frequencies obtained for some typical parameters.

Laser cooled molecules were first trapped optically by \citet{Anderegg2018} who captured CaF molecules in a MOT, cooled them in an optical molasses, and then loaded them into a single-beam far-detuned optical dipole trap with parameters similar to those considered above. They demonstrated that sub-Doppler cooling works in the presence of the dipole trap, and that cooling molecules into the trap increases the number captured by about a factor of five relative to direct capture. Approximately 150 molecules were trapped at a temperature of 60~$\mu$K and a density of $8(2) \times 10^7$~cm$^{-3}$. The trap lifetime was found to be $750(40)$~ms, limited by collisions with background gas. The group went on to show that $\Lambda$-enhanced cooling (see section \ref{sec:sub-Dopp}) also works in the presence of the dipole trap, and that it improves the loading efficiency and final temperature~\citep{Cheuk2018}. They were able to load $1300(60)$ molecules at a temperature of $21(3)$~$\mu$K and a peak density of $6(2) \times 10^8$~cm$^{-3}$, corresponding to a peak phase-space density of about $7 \times 10^{-8}$. They also showed that, in this configuration, the molecules can scatter more than $10^{4}$ photons from the molasses before being lost from the trap, providing an efficient method for imaging single trapped molecules. 

\subsection{Optical tweezer traps}
\label{sec:tweezer}

\begin{figure}[!tb]
    \centering
    \includegraphics[width=0.8\linewidth]{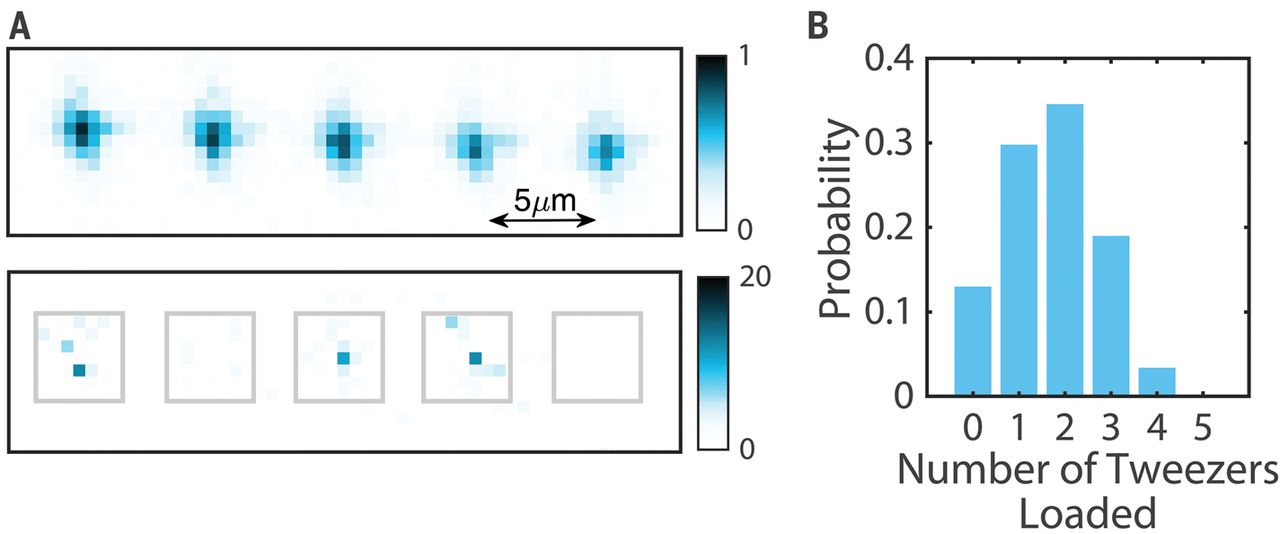}
    \caption{CaF molecules in an array of optical tweezer traps. (A) Images of the array. Upper image is averaged over 500 shots, and lower image is a single shot showing three occupied tweezer traps. Colour scale indicates signal per pixel. (B) Probability versus number of tweezer traps loaded. From \citet{Anderegg2019}. Reprinted with permission from AAAS. }
    \label{fig:tweezer}
\end{figure}

A tweezer trap is an optical dipole trap that is very tightly focussed so that the beam waist is about the same size as the wavelength of the light. In such a tightly confining trap, and in the presence of near-resonant cooling light, rapid light-assisted collisions ensure that only a single particle can be trapped~\citep{Schlosser2001}. This is sometimes known as collisional blockade. An array of tweezer traps with adjustable spacing can be obtained by passing the beam through a spatial light modulator or an acousto-optical deflector driven by multiple rf frequencies~\citep{Bergamini2004, Gauthier2021}. Great strides have been made with atoms in tweezer traps, including the near-deterministic loading of single atoms by tailoring the light-assisted collisions~\citep{Grunzweig2010, Brown2019}, Raman sideband cooling to the motional ground state~\citep{Kaufman2012, Thompson2013}, re-arrangement of tweezer arrays to produce desired configurations~\citep{Endres2016, Barredo2016}, tweezer array optical clocks~\citep{Norcia2019}, and the association of atoms into molecules inside tweezer traps~\citep{Liu2019, Zhang2020}. 

Recently, an optical tweezer array of laser cooled molecules was formed by \citet{Anderegg2019}. A 780~nm beam was focussed to a spot size of 2.3~$\mu$m, and a one-dimensional array of these traps was created using an acousto-optic deflector. Single CaF molecules from the optical dipole trap discussed above were loaded into these tweezer traps in the presence of the $\Lambda$-enhanced molasses, and were detected with 92\% fidelity by imaging their fluorescence onto a camera. Figure~\ref{fig:tweezer} shows both single shot and averaged images of the tweezer array, along with a plot showing the probability distribution of the number of traps loaded. The average loading probability per trap is 34\%. The radial oscillation frequency in the trap was measured to be 35~kHz, and the temperature was $80(20)$~$\mu$K. When there are two molecules in the same trap, the peak density is $3 \times 10^{10}$~cm$^{-3}$  and the peak phase-space density is $3 \times 10^{-7}$. The authors demonstrated that, in the presence of the cooling light, light-assisted collisions occur on a timescale of a few milliseconds, resulting in a collisional blockade that ensures an occupancy of zero or one. 

Using their tweezer trap array, \citet{Cheuk2020} measured inelastic collisions between pairs of CaF molecules prepared in selected quantum states. They used a pair of traps, measured the initial occupancies of the pair by fluorescence imaging, prepared the trapped molecules in a selected sub-level of the rotational ground state using optical pumping and microwave transfer (see section \ref{sec:state_prep_trap}), merged the pair into a single trap so that collisions could occur, then separated the traps again and re-measured their occupancies. They measured rapid two-body loss with no dependence on the choice of hyperfine or Zeeman state, and attributed this loss either to the energetically allowed chemical reaction $\text{CaF}+\text{CaF} \rightarrow \text{CaF}_2 + \text{Ca}$, or to the formation of complexes that are undetected or lost by photoexcitation. This ability to measure collisional processes between molecules at the single particle level is a valuable new tool for physics and physical chemistry.

An important current challenge is to obtain full quantum control over the molecules by cooling them to the motional ground state of the tweezer trap. For atoms, Raman sideband cooling is commonly used~\citep{Kaufman2012, Thompson2013}. This is done by repeatedly applying a two-step process where the first step drives a Raman transition between two internal states and reduces the motional quantum number, and the second step optically pumps back to the original internal state. The first step is coherent and needs to be slow enough to resolve the motional states, while the second step provides dissipation. The application of Raman sideband cooling to molecules in tweezer traps is discussed in \citet{Caldwell2020b}. Strong tensor Stark shifts result in different trap potentials for the two internal states involved in the cooling, resulting in two significant problems: (i) the Raman transition frequency depends on the motional state and (ii) there is an additional contribution to the heating in the optical pumping step as the molecule is suddenly transferred into a different trap. These problems can be solved by applying a magnetic field that is strong enough to uncouple the rotational angular momentum from the spin, so that levels are well characterized by $m_N$, and then choosing a tweezer polarization where the tensor Stark shift is identical for two different $m_N$ levels. To implement sideband cooling in practice, it is beneficial to make the traps as tightly-confining as possible, so that $\omega_r$ and $\omega_z$ are as large as possible. This lowers the mean motional quantum number for a given starting temperature, and makes it easier to resolve the motional states.

\begin{figure}[!tb]
    \centering
    \includegraphics[width=\linewidth]{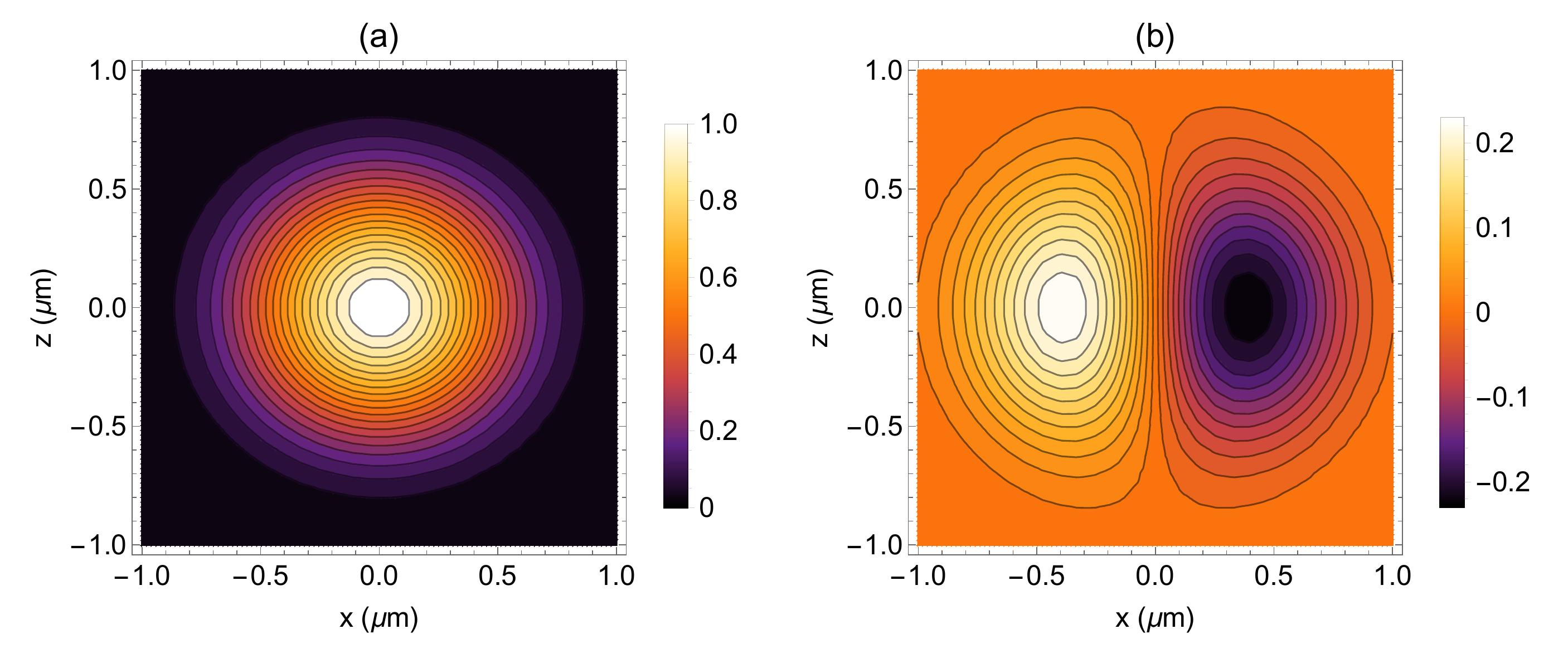}
    \caption{Contour plots in the plane of the focus of an optical tweezer trap, for 780~nm light propagating along $y$ and focussed by a 0.55 NA lens. The input beam is polarized along $x$ and has a $1/e^2$ diameter equal to the lens diameter. (a) Intensity distribution $I(x,z)/I_{\rm max}$. (b) Scaled component of ellipticity in the $z$ direction, $(I(x,z)/I_{\rm max})C_z(x,z)$. }
    \label{fig:tweezer_contours}
\end{figure}

Figure~\ref{fig:tweezer_contours} shows some of the properties of an optical tweezer trap in the plane of the focus. Here, we have assumed that a Gaussian laser beam with a wavelength of 780~nm is focussed by a lens with a numerical aperture (NA) of 0.55. The input beam propagates along $y$, is polarized along $x$ and has a $1/e^2$ diameter equal to the lens diameter. We use the vector Debye integral~\citep{Richards1959} to calculate the electric field components in the region of the trap. Figure~\ref{fig:tweezer_contours}(a) shows the intensity distribution at the focus, $I(x,z)/I_{\rm max}$ where $I_{\rm max}$ is the maximum intensity. The distribution is Gaussian in both directions but is slightly elliptical, having a $1/e^2$ radius of 0.69~$\mu$m along $z$ and 0.74~$\mu$m along $x$. This leads to different trap frequencies in these two directions. Near the focus, the electric field has components along $x$ and $y$, resulting in spatially-varying elliptical polarization about the $z$-axis~\citep{Richards1959}. We characterize this using the quantity $\vec{C}=\text{Im}(\epsilon \times \epsilon^*)$, where $\epsilon$ is a unit polarization vector. The magnitude of $\vec{C}$ gives the degree of ellipticity, with $C=1$ corresponding to circular polarization and $C=0$ to linear polarization, and the direction gives the axis of ellipticity. Figure~\ref{fig:tweezer_contours}(b) shows $(I(x,z)/I_{\rm max}) C_z(x,z)$ in the plane of the focus. We find significant circular polarization components with opposite handedness on either side of the focus. For a molecule with spin, the interaction with the light field depends on the handedness and on the direction of the spin, resulting in a contribution to the ac Stark shift that is proportional to $\alpha^{(1)} m_F C_z I$, where $m_F$ is the projection of the angular momentum onto $z$. In the context of equation (\ref{eq:HStarkAC}), $\vec{C}={\cal P}^{(1)}$, the rank-1 component of the spherical tensor formed from $\epsilon$ and $\epsilon^*$, and its interaction with the rank-1 component of the polarizability tensor ${\cal A}^{(1)}$ results in a vector Stark shift. \citet{Caldwell2020c} have shown how this interaction can be used to create state-dependent optical tweezer traps that can trap pairs of molecules at separations much smaller than otherwise possible. As can be seen from figure~\ref{fig:Polarizability}(c), $\alpha^{(1)}$ dominates the polarizability when the frequency of the light lies between the two fine-structure components of the excited state. In this case, molecules with opposite $m_F$ are trapped on opposite sides of the focus (at the location of the maximum and minimum in figure~\ref{fig:tweezer_contours}(b)). Introducing an additional tweezer at a wavelength where $\alpha^{(0)}$ dominates squeezes the two molecules together. This can be used to enhance dipolar interactions between molecules, which is important for quantum simulation and information processing. A similar outcome can also be achieved by exploiting the tensor polarizability, as described in \citet{Caldwell2020d}.

\section{Applications and future directions}
\label{sec:applications}

Ultracold molecules can be used to search for new physics, explore the physics of many-body quantum systems, store and process quantum information, and probe the quantum mechanical underpinnings of chemistry and collisions.  In this section, we briefly explore some of these exciting applications. We note that the ultracold molecules needed can also be produced by other methods, including atom association~\citep{Ni2008}, optoelectric Sisyphus cooling~\citep{Prehn2016}, and electrodynamic slowing and focussing~\citep{Cheng2016}. Here, we will highlight the particular role that laser cooled molecules can play.  

\subsection{Controlling dipole-dipole interactions}
\label{sec:dipole}

Many of the applications of ultracold molecules hinge on dipole-dipole interactions between two or more molecules.  For two molecules $i$ and $j$, with position vectors $\vec{r}_{i}$ and $\vec{r}_{j}$, the dipole-dipole interaction Hamiltonian is
\begin{equation}
    H_{\rm dd} = \frac{1}{4 \pi \epsilon_{0}} \frac{\vec{d}_{i} \cdot \vec{d}_{j} - 3(\vec{d}_{i} \cdot \hat{e}_{ij})(\vec{d}_{j} \cdot \hat{e}_{ij})}{r_{ij}^{3}}.
    \label{eq:Hdd}
\end{equation}
Here, $\vec{d}_i$ and $\vec{d}_j$ are the dipole moment operators for the two molecules, $\vec{r}_{ij}=\vec{r}_{i} - \vec{r}_{j}$ is their relative displacement, and $\hat{e}_{ij} = \vec{r}_{ij}/r_{ij}$ is a unit vector pointing along $\vec{r}_{ij}$. This form of the interaction is especially useful for molecules which are fixed in place, e.g. molecules in an optical lattice or in tweezer traps. In spherical tensor form, the interaction is
\begin{equation}
    H_{\rm dd} = - \frac{\sqrt{6}}{4\pi\epsilon_0 r_{ij}^3} \sum_{q=-2}^2 (-1)^q C_{-q}^2(\theta_{ij},\phi_{ij})T_q^2(\vec{d}_i,\vec{d}_j).
\end{equation}
In this form, the relative displacement is expressed in spherical polar coordinates $(r_{ij},\theta_{ij},\phi_{ij})$, $C^2$ is the re-normalized spherical harmonic of rank 2 ($C_q^2 = \sqrt{4 \pi/5}\,Y_{2,q}$), and $T^2(\vec{d}_i,\vec{d}_j)$ is the rank-2 spherical tensor formed from $\vec{d}_i$ and $\vec{d}_j$. This form is most useful for freely moving molecules, and also helps to reveal the processes allowed via this interaction.

Let us consider the simple case of a two-level molecule with states $\ket{0}$ and $\ket{1}$  whose energies are 0 and $E_{01}$. The states are chosen so that the transition dipole moment $\vec{d}_{01}=\bra{0}\vec{d}\ket{1}$ is non-zero. In the two-molecule basis $\ket{00}$, $\ket{01}$, $\ket{10}$, $\ket{11}$, the only non-zero matrix elements of $H_{\rm dd}$ are the ones where both molecules change state. The coupling between $\ket{00}$ and $\ket{11}$ is usually very strongly suppressed because the dipole-dipole interaction is tiny compared to $2E_{01}$. Neglecting this, we are left only with the coupling between $\ket{01}$ and $\ket{10}$. The eigenstates are $\ket{00}$, $\ket{11}$, and $\ket{\Psi^{\pm}}=\frac{1}{\sqrt{2}}(\ket{01} \pm \ket{10})$ with energies 0, $2E_{01}$ and $E_{01} \pm E_{\rm dd}$ where
\begin{equation}
    E_{\rm dd} = \frac{1}{4\pi\epsilon_0 r_{ij}^3} \bra{01} \vec{d}_{i} \cdot \vec{d}_{j} - 3(\vec{d}_{i} \cdot \hat{e}_{ij})(\vec{d}_{j} \cdot \hat{e}_{ij}) \ket{10}.
    \label{eq:E_dd}
\end{equation}
In the special case where $\ket{0}$ and $\ket{1}$ have the same value of $m_F$, equation (\ref{eq:E_dd}) simplifies to
\begin{equation}
    E_{\rm dd} = \frac{|\vec{d}_{01}.\hat{z}|^2}{4\pi\epsilon_0 r_{ij}^3}  (1-3\cos^2\theta_{ij}),
\end{equation}
where $\theta_{ij}$ is the angle between $\hat{e}_{ij}$ and the $z$-axis. We see that a pair of molecules prepared in $\ket{01}$ will oscillate back and forth between $\ket{01}$ and $\ket{10}$ at a characteristic frequency $2E_{\rm dd}/h$. The single excitation is exchanged between the two molecules, mediated by the dipole-dipole interaction. As we will see, this exchange interaction can be used to implement a two-qubit gate and to explore aspects of quantum magnetism. 

\begin{figure}[!tb]
    \centering
    \includegraphics[width=\linewidth]{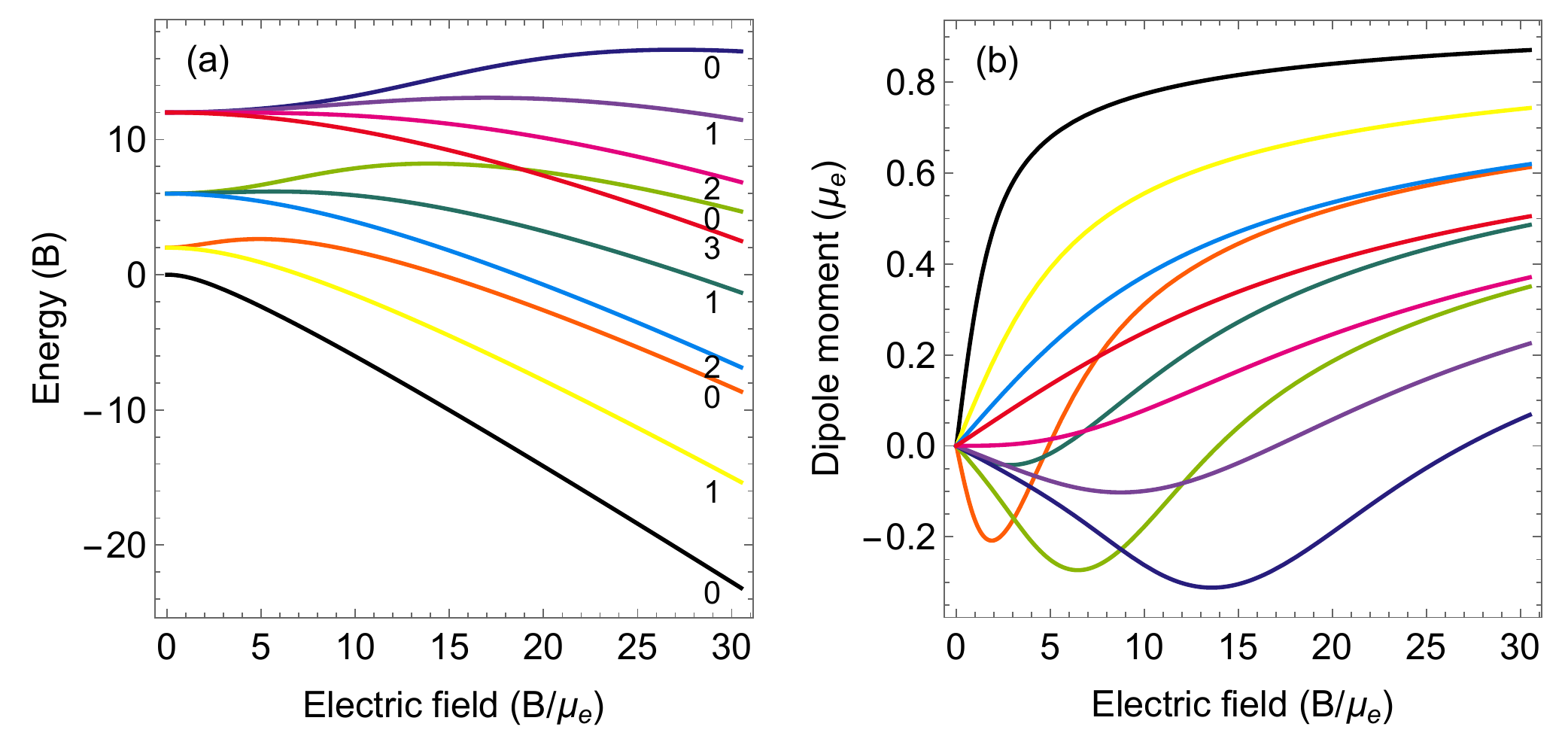}
    \caption{(a) Energy levels of a rigid rotor molecule as a function of applied electric field. The levels are labelled by $|m_N|$. (b) Corresponding dipole moments. }
    \label{fig:starkDC}
\end{figure}

The exchange interaction does not require any applied electric field. Adding an electric field polarizes the molecules, introducing a direct dipole-dipole interaction between them. Taking $\ket{0}$ and $\ket{1}$ to be the eigenstates in the presence of the electric field, $H_{\rm dd}$ now has diagonal matrix elements as well as the off-diagonal ones described above. To determine the induced dipole moments we first calculate the dc Stark shift of the molecule. For a molecule in a $\Sigma$ state, the Stark shift is dominated by mixing of rotational states and all other structure of the molecule can often be neglected. This is sometimes described as the Stark shift of a rigid rotor. In this simple case, the relevant Hamiltonian is 
\begin{equation}
    H_{\rm RR-Stark}=B \vec{N}^2 - \mu_{\rm e}{\cal E}\cos\Theta,
\end{equation}
 where $B$ is the rotational constant, $\vec{N}$ is the rotational angular momentum, $\mu_{\rm e}$ is the molecule-frame dipole moment, ${\cal E}$ is the applied electric field, and $\Theta$ is the angle between the internuclear axis and the electric field direction. Figure~\ref{fig:starkDC}(a) shows the eigenvalues of $H_{\rm RR-Stark}$ as function of ${\cal E}$. The energy, $W_{\rm Stark}$, is expressed in units of $B$, and the electric field in units of $(B/\mu_{\rm e})$, making this a universal plot. The states $\ket{N,m_N}$ and $\ket{N,-m_N}$ remain degenerate in the electric field, so each rotational manifold splits into $N+1$ components labelled by $|m_N|$. Figure~\ref{fig:starkDC}(b) shows the induced dipole moments of the various states, $\langle d \rangle = -d W_{\rm Stark}/d {\cal E}$, as a function of ${\cal E}$. Here, the dipole moments are in units of $\mu_{\rm e}$. At low fields, some of the dipoles are positive and some negative, while at high fields they tend towards $\mu_{\rm e}$ as the molecule becomes fully polarized along the field direction. When one molecule is in state $\ket{a}$ and the other in state $\ket{b}$ ($a,b \in \{0,1\}$), the diagonal elements of $H_{\rm dd}$ are
\begin{equation}
    \bra{a b}H_{\rm dd} \ket{a b} = \frac{\bra{a}d\ket{a}\bra{b}d\ket{b}}{4\pi\epsilon_0 r_{ij}^3}  (1-3\cos^2\theta_{ij}),
\end{equation}
where the dipole moments $\bra{a}d\ket{a}$, $\bra{b}d\ket{b}$ can be read from figure~\ref{fig:starkDC}(b).

Instead of using a static electric field, the dipole moments can also be induced using an oscillating field near resonant with the transition from $\ket{0}$ to $\ket{1}$. For molecules, this is most likely to be a microwave field close to resonance with a rotational transition. In the interaction picture, the new eigenstates are
\begin{equation}
    \ket{\pm} = N \left( \frac{-\Delta \pm \sqrt{\Delta^2 + \Omega^2}}{\Omega} \ket{0} + \ket{1} \right)
\end{equation}
with eigenvalues $\pm\frac{\hbar}{2}\sqrt{\Delta^2+\Omega^2}$. Here, $\Omega$ is the Rabi frequency, $\Delta$ is the detuning from resonance, and $N$ is the normalization factor. These states have dipole moments and the pair of molecules will interact through the dipole-dipole interaction. For example, if we take a microwave field polarized in the $z$ direction, and prepare both molecules in $\ket{+}$, or both in $\ket{-}$, $H_{\rm dd}$ has the diagonal elements
\begin{equation}
    \bra{\pm \pm} H_{\rm dd} \ket{\pm \pm} = \frac{|\vec{d}_{01}.\hat{z}|^2}{4\pi\epsilon_0 r_{ij}^3} \frac{\Omega^2}{\Omega^2 + \Delta^2} (1-3\cos^2\theta_{ij}).
\end{equation}
We see that the interaction can be controlled using the parameters of the microwave field. A more detailed discussion of how microwave fields can be used to control the dipole-dipole interaction between molecules, and how this control can be used to build quantum gates, is given in \citet{Hughes2020}.

\subsection{Quantum simulation}
\label{sec:simulation}

Understanding many-body quantum systems, and engineering new ones, is an important research area with applications in condensed matter physics, nuclear and particle physics, cosmology, chemistry and biology. These systems often exhibit remarkable phenomena, e.g. high temperature superconductivity, but they are difficult to understand because the systems found in nature are often difficult to control and impossible to simulate with a classical computer once more than a few interacting particles are involved.  Our understanding can be improved by engineering a well-controlled quantum system so that it emulates a many-body problem of interest. We derive a model Hamiltonian that captures the main features of the real system, and then build a physical system that implements this model. This is known as quantum simulation~\citep{Feynman1982,Deutsch1985}. Arrays of ultracold atoms and molecules are well suited to this task. Their benefits include control of the particle positions and interaction strengths, the number of accessible dimensions, whether real or synthetic~\citep{Sundar2018}, trap depths, and often the ability to address single particles.  

To date, much of the work in this field has used ultracold atoms trapped in optical lattices to investigate corresponding systems in condensed matter physics.  This has led, for example, to significant insight into the quantum phase transition between a superfluid and a Mott insulator~\citep{Greiner2002}, and models of antiferromagnetism~\citep{Mazurenko2017}.  Atoms interact through the van der Waals interaction, which is short range, or through magnetic dipole-dipole interactions, which are long range but usually quite weak. This limits the scope of accessible problems. By contrast, molecules interact through long-range electric dipole-dipole interactions that can be tuned using static or microwave electric fields, as discussed in section ~\ref{sec:dipole}.  

\begin{figure}[tb]
   \centering
    \includegraphics[width=0.6\linewidth]{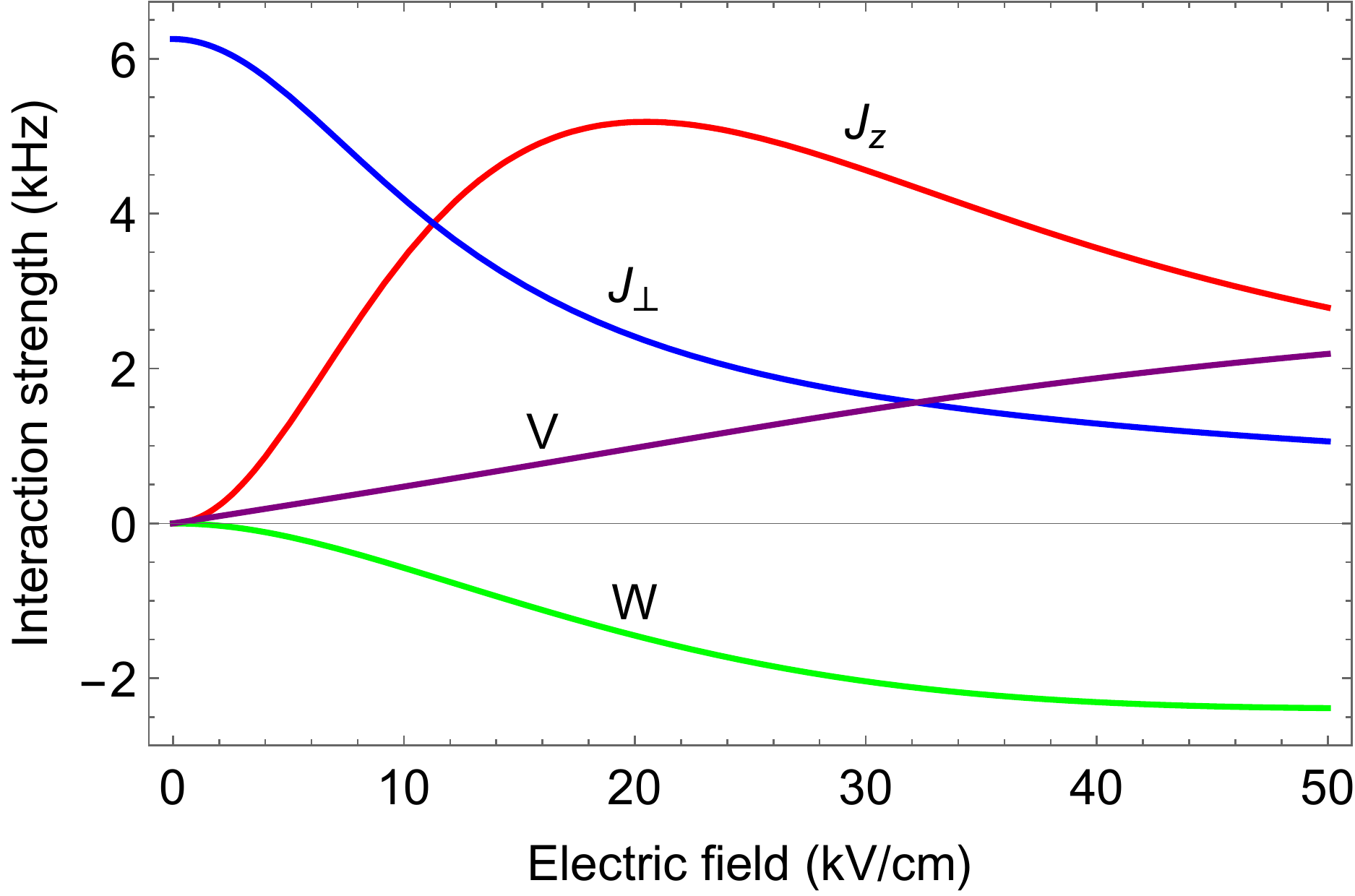}
    \caption{Strength of parameters appearing in the many-body Hamiltonian, equation (\ref{eq:manyBodyH}), as a function of electric field, for CaF molecules in a 1064~nm lattice. }
   \label{fig:manyBody}
\end{figure}

To glimpse the richness of arrays of interacting molecules, let us consider the simple case where we apply a dc electric field along $z$, neglect hyperfine and spin-rotation interactions, and focus only on the states $\ket{\uparrow}$ and $\ket{\downarrow}$ that correlate to the field-free states $\ket{N,m_N}=\ket{0,0}$ and $\ket{1,0}$. We can re-write $H_{\rm dd}$ from equation (\ref{eq:Hdd}) in terms of the spin-1/2 operators acting on molecule $i$, $S^z_i$, $S^{\pm}_i$, and then sum over all pairs of molecules in the lattice, to arrive at the many-body Hamiltonian~\citep{Gorshkov2011, Hazzard2014}
\begin{equation}
    H = \frac{1}{2}\sum_{i\ne j}V_{\rm dd} \left[ J_z S_i^z S_j^z + \frac{J_{\perp}}{2}(S_i^+S_j^- + S_i^- S_j^+) + W(S_i^z n_j + n_i S_j^z) + V n_i n_j\right].
    \label{eq:manyBodyH}
\end{equation}
Here, $n_i$ is the number operator at site $i$, $V_{\rm dd}=(1-3\cos^2\theta_{ij})/r_{ij}^3$, and we have re-expressed the positions of the molecules ($r_{i}$, $r_{j}$) in terms of the lattice constant $a$. The first term in equation (\ref{eq:manyBodyH}), with parameter $J_z$, describes an Ising-type spin-spin interaction, the second term (parameter $J_{\perp}$) describes the exchange of excitations, the third (parameter $W$) is a spin-density interaction where a molecule at one site creates an effective magnetic field along $z$ for a spin on another site, and the fourth term (parameter $V$) describes density-density interactions. We see that a lattice of polar molecules is an extremely rich system, even in this very simple case. The Hamiltonian is long range in 3D, and is anisotropic in both space and spin. In the case of unit filling, the $V$ term is a constant and the $W$ term is similar to a uniform magnetic field, and both can be neglected. The resulting Hamiltonian is the XXZ model. This reduces to the Ising model when $J_{\perp}=0$, to the XX model when $J_z=0$, and to the Heisenberg model when $J_{\perp}=J_z$.

The parameters of the model are related to molecular properties as
\begin{subequations}
    \begin{align}
    J_z &= \frac{(d_{\uparrow}-d_{\downarrow})^2}{4\pi\epsilon_0 a^3}, \\
    J_{\perp} &= \frac{2d_{\uparrow \downarrow}^2}{4\pi\epsilon_0 a^3}, \\
    W &= \frac{(d_{\uparrow}^2 - d_{\downarrow}^2)}{8\pi\epsilon_0 a^3}, \\
    V &= \frac{d_{\uparrow}+d_{\downarrow}}{16\pi\epsilon_0 a^3},
    \end{align}
\end{subequations}
where $d_{\uparrow} = \bra{\uparrow} d_z \ket{\uparrow}$, $d_{\downarrow} = \bra{\downarrow} d_z \ket{\downarrow}$ and $d_{\uparrow\downarrow} = \bra{\uparrow} d_z \ket{\downarrow}$.  Figure~\ref{fig:manyBody} shows how the parameters depend on electric field for CaF molecules in a 1064~nm lattice. The parameters can be significantly larger than single-particle decoherence rates, and their relative strengths are highly tunable, allowing the many-body physics to be explored across a large parameter space. 

Dipole-mediated spin exchange interactions have already been studied in a lattice of ultracold KRb molecules formed by atom association~\citep{Yan2013}, and the application of electric fields will allow other aspects of the model to be explored using this system. Other exotic quantum phases are also within experimental reach~\citep{Buchler2007,Cooper2009,Yao2013}.  As the phase-space density of laser-cooled molecules is increased towards quantum degeneracy (see section \ref{sec:chemistry}), they will be able to contribute to this field. They have much to offer, since they can be detected efficiently, state-selectively, and non-destructively by laser-induced fluorescence imaging, and often have large electric dipole moments. They also often have both electron and nuclear spin degrees of freedom, which provide even more richness and control and can be used to engineer almost any lattice spin model of interest~\citep{Micheli2006}. With the addition of tunnelling between sites, the system models an extended Hubbard Hamiltonian with long-range interactions~\citep{Wall2010, Wall2013}. The ability to explore these models with such control can help solve long-standing problems in many-body quantum physics.

\subsection{Quantum information processing}
\label{sec:information}

Currently, quantum information processing with molecules is nowhere near as advanced as for atoms and ions~\citep{Cirac1995,Wang2016,Ballance2016}. However, molecules offer many intriguing advantages, including long coherence times~\citep{Park2017}, very fast single qubit manipulation using microwave fields, fast two-qubit gates using dipole-dipole interactions, and the potential for electric dipole coupling of the molecules to photonic or solid-state systems. There are many ideas of how to make use of these advantages~\citep{Demille2002, Yelin2006,Andre2006,Karra2016, Ni2018,Yu2019,Albert2020,Sawant2020, Hughes2020}. We consider here two interesting approaches.

The first approach is to use molecules in arrays of tweezer traps. This is attractive because the arrays can be rearranged, pairs can be brought together to implement two-qubit gates, and single-site addressability and readout can be engineered using electric or magnetic field gradients, or by adjusting the laser intensity at a selected site to change the local ac Stark shift.  As discussed in section~\ref{sec:tweezer}, tweezer arrays of molecules have already been formed~\citep{Anderegg2019} and rearrangement protocols can follow those already established for atoms~\citep{Barredo2016}. Important next steps are to: (i) increase the trap frequencies by making smaller traps; (ii) cool the molecules to the motional ground state~\citep{Caldwell2020b}; (iii) control tensor Stark shifts in order to engineer long rotational coherence times in the trap~\citep{Seebelberg2018}; (iv) implement state-dependent traps that can bring pairs of molecules very close together in order to enhance the dipole-dipole interactions between them~\citep{Caldwell2020c, Caldwell2020d}.

A second interesting approach, expounded by \citet{Andre2006}, is to trap molecules close to the surface of a chip where they can be coupled to superconducting microwave resonators fabricated on the same chip, forming a monolithic hybrid quantum processor. The molecules can be trapped using microfabricated electric or magnetic surface traps. The microwave resonator, tuned into resonance with a rotational qubit in the molecule, facilitates transfer of quantum information between the molecule and a microwave photon and can be used to couple selected distant molecules together. Many of the elements needed to realize these ideas are established or emerging. The fabrication techniques needed to make microscopic traps are the same as those used to make atom chips, which are becoming a mature technology. Magnetic trapping of laser-cooled molecules is established, though increases in phase-space density will be needed to load molecules efficiently into microscopic traps. Superconducting resonators with high quality factors have been fabricated on chips and have been coupled to ground-state atoms through the magnetic dipole interaction~\citep{Hatterman2017} and to Rydberg atoms through the electric dipole interaction~\citep{Morgan2020}.

\subsection{Ultracold collisions, collisional cooling, and chemistry}
\label{sec:chemistry}

Another research area where laser-cooled molecules are useful, and where they may enable entirely new areas of study, is the investigation of interactions of molecules at close range, i.e. collisions and chemical reactions~\citep{Heazlewood2021b}.  The control of these interactions through choice of quantum state, molecular orientation, and applied field have been investigated by many authors~\citep{Krems2005, Krems2008, Ospelkaus2010, Bohn2017, Hu2020}. 

A full understanding of the quantum chemistry of molecules often requires an in-depth investigation of their elastic (momentum changing), inelastic (quantum state changing), and reactive (chemical species changing) collisions. 
Measurements of elastic and inelastic cross sections can be sensitive probes of the multi-dimensional potential energy surfaces, which are inherently difficult to calculate accurately. In the ultracold regime, experiment and theory come together in an exciting interplay that is revealing the quantum-mechanical underpinnings of particle interactions.  At normal temperatures, molecules populate a large number of internal quantum states and interactions can often energetically access a vast array of possible pathways between the reactant and product states.  When cold, only a few quantum states are populated and many reaction pathways become energetically inaccessible, proceeding instead via quantum tunneling, entanglement, or resonances~\citep{Chandler2010,Richter2015,Toscano2020}.  In the ultracold limit, interactions involve only one or a few partial waves.  For simple systems consisting of, e.g., pairs of alkali-metal atoms or light atom-molecule systems, it is possible to solve the multi-dimensional Schr\"{o}dinger equation numerically using coupled-channel methods.  For heavier or more complex systems, or at higher temperature, the number of channels needed for convergence quickly becomes computationally prohibitive.  Remarkable progress has been made in bi-alkali systems by instead formulating effective single-channel models based on quantum defect theory~\citep{Idziaszek2010,Gao2010,Kotochigova2010,Frye2015,He2020b}.  Amazingly, by incorporating only a couple of free parameters, such models can often predict dynamical rates and the positions and strengths of collisional resonances in the presence of external electromagnetic fields.  Even more striking is the emergence of universal behaviours in these complex and apparently distinct systems~\citep{Idziaszek2010,Julienne2011}.  Expanding these models to predict the behavior of new ultracold species produced using laser cooling is a likely next step.

Collisional cooling will almost certainly be needed to increase the phase-space density of laser-cooled molecules towards quantum degeneracy. One option is sympathetic cooling with ultracold atoms~\citep{Lim2015, Son2020}, and another is evaporative cooling of the molecules~\citep{Valtolina2020}. In both cases the ratio of elastic to inelastic collision rates is critical - the elastic collisions are needed for thermalization, whereas inelastic collisions typically lead to trap loss. To establish good collisional cooling routes it is important to study these elastic and inelastic processes and learn to control them. There has been a great deal of theoretical work on sympathetic cooling of molecules~\citep{Soldan2002, Krems2003, Lara2006, Wallis2011, Gonzalez-Martinez2011,Tscherbul2011}, and recent experiments in this direction have begun to study collisions between laser-cooled molecules and atoms~\citep{Jurgilas2021}. Evaporative cooling of molecules can be assisted by using an electric field to introduce dipolar interactions that enhance elastic collision rates. Indeed, when the dipoles become large enough, the scattering enters a universal regime and the elastic cross section can be exceptionally large~\citep{Bohn2009}. Amazingly, inelastic and reactive collisions are very strongly suppressed at electric fields where the energy of a pair of molecules prepared in one configuration (e.g. both in $(N,m_N)=(1,0)$) crosses that of a different configuration (e.g. one in $(0,0)$ and the other in $(2,0)$). As the molecules approach from any angle, the dipole-dipole interaction turns the crossing into an avoided crossing, resulting in a large repulsive barrier at long range, which prevents the molecules getting close enough for inelastic or reactive collisions. This electric shield has been studied theoretically~\citep{Avdeenkov2006, Quemener2016} and recently demonstrated experimentally~\citep{Matsuda2020}. Theoretical work also shows how to shield against unwanted collisions using a microwave field~\citep{Karman2018}. A remarkable recent experiment with CaF molecules in tweezer traps has demonstrated such microwave shielding, showing a reduction in inelastic loss by a factor of six~\citep{Anderegg2021}. 

A striking example of experimental progress in ultracold chemistry, is the controlled formation of a single molecule from two individual atoms~\citep{Liu2018,Zhang2020,He2020}.  Another example is the complete mapping of a chemical reaction, including direct observations of reactants, transient intermediate complexes, and products~\citep{Hu2019}.  A third is the study of collisions between laser-cooled molecules in optical tweezers at the single particle level~\citep{Cheuk2020}.   Laser cooling  diversifies the set of species and properties available for studying chemistry at ultracold temperatures. Laser cooling of chemically important diatomic species, such as CH and NH, and more complex polyatomic species, will enrich this topic. It may eventually be possible to fully understand and control the emergence of chemical complexity in cold, low density environments like those found in the interstellar medium and planetary atmospheres~\citep{Smith2000,Herbst2009}.

\subsection{Probing fundamental physics}
\label{sec:fundamental}

Laser cooled molecules can be used to search for new particles and forces beyond the standard model of particle physics~\citep{DeMille2017, Safronova2018}.  This can be done by making ultra-precise measurements that test the discrete symmetries of nature -- charge conjugation (C), parity (P), and time-reversal (T).  Laser-cooled molecules can greatly increase the precision of such measurements.  A great example is the ongoing search for the electric dipole moment (EDM) of the electron, which is associated with new sources of charge-parity (CP) violation required to explain the observed matter / anti-matter imbalance in the Universe.  CP violation in the standard model fails, by many orders of magnitude, to explain this imbalance. Many extensions of the standard model introduce new sources of CP violation which naturally grant the electron an electric dipole moment large enough to measure.  Precise measurements using heavy dipolar molecules are one of the best ways to verify, or eliminate, these theories experimentally~\citep{Hinds1997}.  Intriguingly, state-of-the-art experiments~\citep{Cairncross2017,Andreev2018} now probe deep into the parameter space of many standard model extensions.  Currently, the most sensitive experiments use molecules that combine the inherent sensitivity of a heavy atom together with the high polarizability of a polar molecule.  All currently use molecules at temperatures above 1~K. Laser cooling can hugely increase the coherence time in these experiments, and can facilitate entirely new experimental techniques~\citep{Tarbutt2013, Aggarwal2018,Lim2018,Augenbraun2020,Fitch2020b, Hutzler2020}.  Such experiments are projected to increase experimental sensitivity to the EDM by one or two orders of magnitude over the coming decade.  

A fascinating new avenue is the use of laser cooled polyatomic molecules for testing T-violating effects such as electric dipole or magnetic quadrupole moments~\citep{Augenbraun2020,Hutzler2020,Maison2019,Denis2020}.  Molecules such as YbOH~\citep{Kozyryev2017b} and HgOH~\citep{Mitra2021} have high sensitivity to T-violating effects, are laser coolable, and provide robust control of systematic errors.  Due to their closely spaced parity doublets ($l$-doubling, see section~\ref{sec:polyatomics}) these molecules can be fully polarized in a small electric field and provide internal co-magnetometry. Experiments with these molecules may ultimately be limited only by the available coherence time, which for YbOH is $\approx$1~s due to the relatively high energy and anharmonicity of the $l$-doublet used~\citep{Augenbraun2020c}. Even higher sensitivities may be accessible by using systems with lower energy parity doublets, such as isoelectronic \emph{nonlinear} polyatomic molecules like YbOCH$_{3}$~\citep{Augenbraun2020c}.  Here, parity doublets arising from rigid-body rotation ($K$-doublets) are even more closely spaced and have lower overall energies, leading to longer coherence times and full polarization in even smaller electric fields.  Another benefit is that the chemical substitution of H by CH$_{3}$ is expected to improve the Franck-Condon factors at the expense of only slightly increasing the mass, which may make laser cooling easier.  Measurements of electric dipole moments with these systems may ultimately probe new physics with equivalent energies at the PeV scale~\citep{Kozyryev2017}.

Molecules are also being used to measure the nuclear anapole moment, a signature of P-violation due to interactions in nuclei involving the weak force~\citep{Cahn2014,Altuntas2018}. These experiments could also benefit from laser cooling. Also of interest is the parity-violating energy difference between left- and right-handed chiral molecules, which was predicted long ago but has never been measured~\citep{Quack2008,Tokunaga2013}.  Noting the recent progress with large molecules, it now seems feasible to apply laser cooling to chiral species~\citep{Isaev2018,Augenbraun2020b}.

Molecules can also be used to test the hypothesis that fundamental constants may depend on time, space, or the local density of matter. Such variations are predicted by grand-unified theories and by some theories of dark matter and dark energy~\citep{Olive2008,Uzan2011}.  Transitions in molecules depend mainly on two fundamental constants, the fine-structure constant $\alpha$ and the proton-to-electron mass ratio $m_{\rm p}/m_{\rm e}$.  For example, rotational and vibrational transition frequencies scale as $(m_{\rm p}/m_{\rm e})^{-1}$ and $(m_{\rm p}/m_{\rm e})^{-1/2}$, respectively.  Furthermore, when the transition energy results from a near cancellation between two large contributions of different origin, certain transitions will exhibit even greater sensitivities~\citep{Chin2009}.  Laboratory experiments can be compared to astronomical observations to probe variations on cosmological scales~\citep{Safronova2018}, as has been done using beams of OH~\citep{Hudson2006} and CH~\citep{Truppe2013}.  Ultra-precise laboratory measurements over a timescale of a few years test the present-day drift or oscillation of the constants. To date, the most stringent limits come from atomic clock measurements~\citep{Shwarz2020, Lange2021}.
Clocks based on transitions in ultracold molecules are currently under development. For example, a direct limit on the time variation of $m_{\rm p}/m_{\rm e}$ has been obtained using ultracold KRb molecules~\citep{Kobayashi2019}, a fountain of ultracold NH$_{3}$ molecules has been demonstrated, and a clock based on a vibrational transition in Sr$_{2}$ molecules in an optical lattice is being developed~\citep{Kondov2019}.  Clocks based on vibrational transitions in laser-cooled molecules are an attractive complementary approach~\citep{Kajita2018}.

\section{Concluding remarks}

The extension of laser cooling and trapping methods from atoms to molecules has made huge progress over the last decade. The field continues to expand rapidly and is moving from the development of techniques towards many exciting applications. We anticipate a great number of developments over the next few years. These are likely to include new methods to slow molecules more efficiently, a variety of new species being cooled and trapped, increases in phase-space density by collisional cooling together with shielding of unwanted collisions, sideband cooling to bring molecules to the ground-state of a trap, control of dipole-dipole interactions to entangle molecules, improvements to hyperfine and rotational coherence times in magnetic and optical traps, small-scale quantum simulations, and the first tests of fundamental physics using laser cooled molecules. The continued flourishing of this field will help to link molecular physics with numerous other sciences including quantum science and technology, condensed matter physics, chemistry, metrology, high energy physics, and cosmology. 

\section*{Acknowledgements}

We thank H\'el\`ene Perrin, Daniel McCarron, Stefan Truppe, and Ben Sauer for their careful reading and helpful feedback on this manuscript.
We acknowledge generous support from EPSRC (grant EP/P01058X/1), STFC (grant ST/S000011/1), the John Templeton Foundation (grant 61104), the Gordon and Betty Moore Foundation (grant 8864), the Alfred P. Sloan Foundation (grant G-2019-12505), and the Royal Society.

\bibliography{references}

\end{document}